%% file: QweakNIM_pdffigs.tex
\documentclass[twocolumn,3p,number,sort&compress,longtitle,letterpaper]{elsarticle}

\usepackage{natbib}
\usepackage{amsmath,amssymb,amsthm}

\usepackage{lineno}
\usepackage{hyperref}
\hypersetup{colorlinks, linkcolor=red}

\usepackage{epstopdf}

\usepackage[printonlyused,footnote,nohyperlinks]{acronym}
\usepackage{multirow}
\usepackage{color}
\usepackage{graphicx}
\usepackage{latexsym}

\usepackage{epsfig,lineno,textcomp,subfigure,units}

\newcommand{\qweak}{${Q \rm{_{weak}}}$ }
\newcommand{\moller}{M\o ller}
\newcommand{\muA}{$\mu $A }  

\newcommand{\Qweak}{$Q{\rm{_{weak}}}$ } 
\def\s2tw{$\mathrm {\sin^2 \theta _w}$}

\def\Qwp{${Q^{p}_w }$}

\newcommand{\cer}{\v{C}erenkov }
\newcommand{\expec}[1]{\langle #1 \rangle}
\newcommand{\range}[3]{#1 $\leq$ #2 $\leq$ #3} 

\begin{document}

\begin{frontmatter}

\title{The  \Qweak Experimental Apparatus \\
}


\input{authors}

\clearpage
\newpage
\input{abstract}

\begin{keyword}
parity violation \sep electron scattering \sep high luminosity \sep
liquid hydrogen target  \sep particle detectors
\end{keyword}

\end{frontmatter}

\clearpage
\tableofcontents
\vspace{1cm}


\input{acronyms}


\section{Introduction}
\label{sec:Intro}

\subsection{Motivation}
\label{sec:Motivation}

The \qweak\ experiment was designed~\cite{proposal} to perform the first determination~\cite{QweakPRL} of the proton's weak charge, \Qwp,  which is the neutral-weak analog of the proton's electric charge. While measurement of this fundamental property of the proton is interesting in its own right, it can also be related~\cite{Erler1} to the weak mixing angle \s2tw and will provide the most sensitive measure of the $Q^2$ evolution (running) of \s2tw below the $Z$-pole. As such it provides a sensitive test for new physics beyond the 
\ac{acro-sm}. When combined with precise experiments on other targets, such as can be found in atomic parity-violating  measurements on $^{133}$Cs~\cite{APV1}, the axial electron, vector quark coupling constants   $C_{1i} = 2 g^e_A g^i_V$ can be extracted and used (for example) to form the weak charge of the neutron as well~\cite{QweakPRL}. 

To accomplish these goals, a precise measure of the \ac{acro-pves} 
asymmetry $A_{ep}$  from unpolarized hydrogen must be performed at low 4-momentum transfer squared ($-Q^2 $). The asymmetry $A_{ep}$ is the difference over the sum of elastic $\vec{e}p$ cross sections measured with longitudinally polarized electrons of opposite helicity. After a small correction for the one energy-dependent electroweak radiative correction that contributes at forward angles~\cite{WallyNew}, in the 
forward-angle limit the asymmetry can be cast in the simple form

\begin{equation}
A_{ep} / A_0 = Q_{W}^{p}+Q^{2}F^{p}(Q^{2},\theta) ,
\label{asymmetry_eqn}
\end{equation}
where
$A_0 = -G_{F}Q^{2} / \left( {4\pi\alpha\sqrt{2}} \, \right)$, 
 $G_F$ is the Fermi constant, and $\alpha$ the fine structure constant. 
The second term $Q^2 F^p$ contains the nucleon structure defined in terms of electromagnetic, neutral-weak, and axial form factors. $F^p$ can be determined~\cite{2007C1paper} from existing \ac{acro-pves} data at modestly higher $Q^2$~\cite{SAMPLEbkwrd, SAMPLE, Happex1, Happex1p1, Happex1p2, Happex2He, Happex3, G0forward, G0backward, PVA41, PVA42, PVA43}, and is 
suppressed at lower $Q^2$ relative to \Qwp\; by the additional factor of $Q^2$. 

The strategy employed in the \Qweak experiment was to perform the most precise $\vec{e}p$  asymmetry measurement to date~\cite{QweakPRL} at a $Q^2$ four times smaller than any previously reported $\vec{e}p$ experiment, to ensure a reliable (short) extrapolation to threshold where the intercept \Qwp\; of Eq.~\ref{asymmetry_eqn} is the quantity of interest. As mentioned above, the nucleon structure term is also smaller at smaller $Q^2$. The fundamental challenge is that the expected \ac{acro-sm}
asymmetry at the small $Q^2$ of the experiment (0.025 GeV$^2$) is only $-230$ ppb, and the proposed goal of a $\sim$2.5\% asymmetry measurement implies that the experiment must achieve an overall uncertainty of scale 6 ppb. The small $Q^2$, small asymmetry, and ambitious uncertainty goal led to an experiment which pushed boundaries on many fronts, as described below.

\subsection{Overview of the Experiment}
\label{sec:Expt}

\begin{figure*}[!hhhbtb]
\begin{center}
\includegraphics[width=\textwidth]{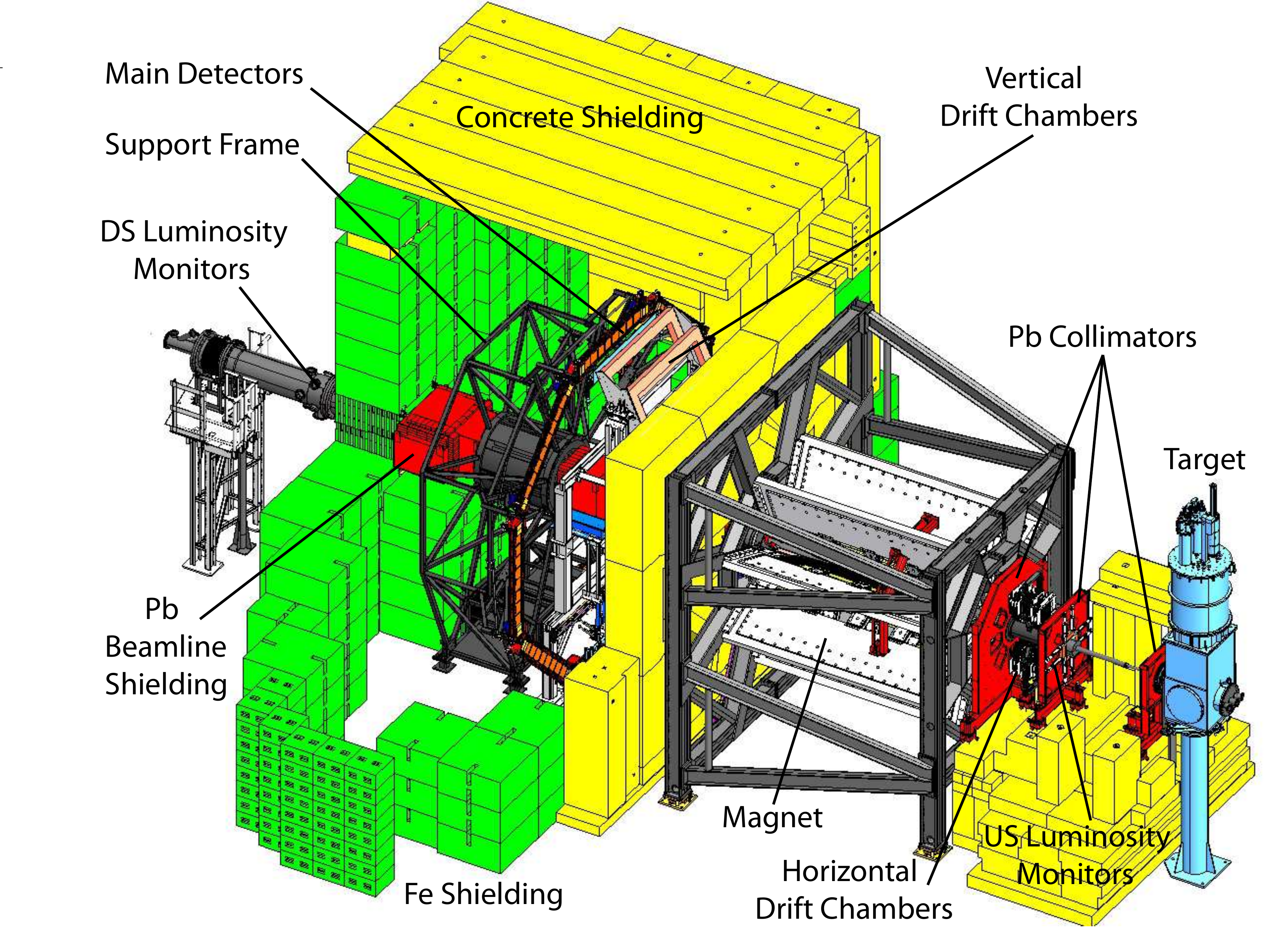}
\end{center}
\caption{\label{fig:QweakApparatus}
\ac{acro-cad}  view of the experimental apparatus. The beam is incident from the right. The key elements include the liquid hydrogen target scattering chamber, a triple collimator system, a resistive 8-fold symmetric toroidal magnetic spectrometer, and eight \v{C}erenkov detectors for the measurement of scattered electrons.  Tracking chambers are illustrated
just upstream of the \v{C}erenkov detectors, as well as just upstream of the third collimator.
Portions of the extensive steel and concrete shielding are also shown. The experiment had two operating modes: a low-current calibration mode for $Q^2$ acceptance mapping and background measurements, and a high-current production mode for the asymmetry measurement. The tracking detectors were only used during low current running (and were retracted during high current running). 
}
\end{figure*}

\begin{figure*}[!hhhbtb]
\begin{center}
\includegraphics[angle=-90,width=\textwidth]{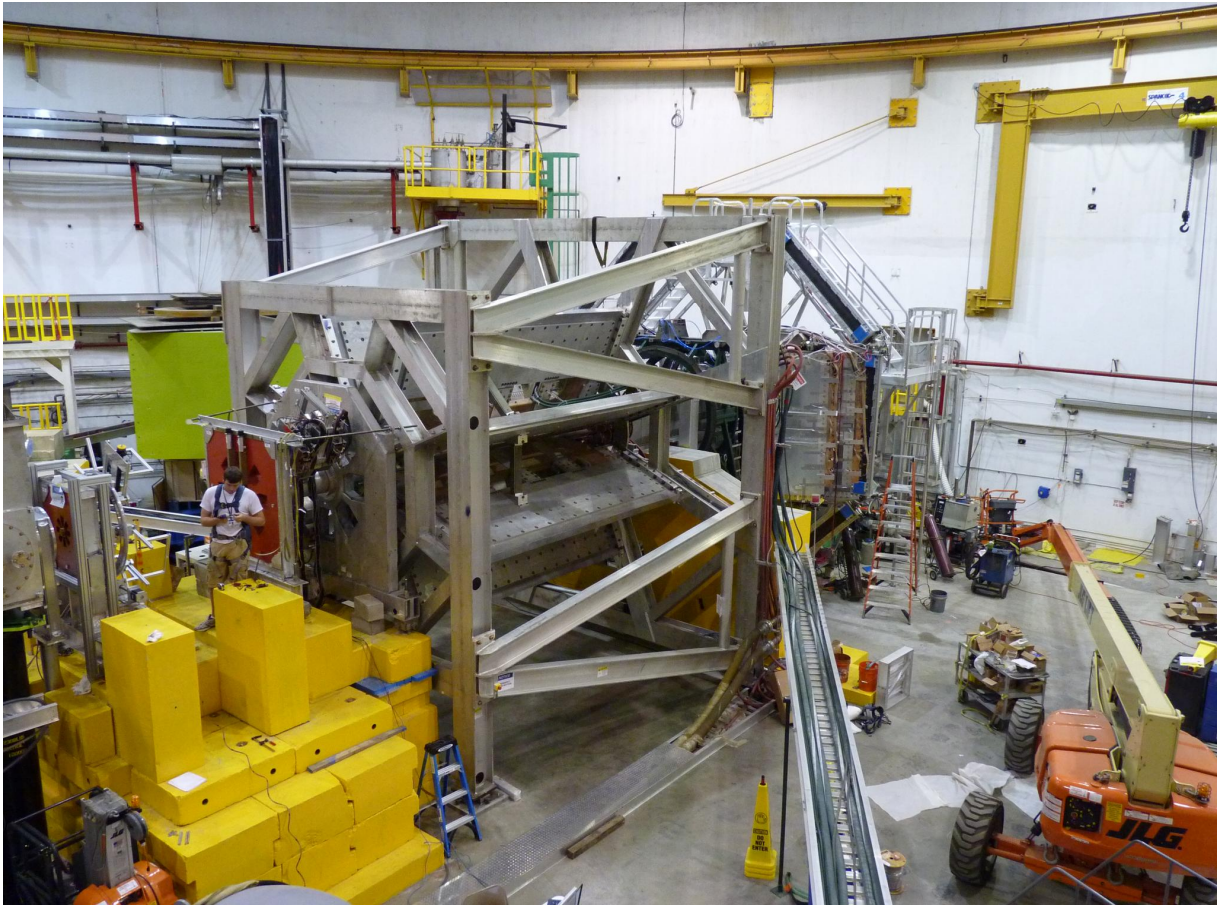}
\end{center}
\caption{\label{fig:Installation}
The \Qweak experimental apparatus during installation, before most of the equipment was covered with shielding. The target scattering chamber is on the left, followed by the first two lead collimators on either side of the man in the figure. A third lead collimator faced with aluminum sits just in front of the   large frame in the center of the picture, which supports the eight magnet coils.  
Three of the eight quartz detector bars appear just downstream and to the right of the magnet as skinny, long 
bars. }
\end{figure*}

A custom apparatus (see Figs.~\ref{fig:QweakApparatus} and \ref{fig:Installation}) was built and installed in Hall C at \ac{acro-jlab}~\cite{CEBAF} to provide the high luminosity, large acceptance, and systematic control required for the \Qweak experiment. Several improvements (see Sec.~\ref{sec:Source}) were made in the accelerator's source and injector to meet the requirements of the experiment, which  
employed a 180 $\mu$A beam of 1.16 GeV, $\sim$89\% longitudinally polarized electrons. Improvements to the beamline instrumentation were made in the polarized source and the Hall C beamline (see Sec.~\ref{sec:Beam:Beamline}). The incident beam polarization was measured in two independent polarimeters (Sec.~\ref{sec:Components:Polarization}).

 Electrons scattered from a 34.4 cm liquid hydrogen target (Sec.~\ref{sec:Target}) were detected in eight synthetic quartz \v{C}erenkov detectors each 200 cm $\times$ 18 cm $\times$ 1.25 cm thick (Sec.~\ref{sec:maindetectors}) arrayed in an azimuthally symmetric pattern about the beam axis, which covered 49\% of 2$\pi$ in the azimuthal angle $\phi$.  
The eight-fold azimuthal symmetry minimized and helped characterize effects arising from \ac{acro-hc} beam motion as well as residual transverse polarization in the beam. 
A carefully tailored  triplet of lead collimators (Sec.~\ref{sec:Components:Collimator}) restricted the scattering angular acceptance to \range{5.8$^\circ$}{$\theta$}{11.6$^\circ$} and suppressed backgrounds. A resistive toroidal magnet (Sec.~\ref{sec:Components:Spectrometer}) between the target and the detectors separated the elastic electrons from inelastic and M\o ller electrons. In conjunction with the collimation system,  the magnet also separated  elastically scattered electrons from  direct line-of-sight (neutral) events originating in the target. 

A number of ancillary detectors helped characterize backgrounds and establish \ac{acro-hc} beam properties in the experiment. 
A tracking system (Sec.~\ref{sec:Tracking}) consisting of drift chambers before and after the magnet was deployed periodically to verify the acceptance-weighted central kinematics of the measurement, and to help
study backgrounds. The electronics and data acquisition system are described in Sec.~\ref{sec:Components:DAQ}. The extensive simulations performed for the experiment, as well as the analysis scheme, are described in Sec.~\ref{sec:Software}.
The parameters of the experiment are summarized in Table~\ref{tab:kinematics}.

\begin{table}[hbtbp]
\begin{center}
\begin{tabular}{ll}
Quantity & Value   \\ \hline
Beam energy & 1.16 GeV \\
Beam polarization & 89\% \\
Target length & 34.4 cm \\
Beam current & 180 $\mu$A \\
Luminosity & 1.7x10$^{39}$ cm$^{-2}$ s$^{-1}$ \\
Beam power in target & 2.1 kW \\
$\theta$ acceptance& $5.8^\circ - 11.6^\circ$ \\
$ \phi$ acceptance & $49$\% of $2\pi$  \\
$Q^2$ & 0.025 GeV$^2$ \\
$\Delta \Omega_{\rm elastic}$ & 43 msr  \\
$\int |\vec B| dl$ & 0.9 T $\cdot$ m  \\
Total detector rate  & 7 GHz \\
\end{tabular}
\end{center}
\caption{Typical parameters characterizing the second half of the experiment.}
\label{tab:kinematics}
\end{table}

\subsection{Asymmetry Considerations}
\label{sec:Asymmetry}

The current from the 
\ac{acro-pmt} at each end of the eight quartz \cer detectors was read out, converted to a voltage, and digitized. 
The raw asymmetry measured for a given detector \ac{acro-pmt} is provided by the following expressions:
\begin{eqnarray}
A_{\rm raw} & = & \frac{Y^{+}-Y^{-}}{Y^{+}+Y^{-}} \label{eqn:Ayield} \\ 
 & = & P\left( \frac{f_p}{R} A_{ep} + \sum_b f_b{A_b} \right) \nonumber \\
 & & + A_{\rm beam} + A_T + A_{\epsilon}.
\label{eqn:AsyMeas}
\end{eqnarray}
Here $Y^{\pm}$ is the integrated signal yield seen in a given \ac{acro-pmt}  for a right-handed ($+$) or left-handed ($-$)
electron beam helicity state, 
normalized to the measured beam charge. $A_{ep}$ is the elastic $\vec{e}p$ asymmetry the experiment was designed to provide.
The factors $f_p = 1-\sum_b f_b $ and 
$f_b = {\expec{Y_b}}/{\expec{Y}}$ represent
the fractional contributions (dilutions) of  elastic $ep$ events and  background events to the total yield, respectively.
 $P$ is the beam polarization, the $A_b$ are various background asymmetries, and $A_{\rm beam} = A_{\rm beam}(E,X,Y,X^\prime,Y^\prime)$ is the false asymmetry due to \ac{acro-hc} changes in the beam properties. The latter includes yield changes due to beam position $(X,Y)$, beam angle $(X^\prime, Y^\prime)$, and beam energy $(E)$ on target. The  beam charge asymmetry was 
 reduced with an active feedback loop. 
$A_T$ accounts for potential contributions from transverse polarization components in the nominally longitudinally polarized beam. 
The $A_{\epsilon}$ term accounts for 
electronic contributions from potential helicity signal leakage into the 
\ac{acro-daq} electronics or the detector signal pedestal. 
As described in~\cite{QweakPRL}, the factor $R$ accounts for small 
corrections due to the effects of bremsstrahlung, light variation and nonuniform $Q^2$ distribution across the detectors, the finite precision of the $Q^2$ determination, and transforming from $\langle A(Q^2 )\rangle$ to $A(\langle Q^2 \rangle)$. 

Using Eq.~\ref{eqn:AsyMeas} as a basis, the experiment was designed to make precise measurements of the beam polarization, the momentum transfer, and the scattered electron yield. The experiment was also designed to highly suppress backgrounds and \ac{acro-hc} electronic and beam effects.
Components were included in the experiment to allow ancillary measurements of background asymmetries and yields, as well as \ac{acro-hc} beam properties.

\begin{figure}[!hhhbtb]
\begin{center}
\includegraphics[width=0.51\textwidth]{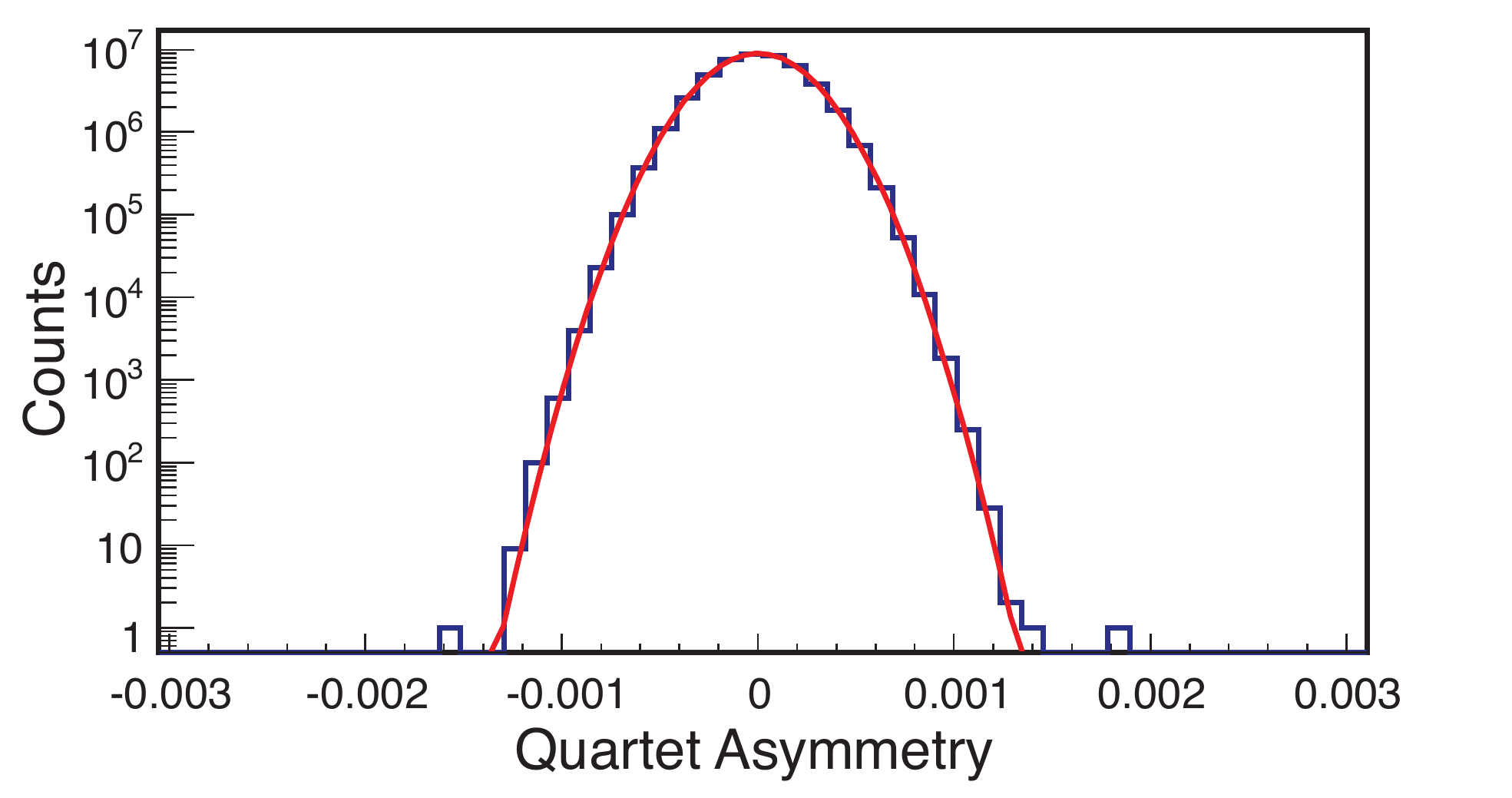}
\end{center}
\caption{\label{fig:DeltaA}
Measured asymmetry (blinded) of helicity quartets
   (see text) accumulated over several days.
 The smooth curve is a Gaussian fit, with 
 an \acs{acro-rms} width of 230 ppm. }
\end{figure}

The  statistical accuracy $\Delta A/A$ achievable in the experiment  can be expressed in terms of the asymmetry width $\sigma_A$ measured over helicity quartets, the total number of helicity quartets $N$, the expected asymmetry $A$, and the beam polarization $P$ according to 
\begin{equation}
\frac{\Delta A}{A} = \frac{\sigma_A}{ A P \sqrt{N}}.
\end{equation}
Helicity quartets refer to the pattern of beam helicity states used in the experiment: either $(+--+)$ or $(-++-)$. Assuming an efficiency of 50\%, and the helicity reversal rate of 960/s used in the experiment, $N$ is about $10^7$ quartets per day. The asymmetry width is the sum in quadrature of  contributions from the statistics per quartet 
accumulated in the detectors (215 ppm corrected for the 70 $\mu$s helicity reversal switching time, 42 $\mu$s gate delay, and 10\% detector resolution), the beam current monitor resolution ($\sim$43 ppm), and the  width from noise (density fluctuations) in the liquid hydrogen target ($\sim$55 ppm). $\sigma_A$ was typically 225$-$230 ppm in the experiment (see Fig.~\ref{fig:DeltaA}), and was dominated by counting statistics. Under these conditions 270 days are required to reach a statistical accuracy $\Delta$A/A of 2.1\%. 

\subsection{Beam Helicity Reversal}
\label{sec:Helicity}

A crucial tool to suppress systematic effects and facilitate a clean extraction of the physics asymmetry in the \Qweak  experiment was fast helicity reversal. Since the detector signals can, and do, 
 fluctuate, the faster the helicity reversal, the more accurate is the description of the experimental apparatus as a linear measurement device. Possible signal changes include slow gain drifts, target density fluctuations, and beam drifts. 

The performance of the Pockels cell used to reverse the beam helicity in the injector (see Sec.~\ref{sec:pockel}) was upgraded for the \qweak  experiment. The helicity switching time was reduced from 500 $\mu$s to 70 $\mu$s and the helicity reversal rate was increased from 30 to 960 per second. 
A quartet helicity reversal pattern was used to remove  linear drifts in the detector signals, and the fast reversal made the approximation of relatively slow random fluctuations as linear drifts valid. 

The 
asymmetry was calculated for each
helicity quartet using Eq.~\ref{eqn:Ayield}. 
When averaged over long time periods 
the small remaining asymmetries due to non-linear drifts should have random signs and average out. 
Any remnant of these drifts and fluctuations (along with other sources of unwanted excess noise) would increase the
width of the measured asymmetries. Therefore,
the health and efficiency of the experiment can be assessed by examining
the difference between the observed asymmetry width, and that expected from the sum in quadrature of counting statistics, the beam current monitors, and the target. 

\subsection{Logistics of the Experiment}
\label{sec:Logistics}

The experiment 
was performed in two very different modes.  The primary measurement of the $\vec{e}p$ asymmetry utilized beam currents up to 180 $\mu$A with corresponding rates in each of the eight quartz detector bars of almost 0.9 GHz. 
The quartz bars were read out with \ac{acro-pmt}s on each end of each bar. The \ac{acro-pmt}s were fitted with low-gain bases  and the current from the bases was integrated. This part of the experiment is referred to as integrating  
mode.

A number of smaller blocks of time dispersed over the main measurement were devoted to what is referred to below as event  
mode, which was devoted to measurement of $Q^2$ and background characterizations. During this portion of the experiment, the beam current was reduced over six orders of magnitude to 50$-$100 pA, and tracking chambers (horizontal and vertical drift chambers) were inserted into the scattered electron acceptance. Trigger scintillators were also placed in front of the main detector quartz bars. High-gain bases were installed on the quartz bar \ac{acro-pmt}s to permit counting of individual pulses. The event-mode electronics and \ac{acro-daq} were distinct from those used in integrating mode. 

The experiment was also divided into distinct data collection periods. During an initial setup period of several months, various parts of the experiment were debugged. A short commissioning period (early Feb. 2011) took place once all the equipment was finally in place and functioning. Those results are presented in ~\cite{QweakPRL}. Following this a period of several months of production running from February-May 2011  (referred to below as Run 1) took place. This was followed by a scheduled six month accelerator down period during which some parts of the experiment were opportunistically improved in response to the lessons learned from Run 1. After that a highly efficient  period (Run 2) from November 2011 to May 2012 occurred during which most of the data for the primary (integrating mode) measurement were acquired. It is the configuration of the experiment that existed during this final Run 2 period that is described in this article.

\section{Polarized Source}
\label{sec:Source}

Parity-violation experiments have higher demands from the accelerator source than typical experiments performed at \ac{acro-jlab}, with \Qweak  being the most demanding to date~\cite{Poelker:2008zz}.  In fact, the polarized source is considered part of the experimental apparatus due to the stringent (nanometer scale) requirements placed on 
 \ac{acro-hc} differences in beam parameters.
In addition, the \Qweak experiment needed a higher helicity reversal rate and a higher beam current 
than previous experiments.  These requirements led to the development of a new high-voltage switch for the Pockels cell that could provide spin flipping at 960/s~\cite{poelker} and construction of a new higher-bias-voltage photogun~\cite{adderley}. 

Conceptually, the system is rather simple~\cite{sinclair}.  Circularly polarized laser light is incident on a photocathode, producing electrons that are accelerated in an electrostatic field.  The helicity of the photons is transferred to the electrons. A schematic   of the polarized source in the context of the accelerator and experimental hall is shown in  Fig.~\ref{fig:sourcetable}.

\subsection{Helicity Signal}
\label{sec:Source:HelicitySignal}

\begin{figure*}[!hbtbhbtb]
\centerline
{\includegraphics[width=\textwidth,angle=0]{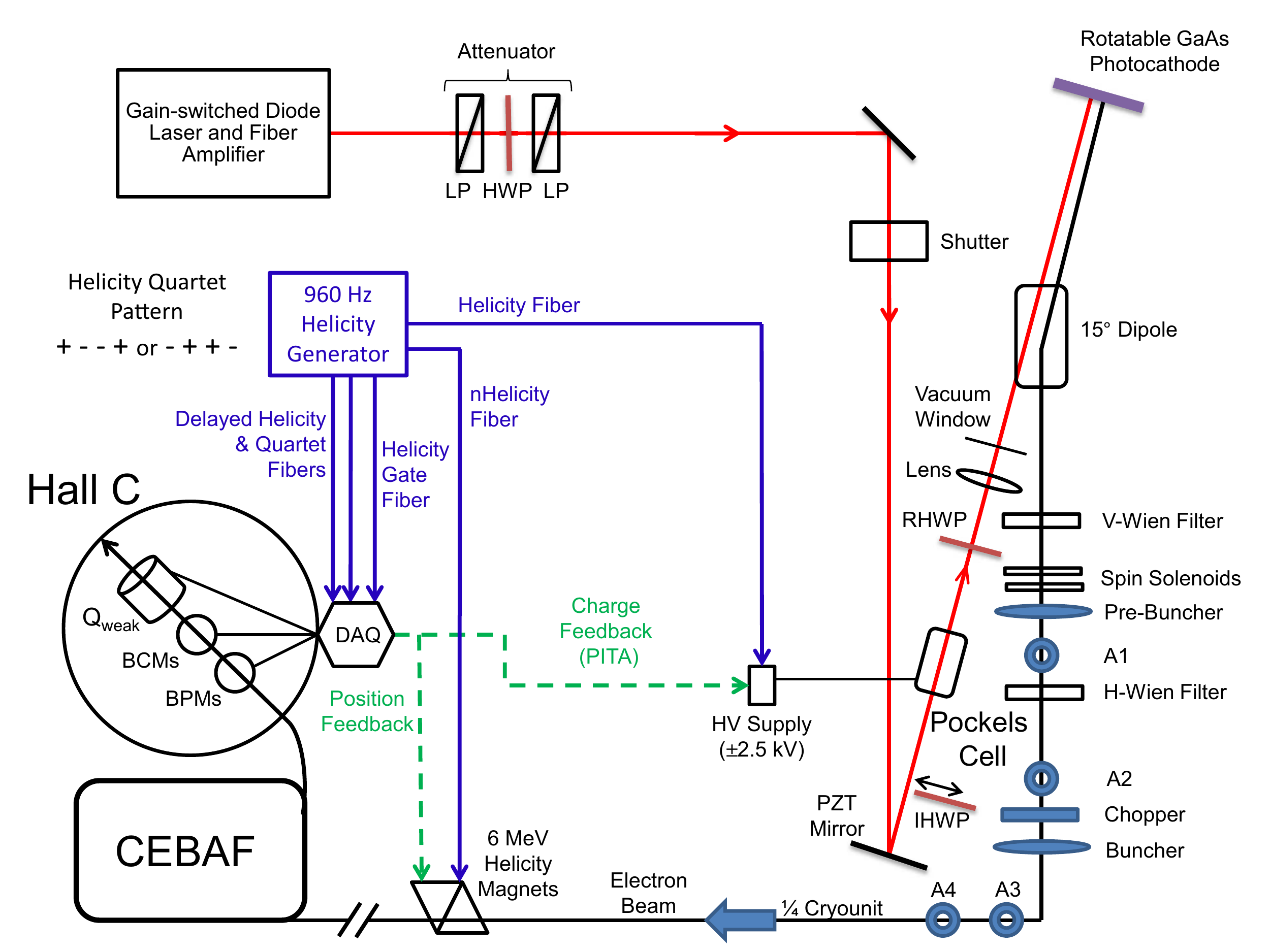}}
\caption{\label{fig:sourcetable} A schematic of the polarized injector components (see text) used for the \Qweak experiment.}
\end{figure*}

A helicity board located in the injector service building in
an electrically isolated \ac{acro-vme}  crate generated five fiber signals:
Helicity, nHelicity, Delayed Helicity, Quartet, and Helicity Gate as
illustrated in Fig.~\ref{fig:sourcetable}.
The Helicity Gates were produced with a frequency of  960.015~Hz, and thus a period of 1041.65~$\mu$s.  
The Helicity signal was used to switch the Pockels cell high voltage.
The nHelicity signal (complementary to the Helicity signal) was used
to control the helicity magnets. This way the helicity board always drew the same current regardless of the helicity state and further protected against any electrical pickup. 
 In addition, great care was taken within the injector to isolate the reversal signal from cables and ground paths that run throughout the accelerator/endstation complex, as even a weak coupling can result in a significant and varying false asymmetry. 

The Delayed Helicity signal was sent to the \ac{acro-daq} and was delayed by eight Helicity Gates, {\it i.e.},
it reported the state of the electron beam helicity eight Helicity Gates in the past. 
This technique 
 provides strong protection from electrical pickup that might occur if real-time decoding was used. 

The helicity patterns were generated in quartets
of four Helicity Gates, where the first and fourth gates had the same helicity, and the second and third had the opposite helicity as the first gate.
The helicity of the first gate in each quartet was determined using a 30-bit pseudo-random algorithm. 
The Quartet 
signal was true at the beginning of each new pattern, and was also sent to the \ac{acro-daq}.

\begin{figure}[hbt]
\begin{center} 
\includegraphics[width=0.475\textwidth]{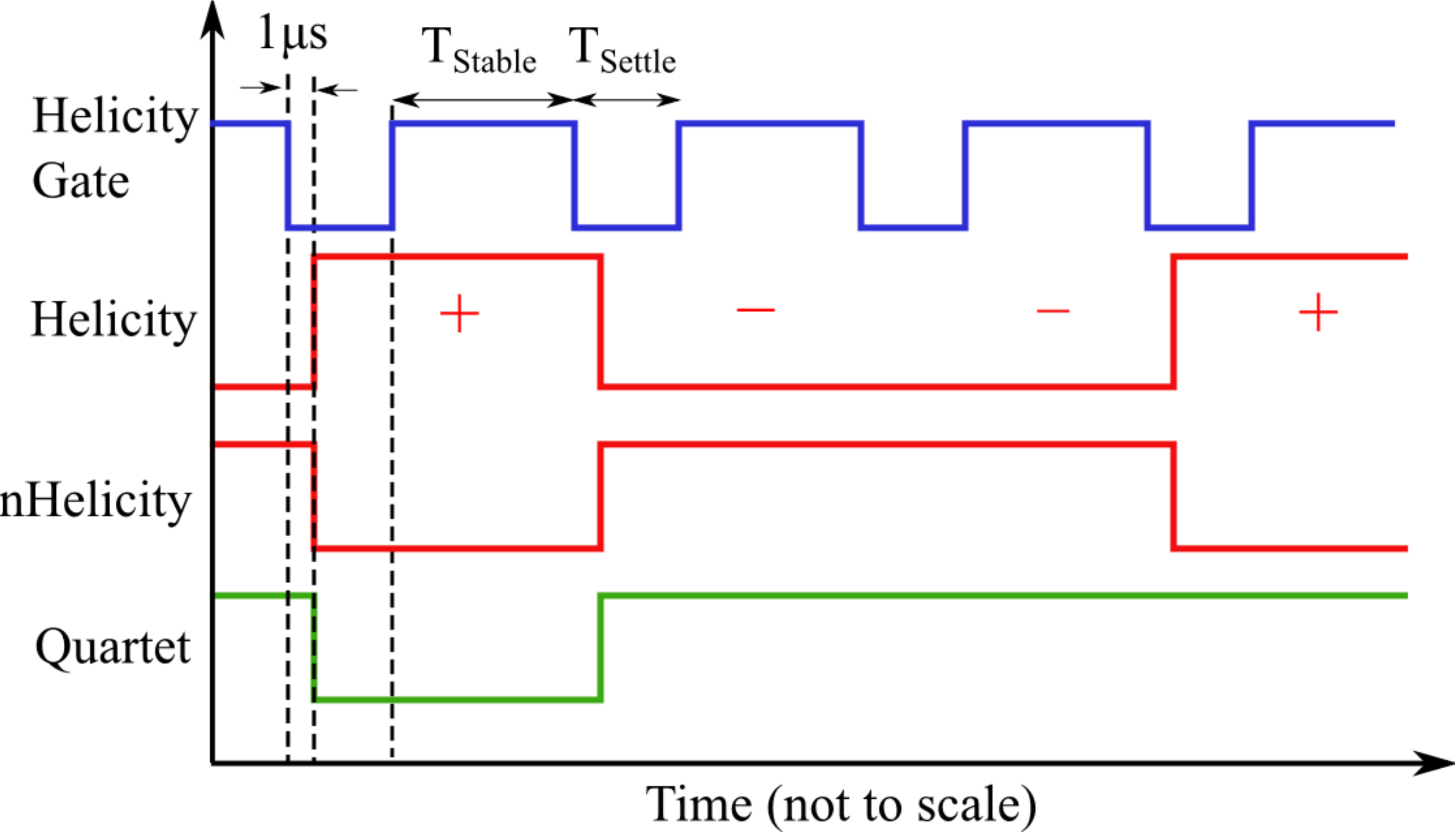}
\caption{ \label{fig:helicity_timing} 
Timing diagram of the helicity signals from the polarized source.
See text for details.
The scale of the horizontal axis is exaggerated to show details of the signal timing. 
}
\end{center}
\end{figure}

The Helicity Gate signal sent to the \ac{acro-daq} was defined by the 70~$\mu$s period ``$T_{\rm Settle}$" during which the Pockels cell high voltage would change.
The remaining 971.65~$\mu$s indicated a period of stable helicity ``$T_\text{\rm Stable}$."  
The helicity board generated the $T_{\rm Settle}$ signal in the Helicity Gate train 1.0~$\mu$s before all other signals.
The relative timing of the helicity signals is depicted schematically in Fig.~\ref{fig:helicity_timing}.

\subsection{Laser and Pockels Cell}
\label{sec:pockel}

The laser light was provided by a gain-switched 
\ac{acro-rf} 
pulsed diode operating at 1560~nm,  amplified in a fiber amplifier, and then frequency-doubled 
to 780~nm in a
lithium niobate 
 crystal.  Three lasers operating at a repetition rate of 499 MHz were used to individually supply beam to each of the three experimental halls at \ac{acro-jlab}.  The beams  were combined~\cite{sinclair} using a polarizing beam-splitter for the high-current halls and a partially transmissive mirror for the low-intensity hall. 
 A consequence of this arrangement was that the \qweak  Hall C beam had opposite polarization to the others.

The linearly polarized laser beams 
passed through a Pockels cell (an optical element with birefringence dependent on applied voltage) with its fast axis at 45$^\circ$.  
At $\sim$2.5 kV, the Pockels cell functioned as a quarter-wave plate and the laser light  emerged with circular polarization.  Reversing the voltage reversed the birefringence of the crystal and therefore the helicity of the laser beams.

A potentially serious source of systematic error can arise from changes in the beam properties, such as position, angle, and energy that are correlated with the polarization of the beam.  Sources of   \ac{acro-hcba} 
are dealt with  by minimizing the effects as much as possible, and  by measuring the beam parameters in the experimental hall and correcting the measured asymmetry for them (Sec.~\ref{sec:Beam:BMOD}).

\begin{figure}[hbt]
\begin{center} 
\includegraphics[width=0.475\textwidth]{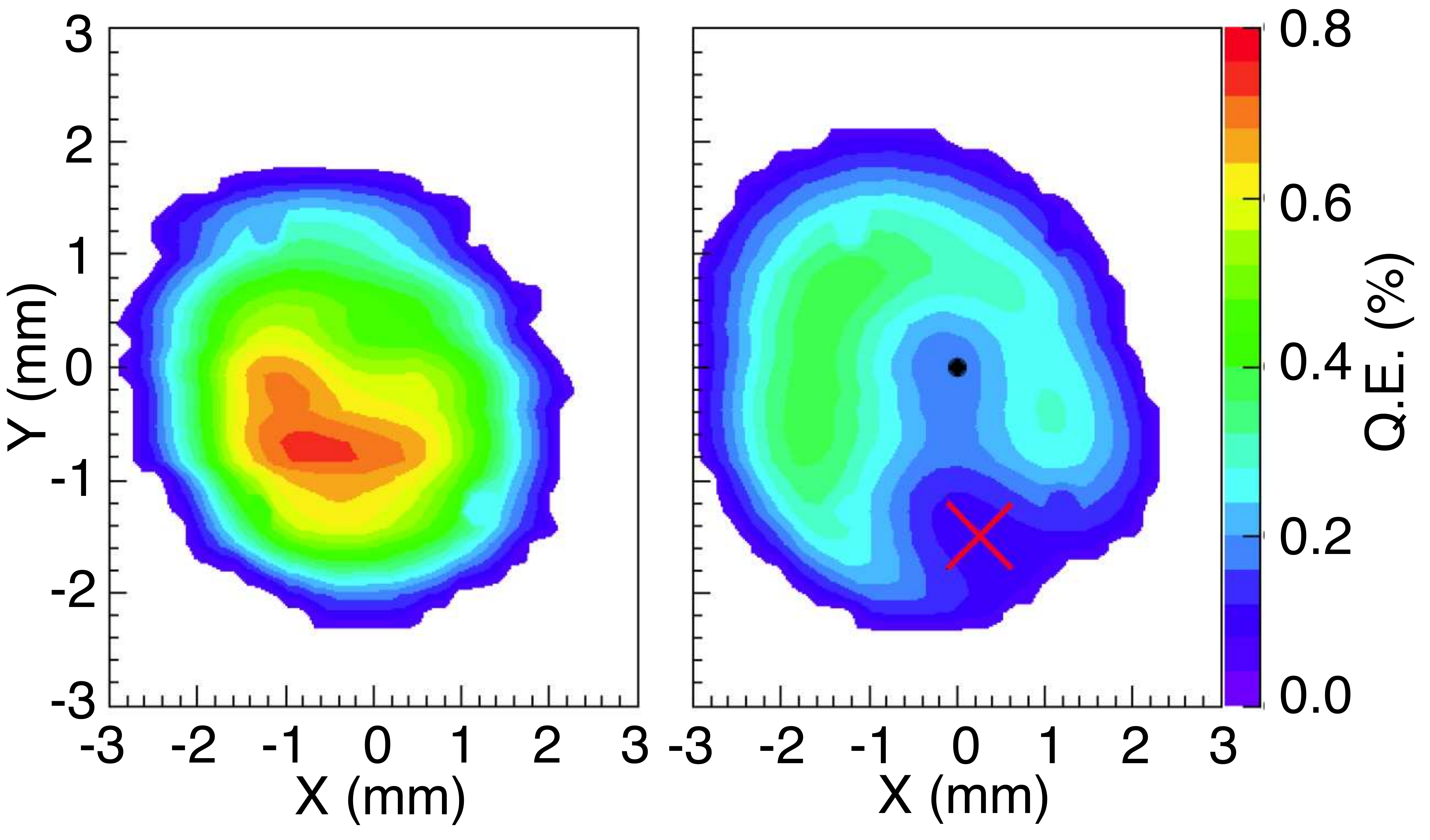}
\caption{ \label{fig-QEProfiles} 
QE profile scans before (left) and after (right) a four week period of high current running. In the latter figure, the ``dot" shows the electrostatic center and the ``X" shows the spot from where 180~$\mu$A beam was delivered to the experiment. The active area was $\sim$5~mm in diameter and the laser spot size was 1~mm in diameter.
} 
\end{center}
\end{figure}

\ac{acro-hcba}s in this experiment were minimized~\cite{paschke} by carefully aligning the optical elements, particularly the Pockels cell.
The \ac{acro-hc}  position differences, measured at the first \ac{acro-bpm} that the electron beam encountered after leaving the photocathode, were the smallest yet measured at \ac{acro-jlab} ($\le 20$ nm).
Illumination of the photocathode using laser beams with a Gaussian spatial profile leads to preferential \ac{acro-qe} degradation at the center of the laser spot location.  After many hours of use, a ``\ac{acro-qe} hole" forms at the photocathode, and  the spatial distribution of the electron beam changes accordingly, with more beam produced at the edges of the laser spot, where \ac{acro-qe} remains high.  This gradual evolution of the electron beam spatial distribution causes an increase in measured position differences.   
The development of a typical \ac{acro-qe} hole is illustrated in Fig.~\ref{fig-QEProfiles}.

The \ac{acro-ihwp} 
was the last optical element before the Pockels cell, and its  function was to reverse the polarization of the beam without changing the trajectory through the Pockels cell.  It was alternately inserted and removed approximately every 8 hours. 
This slow helicity reversal was used to cancel  \ac{acro-hcba}s 
related to lensing or steering by the Pockels cell crystal, by forming the difference of asymmetries measured with the \ac{acro-ihwp} in and out.

\begin{figure}[hbt]
\begin{center} \hspace*{-1.5cm}
\includegraphics[width=0.63\textwidth]{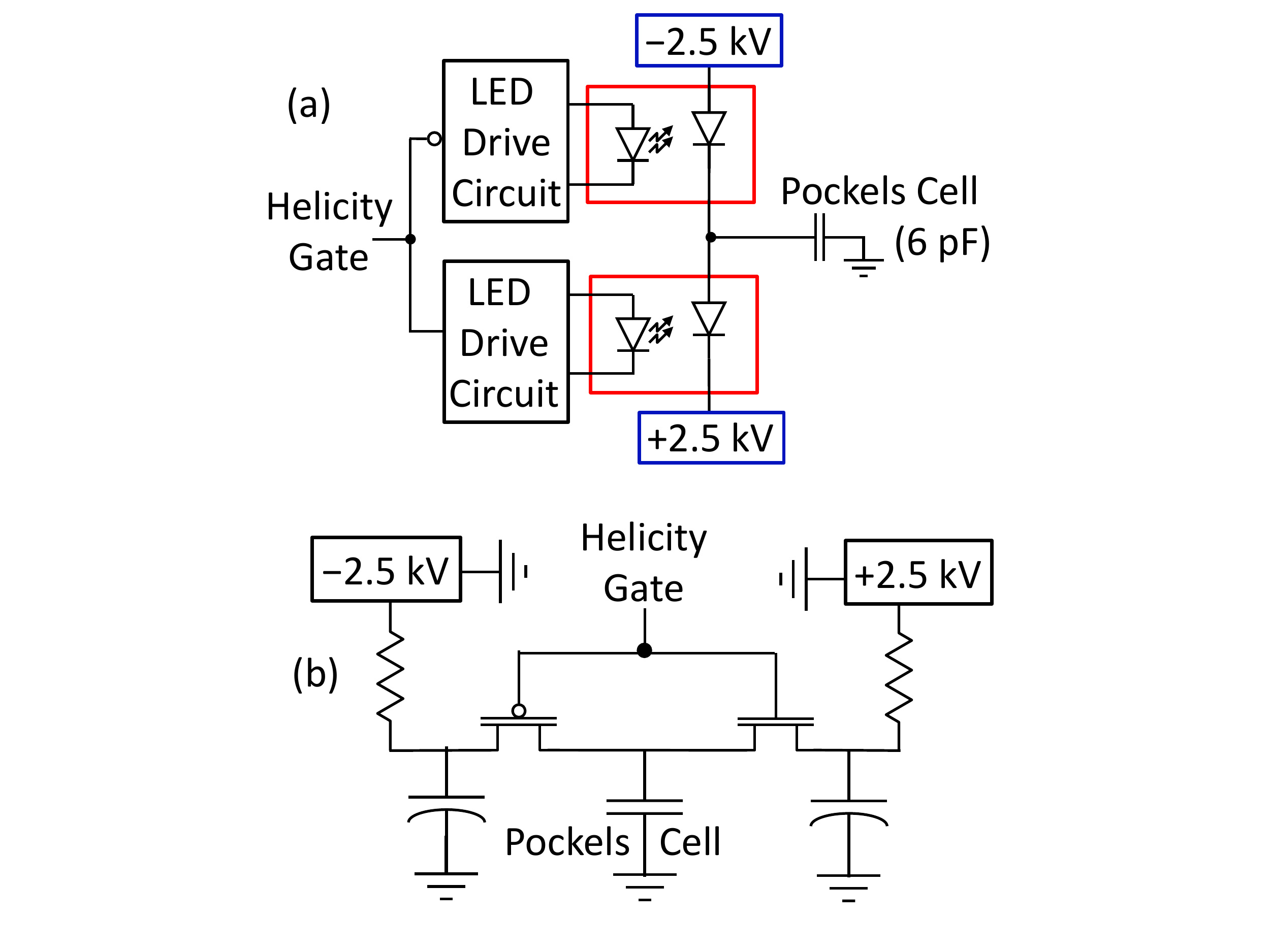}
\caption{ \label{fig-PCSwitch} 
(a) Fast Pockels cell high-voltage switch developed for the \qweak experiment. The Pockels cell can be viewed as a strictly capacitive device of 6~pf.  Because these opto-couplers are so small, they could be mounted directly to the top of the cell.  This helped limit the cable capacitance and stray inductance. 
(b) A diagram of the old Pockels cell HV circuit, which incorporated fast HV transistor switches.
} 
\end{center}
\end{figure}

The faster Pockels cell high-voltage switch~\cite{poelker} developed for the \Qweak\ experiment was constructed using high-voltage optical diodes~\cite{OC100} that ``reverse conduct'' when light is applied.   The diodes were fast enough to switch the $\sim$2.5~kV required within about 60~$\mu$s, 
by shining light from \ac{acro-led}s on them. This had the additional advantage of providing electrical isolation to prevent leakage of the helicity signal into the electronics.  The voltage was ramped up in stages over the transition to 
minimize  induced oscillations, or ``ringing.''  
The new switch had much lower capacitance than previous \ac{acro-mosfet} switches~\cite{Behlke} and virtually eliminated issues that 
previously resulted from voltage droop.  In order to ensure that the transition was complete,  70~$\mu$s were allowed to elapse before data-taking was resumed. 
This represented a 6.72\% dead time from helicity reversal at 960/s. 
Simple schematic diagrams illustrating the difference between the new and old switching schemes are provided in Fig.~\ref{fig-PCSwitch}.

Strained-superlattice photocathodes exhibit ``\ac{acro-qe} anisotropy," which is terminology that describes photoelectron yield that varies with the orientation of the incident linearly polarized light.  The \ac{acro-qe} of typical strained-superlattice photocathodes can vary by $\sim$4\%, depending on the orientation of incident linearly polarized light.  
Although great lengths were taken to provide 100\% circularly polarized light, in practice perfect circular polarization was not achieved. 
Furthermore the residual linear polarization component varied across the beam spot, giving rise to higher-moment effects such as helicity-correlated position differences. To address this issue, the \ac{acro-rhwp} was used to rotate the residual linear polarization and provide equal \ac{acro-qe} for the two helicity states. In practice   a small residual sensitivity to asymmetric linear polarization was allowed, so that asymmetric shifts in the Pockels cell voltages could be used to counteract
effects from downstream elements. A different orientation of the \ac{acro-rhwp} was required when the \ac{acro-ihwp} was inserted. More details on the optimization of the polarized source can be found in~\cite{Silwal}.

The final element that the laser beam encountered before the vacuum window and the photocathode was a lens which served both to determine the size of the laser spot on the photocathode and, by virtue of a remote motion mechanism, to move the position of the spot on the photocathode. The effect of the vacuum window birefringence on the laser polarization was minimized by rotating the photocathode.

\subsection{Photocathode and Gun}
\label{sec:photogun}

The photocathode was a p-doped strained-superlattice GaAs/GaAsP  wafer which allows spin-selective promotion of electrons to the conduction band by photons with energies slightly larger than the semiconductor band gap.  
The surface of the photocathode was activated with 
Cesium and NF$_3$ to 
reduce the work function and obtain the negative electron affinity 
required to extract an electron beam.  During \Qweak\!\!, the photocathode was ``reactivated" once during Run 1, and once during Run 2.
As mentioned above, the photocathode exhibited a QE anisotropy of 4\%. 
This \ac{acro-pita}  
was responsible for most of the \ac{acro-hcba}s, particularly as there were analyzing gradients in the photocathode and polarization gradients in the beam.  The \ac{acro-pita} effect was used for charge feedback.  Small changes to the Pockels cell voltage for each helicity state were used to change the amount of linear light in the laser beam such that, once analyzed by the photocathode, the number of electrons was the same in each state.

A new 
``inverted electron gun''~\cite{adderley} was developed  for the \Qweak\ experiment. This design utilized a compact, tapered ceramic insulator that extended into the vacuum chamber which 
increased the distance between biased and grounded parts of the gun and reduced the amount of metal biased at high voltage.  
Electrons leaving the photocathode experienced a field strength of  $\sim$2~MV/m and the field strength within the cathode/anode gap was $\sim$5~MV/m.  Although the maximum field strength inside the gun was $\sim$9~MV/m, the gun operated reliably at 130~kV without measurable field emission. 
  The experiment ran consistently at  beam currents of $\sim$180~$\mu$A, significantly higher than has been delivered previously at \ac{acro-jlab}.  
Space-charge induced emittance growth at this current is significant, and beam loss at the injector apertures A1$-$A4 (see Fig.~\ref{fig:sourcetable}) would have been difficult to eliminate using the previous photogun~\cite{poelker} operating at 100~kV. Beam loss during \qweak was typically 3\% or less, while operating an injector bunching cavity at relatively modest field strength. 

\begin{figure}[htb]
 \centerline{\includegraphics[width=0.5\textwidth,angle=0]{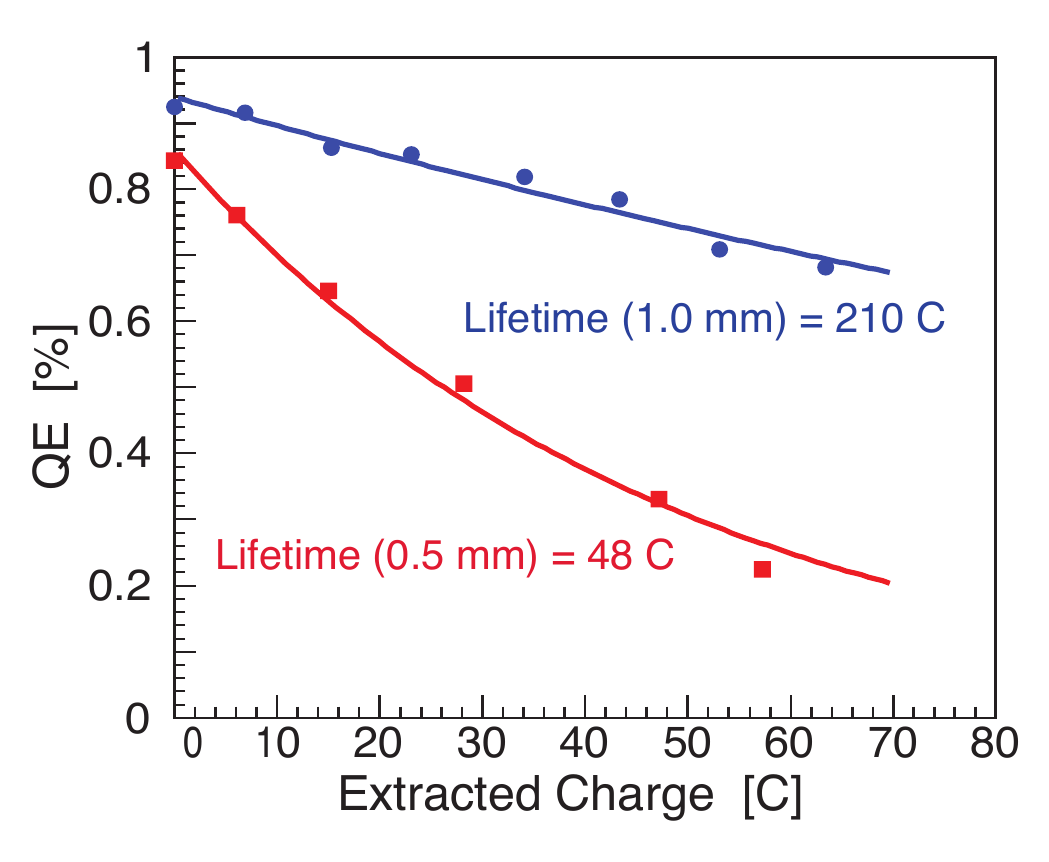}}
\caption{\label{fig:lifetime}
The measured \ac{acro-qe} is plotted against the charge extracted  from the photocathode.
The fitted curves represent $1/e$ charge-lifetimes of a spot on the photocathode, before (square points)  and after (circles) the spot size was increased by changing to a lens with a longer focal length.}
\end{figure}

During Run 1 only  modest $1/e$ photocathode charge-lifetimes of $\sim$50~C per laser spot were achieved.  Several different spots could be utilized before the photocathode required a reactivation cycle.  Ion back-bombardment is the predominant mechanism degrading the photocathode \ac{acro-qe}
 during electron emission~\cite{sinclair}.   This effect was mitigated by replacing the 1.5 m focal length lens with one of 2.0 m, thus increasing the \ac{acro-fwhm} of the laser spot from 0.5~mm to 1.0~mm and distributing the ion damage over a bigger region~\cite{lifetime}.  
Cathode lifetimes of $\sim$200 C were  achieved during Run 2 with the larger laser spot.  
Figure~\ref{fig:lifetime} shows $1/e$ fits to the daily-measured \ac{acro-qe} 
against the extracted beam charge obtained in both configurations.

\subsection{Injector}
\label{sec:injector}

The purpose of the injector section is to accelerate the beam to relativistic velocities 
synchronized with the \ac{acro-cebaf} linacs.
For a high-current beam, longitudinal bunching is also required.

The apertures  and the chopper~\cite{chopper} were used to limit the emittance of the beam by trimming the transverse and temporal (longitudinal) dimensions, respectively. Halfway between the photocathode and the chopper (in the 130 keV region), the pre-buncher prepared the longitudinal component of the beam for the chopper.  The chopper was a pair of 499 MHz 
RF deflecting cavities phased to sweep the beam in a circle with a revolution frequency of 499 MHz.  The chopper aperture was kept open at its widest extent, approximately $20^\circ$,  throughout the experiment.  Losses could be significant in these areas since the space-charge of the high-current beam caused the beam to expand in all dimensions.  
In practice, a significant increase in the width of the  charge asymmetry distribution measured by the experiment indicated when  the beam trajectory in the injector needed to be tuned to minimize interception.

The two-Wien spin flipper was composed of a vertical Wien filter followed by two solenoids and then a horizontal Wien, described in detail in Ref.~\cite{Adderley:2011ri}. Ideally the system reverses the polarization of the electron beam in the injector without changing the optical focusing properties of the system, by reversing the current in both solenoids. 
This reverses the correlation between the helicity of the laser and photo-produced electrons, and the electrons that ultimately arrive in the experimental hall.  This canceled polarization induced \ac{acro-hcba}s, 
most notably those related to differences in the beam spot size, which, unlike the trajectory, were not directly measured in the experiment. The two-Wien system was changed monthly throughout the experiment.

The helicity magnets   were a set of four air-core dipole magnets placed in the 6 MeV region of the injector beamline  to kick the beam differentially for each helicity state. They were used  
to control \ac{acro-hc} position ($X, Y$) and, less effectively, angle ($X^\prime$, $Y^\prime$) differences.
Sensitivities were determined periodically and corrections were applied approximately daily to reduce the differences.

There were two \ac{acro-rf} 
cavities in the injector used for velocity bunching the beam: the buncher and the pre-buncher (see Fig.~\ref{fig:sourcetable}). They were run with a phase offset so that the beam pulse would arrive at the zero-crossing potential and feel an accelerating force at the back of the bunch and a decelerating force at the front of the bunch.  Despite using these cavities, there were still issues related to beam blowup from the large space charge, such as large-halo scraping on apertures in the experimental hall.   In order to deal with this during Run 2, the injector was given a special (``M56'') tune in order to further bunch the beam.  The injector chicane was tuned to give a standard magnetic compression for a relativistic beam, equivalent  in effect to velocity bunching at low energy.  The quadrupole magnets in the chicane were set to give the lower-energy electrons, which were also towards the back of the bunch, a shorter path through the chicane, allowing them to catch up.  This procedure resulted in a factor of 2 reduction in bunch length arriving at the accelerator.

\section{Beam Transport and Diagnostics}
\label{sec:Beam:Beamline}

\subsection{Accelerator}
\label{sec:accelerator}

At the time of the \Qweak experiment, the 
\ac{acro-cebaf} accelerator~\cite{CEBAF} consisted of a single-pass 62 MeV injector (see Sec.~\ref{sec:injector}), and two 548 MeV superconducting  
linacs joined by recirculating arcs allowing one to five pass beam.  The two linacs are typically operated at equal energy, with the injector at 11.25\% of the North Linac energy.  Except for a brief period of 2-pass, 1.16 GeV operation at the start of Run 2, the experiment used 1-pass beam of 1.16 GeV. The 2-pass running provided a useful check for the experiment as an independent ($g-2$) helicity reversal relative to 1-pass operation.

A 
resistive
copper cavity accelerated the beam leaving the polarized source from 130 keV to 500 keV. A pair of \ac{acro-rf} 
superconducting cavities 
accelerated the beam from 500 keV  to 6 MeV, and
contain an RF skew quadrupole term which couples the $x$ and $y$ beta functions.  
The excellent normalized emittance provided by the photogun is therefore degraded, typically by
an order of magnitude. 

After acceleration from 6 MeV to 62 MeV the beam passed through a region used to match the transverse optics to the North Linac.  This match generally preserves normalized emittance. After the matching region a chicane is used 
to avoid recirculated beams of higher energy.  The dispersion in the chicane allowed for injector energy feedback.  The injector and higher-energy beams were rendered co-linear in a dipole at the start of the North Linac.  The beam for \Qweak then went through the North Linac, was separated vertically from other energy beams, went through a 180$^\circ$ arc, was merged with other energy beams from the other arcs, and passed into the South Linac.  It was extracted from among the other beams by a 499 MHz \ac{acro-rf} separator and a series of septum magnets.  The beam was then directed into the Hall C arc, consisting of (quads and) eight 3~m long dipole magnets which deflected the beam 34.3$^\circ$ to  experimental Hall C.  A transverse optics matching region before the Hall C arc was used to restore the desired beam envelope functions to design.  

A fast feedback system  minimized  excursions in both planes at the  entrance and exit of the Hall C arc and at the high-dispersion \ac{acro-bcm}. 
Four air-core correctors and the last \ac{acro-rf} zone in the South Linac were the actuators for this feedback system with $\sim$1 kHz response.  

A second optics adjustment region between the Hall C arc and the Compton polarimeter  prepared the beam waist needed for polarimetry.  After the Compton polarimeter there was a final array of quadrupoles to match the beam to the LH$_2$ target  and finally a set of 
raster magnets to diffuse the $\sim$200~$\mu$m beam profile at the cryotarget to (typically) 4$\times$4~mm$^2$. 

The raster reduced the effects of target boiling (see Sec.~\ref{sec:Target:Performance}) and prevented 
the beam from burning through the target windows.  
The beam spot on the target traced a uniform, square Lissajous pattern generated by two air-core magnets driven by triangular waves with fundamental frequencies of $\nu_1$$=$24.960 and $\nu_2$$=$25.920 kHz. The raster pattern repeated with a frequency of $\left( \nu_2 - \nu_1 \right)$$=$960.000 Hz, so each of the 960.015/s Helicity Gates integrated over a nearly
complete raster pattern.  If the raster period was substantially longer than the Helicity Gate period, then each helicity event would integrate over a different portion of the target face and  introduce additional noise in the detector asymmetries. This was verified in a set of test runs with 160 $\mu$A beam current. The asymmetry width measured for the 960 Hz raster patterns was 239 ppm, and increased for raster patterns at 480 Hz (240 ppm), and 240 Hz (253 ppm).  

At the exit of the experimental hall, the transition to the beam-dump tunnel was redesigned to withstand the greater power density of the beam used in the experiment. The window separating the upstream beamline vacuum from the helium-filled downstream beamline to the beam dump consisted of two hemispherical ($r=38$ cm) aluminum 2024-T6 windows (0.76 mm thick upstream, 0.51 mm thick downstream) separated by 2.3~cm of water circulated through a chiller.

\subsection{Beam Current Measurement}
\label{sec:Beam:BCM}

The \qweak experiment employed six \ac{acro-rf} cavity  
\ac{acro-bcm}s. 
They were located upstream of the target at distances of 16~m (\ac{acro-bcm}5, \ac{acro-bcm}7 \& 8), 13.4~m (\ac{acro-bcm}1 \& 2), and 2.7~m (\ac{acro-bcm}6). Calibrations of the \ac{acro-bcm}s between 1--180 $\mu$A  were performed using a Parametric Current Transformer~\cite{unser} (Bergoz Unser monitor) in the Hall C beamline.   After calibration, the \ac{acro-bcm} linearity 
was observed to be better than 0.5$\%$ between 20--180 $\mu$A.
At the extremely low beam currents used for the event mode of the experiment (10~ nA to 1 $\mu$A),  a Faraday cup in the injector was used for calibration.

The \ac{acro-bcm}s  provided stable, low noise, 
 continuous (non-invasive)  beam current measurements. 
To avoid radiation damage, the sensitive \ac{acro-bcm} electronics were located outside the experimental hall. 
\ac{acro-bcm}1, \ac{acro-bcm}2, and the Unser used analog receivers. Digital receivers developed for the \Qweak experiment were used with four additional new \ac{acro-bcm}s. The \ac{acro-bcm} cavities were tuned to the third harmonic of the beam frequency (1497 MHz), and temperature stabilized at 43$^\circ$~C to preserve the tune. 
The analog receivers frequency downconverted the cavity outputs to 1 MHz, and then used \ac{acro-rms}-to-DC converters to demodulate the signals.  The digital receivers downconverted to 45 MHz, and then digitally sampled and processed the signals.  In both cases, voltage levels proportional to beam current and band-limited to $\sim$100 kHz were provided  to the 18-bit sampling 
\ac{acro-adc}s~\cite{qwads} described in Sec.~\ref{sec:Components:DAQ:CurrentMode}.

Two metrics were used to assess the performance of the \ac{acro-bcm}s. The most useful was the width of the  
\ac{acro-dd} of asymmetries derived from a pair of \ac{acro-bcm}s, because fluctuations in the beam charge canceled, resulting in a metric sensitive only to the instrumental resolution of a \ac{acro-bcm} pair. For example, the \ac{acro-dd} of \ac{acro-bcm}s 7 \& 8 is 
\begin{equation}
DD_{78} = \frac{Q^+_7 - Q^-_7}{Q^+_7 + Q^-_7} - \frac{Q^+_8 - Q^-_8}{Q^+_8 + Q^-_8},
\end{equation}
where $Q^j_i$ denotes the charge measured by \ac{acro-bcm} i for beam helicity j. The resolution of an individual \ac{acro-bcm} was taken as its \ac{acro-dd}$/\sqrt{2}$. The other \ac{acro-bcm} performance metric was the width of the main detector asymmetry as defined in Eq.~\ref{eqn:Ayield} and discussed in Sec.~\ref{sec:Asymmetry}. However, the \ac{acro-bcm} resolution was only one of several effects contributing in quadrature to that width.

\begin{figure}
 \centerline{\includegraphics[width =0.5\textwidth,angle=0]{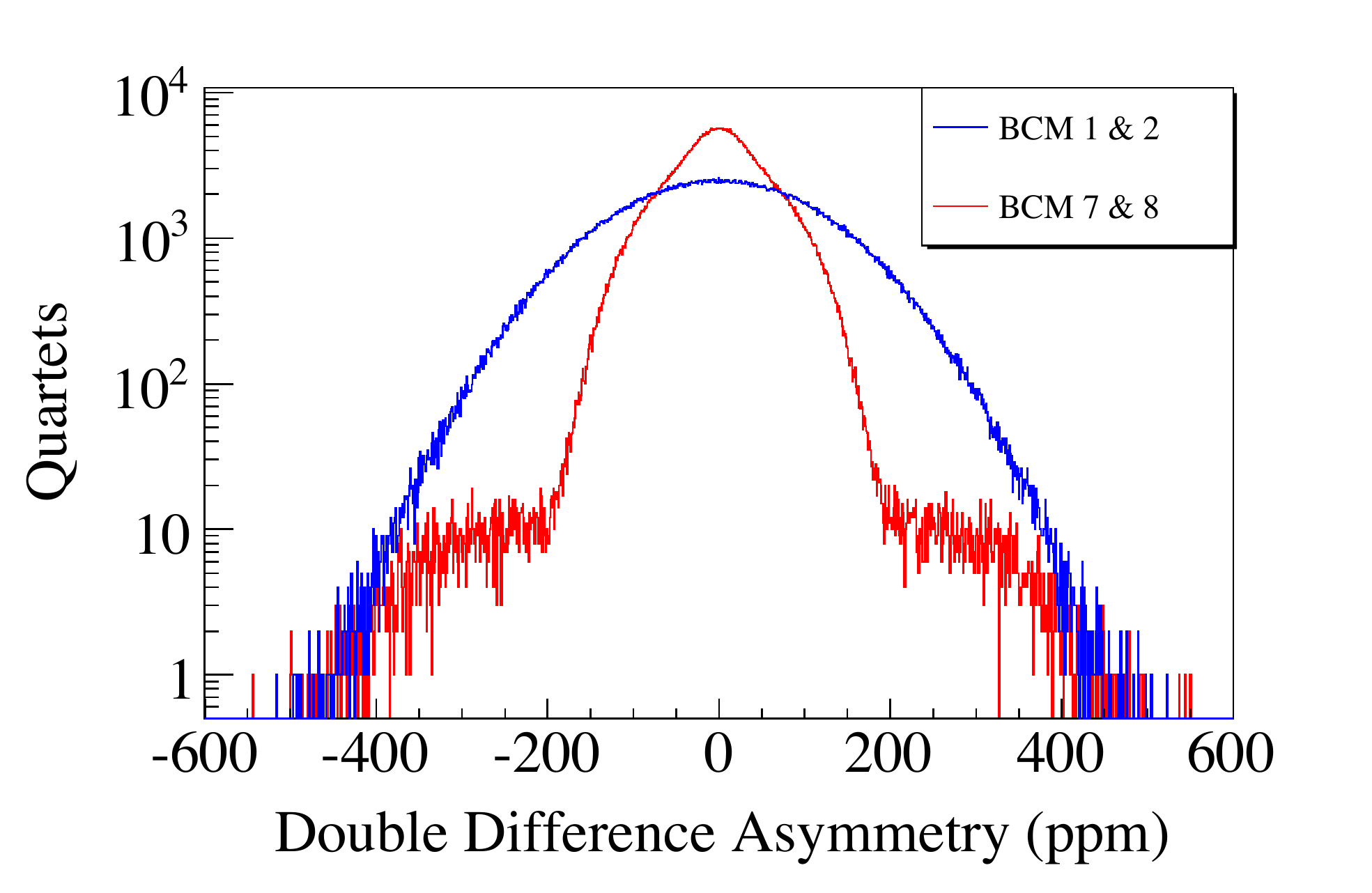}}
\caption{\label{fig:bcmPlots}
Charge asymmetry double difference plots for two pairs of \ac{acro-bcm}s  from a typical 1 hour run. The \ac{acro-dd}$_{12}$ (broad curve) has an \ac{acro-rms}  width of 115 ppm. The \ac{acro-dd}$_{78}$ (skinny curve with shoulders) has an \ac{acro-rms} width of only 57 ppm. }
\end{figure}

During Run 1 the experiment relied primarily on \ac{acro-bcm}1 \& 2, which provided a \ac{acro-dd}$_{12}$ of 100--140 ppm (see Fig.~\ref{fig:bcmPlots}). 
\ac{acro-bcm}5 \& 6 were connected to prototype digital receivers that were located in the experimental hall and sustained radiation damage.  
Between Run 1 and 2, \ac{acro-bcm}7 \& 8 were added and the new digital receivers for \ac{acro-bcm}s 5-8 were installed outside the hall.
The new electronics took advantage of improved digital signal processing techniques and utilized 18-bit, 1 MHz \acp{acro-dac} to generate the output voltage. Finally, air-core coaxial cables were replaced with Heliax~\cite{Heliax} cables. 
The resulting \ac{acro-dd}$_{78}$ was typically only $\sim$60 ppm, so \ac{acro-bcm}8 was used for charge normalization in Run 2.

\begin{figure}[!htb]
\centering
\hspace*{-.5cm}
\includegraphics[width =0.475\textwidth,angle=0]{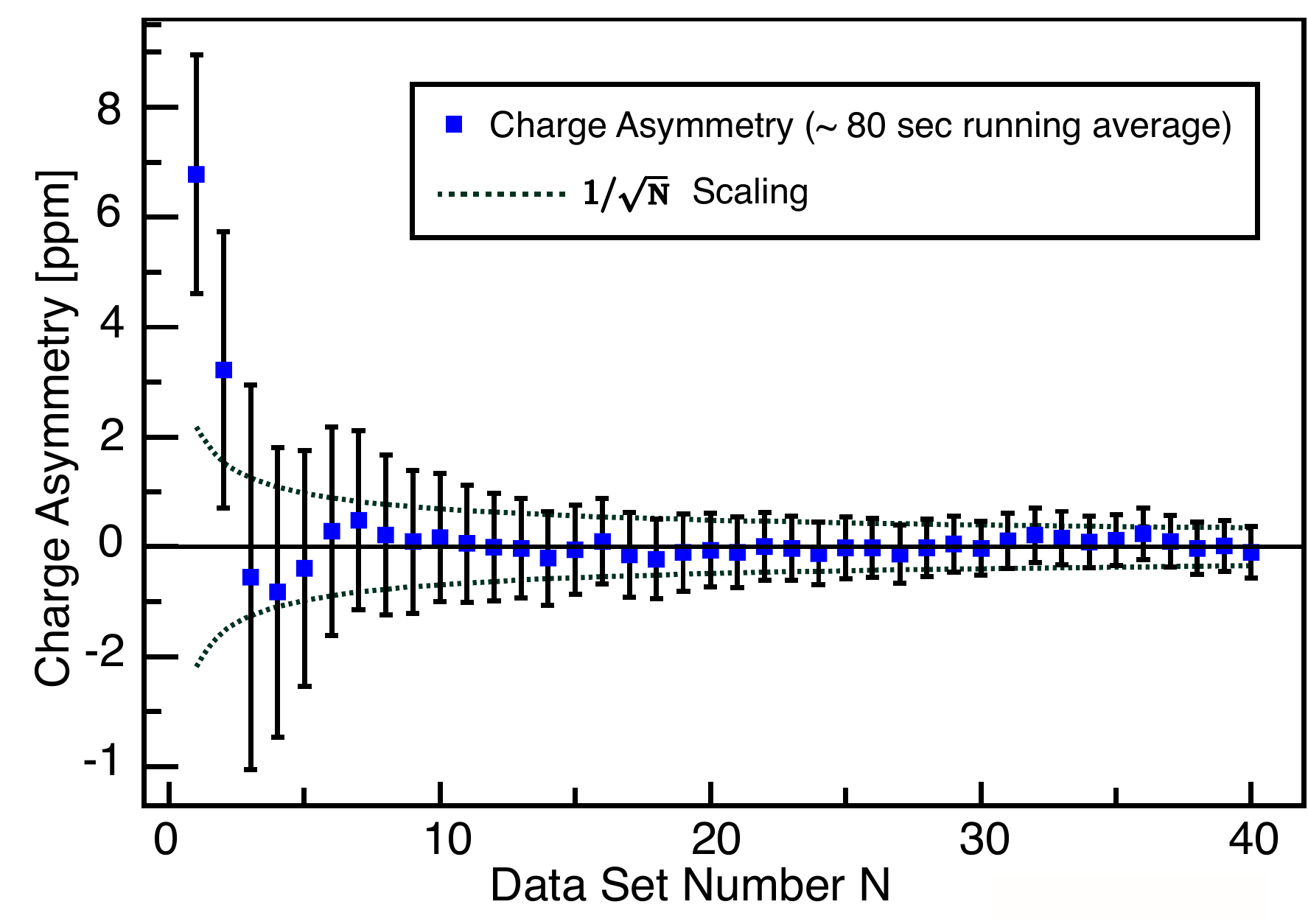}
\caption[Charge feedback applied over 80 s time intervals.]{\label{fig:charge_feedback}Running charge asymmetry from a typical 1 hour run showing the effect of charge feedback applied over 80 s intervals~\cite{Rakitha}. The horizontal axis shows the number of 80 s intervals. 
The dotted  
curve is the 
${\rm {1}/{\sqrt{N}}}$ statistical convergence.
} 
\end{figure}

Assuming a detector non-linearity of 1\%, the experiment required that the overall helicity-correlated 
\ac{acro-ia}  
be kept below 100 ppb in order to limit this contribution to the uncertainty in the asymmetry measurement to $<1$ ppb. Since the measured \ac{acro-ia} was typically a few ppm over a 1 minute interval, 
an active charge feedback system was used.
The cumulative \ac{acro-ia} was measured in 80~s intervals (see Fig.~\ref{fig:charge_feedback}), and the feedback scheme adjusted the Pockels cell voltages (\ac{acro-pita} offsets) at the polarized source (see Sec.~\ref{sec:photogun}) to null it. Over a typical month of running, the \ac{acro-ia} was typically only 40~ppb~\cite{Rakitha}.

\subsection{Beam Position and Angle}
\label{sec:Beam:BPM}

Continuous beam position monitoring in the experiment was carried out using stripline monitors~\cite{Powers:1996ai, Musson:2009zz} equipped with two $+/-$ pairs of perpendicular antennas tuned to the \ac{acro-rf} structure of the beam. The readout of each $+/-$ pair was multiplexed using switched-electrode electronics every 4.2~$\mu$s to eliminate the effects of gain differences in the electronics. Each of the four antennas from each 
\ac{acro-bpm} was read out for each helicity state of the beam into 18-bit sampling \ac{acro-adc}s custom-built~\cite{qwads} for this experiment's faster reversal rate, described in Sec.~\ref{sec:Components:DAQ:CurrentMode}. There were 24 \ac{acro-bpm}s read out in the injector beamline, and 23 in the Hall C beamline. The \ac{acro-bpm}s were used with beam currents between 50~nA and 180 $\mu$A.

The beam position and angle at the LH$_2$ target were determined~\cite{Buddhini} from a linear least squares fit of 4 or 5  \ac{acro-bpm}s in a magnetic field-free drift region between 1.5 m and 10.5 m upstream of the target.  
Using two \ac{acro-bpm}s as a reference, the offsets of the remaining \ac{acro-bpm}s in front of the target were adjusted by $\sim$1~mm to bring them into agreement. These offsets were stable over the 2 years of the experiment at the 25 micron level.
 Typical position resolutions of  1~$\mu$m (1.7~$\mu$m) were achieved with 5 (4) \ac{acro-bpm}s using methods  similar to the \ac{acro-dd} technique described in Sec.~\ref{sec:Beam:BCM}, implying $\sim$1~nm scale resolution in an hour. Likewise, angle resolution at the target was typically 150 nrad.

A slow (1~s update) position lock was implemented to maintain the desired beam position and angle  on the target, using this calculated target position in conjunction with pairs of corrector magnets upstream of the \ac{acro-bpm}s.

\subsection{Beam Energy}
\label{sec:Beam:Energy}

Two types of beam-energy measurements were required for the experiment: an absolute beam-energy measurement for the incident energy and  $Q^2$ determination, and the energy-asymmetry measurement at the target to remove false asymmetries generated by \ac{acro-hc} energy fluctuations.

Position sensitive 3-wire scanners (harps~\cite{Yan:1995nc}) located before, in the middle, and after the Hall C arc were used for  invasive and therefore infrequent energy measurements accurate to $\sim$$10^{-3}$.  
These were carried out by utilizing the Hall
C arc beamline  as a spectrometer~\cite{Yan:1993fd} according to 
\begin{equation}
{ {p} {\rm (GeV/c)} = \frac{0.3}{\sin{\Theta}} \int B({\rm T}) \; dl({\rm m}), }
\end{equation}
where $\Theta$ is the angle (34.3$^\circ$) by which the electron beam bends in the arc and $\int B dl$ is the magnetic field integral over the eight 3 m long dipoles in the arc beamline~\cite{Yan:1992}. With the arc quadrupole and corrector magnets off during these energy measurements, the momentum dispersion is 12 cm/\% at the end of the arc. These invasive energy measurements were used to benchmark continuous (non-invasive) energy measurements with a relative accuracy of $\sim$100 ppm obtained using \ac{acro-bpm}s along the Hall C beamline in conjunction with knowledge of the arc optics and dipole magnetic fields.

The \ac{acro-hc} beam-energy asymmetry at the target was determined using the \ac{acro-bpm} (\ac{acro-bpm}3C12) located in the region of highest dispersion (typically 4 cm/\%) in the Hall C arc. The horizontal ($X$) beam-position differences measured at \ac{acro-bpm}3C12 are sensitive to position, angle, and energy differences. Therefore, 
relative energy differences at the target were obtained from
\begin{equation}
\label{eq:energy_equation}
 \frac{\Delta \textit{P}}{\textit{P}} = \frac{\Delta X_{\rm 3C12}}{411} - \frac{\Delta X_{\rm target}}{596} + \frac{\Delta X^\prime_{\rm target} }{0.443},
\end{equation}
where the subscripts indicate the beam position differences (in cm) at 3C12/target, $X^\prime_{\rm target}$ represents the (horizontal) beam angle in $X$ at the target (in radians) and the denominators on the right of Eq.~\ref{eq:energy_equation} account for the first-order transport matrix elements for beam propagation between 3C12 and the target.

\subsection{Beam Modulation}
\label{sec:Beam:BMOD}

As described above in Eq.~\ref{eqn:AsyMeas} and Sec.~\ref{sec:pockel}, 
unwanted \ac{acro-hc} changes in  
the transverse beam positions $X$ (horizontal) and $Y$ (vertical), beam angles $X^\prime$ and $Y^\prime$, and incident energy $E$ on the target give rise to false asymmetries (\ac{acro-hcba}).
These \ac{acro-hcba}s $A_{\rm beam}(E,X,Y,X^\prime,Y^\prime)$ can be heavily suppressed with careful tuning at the polarized source and a symmetric detector array. However, the residual effects must be measured and controlled. $A_{\rm beam}$ is determined using the following expression:

\begin{equation}
A_{\rm beam} = \sum_{i = 1,5}
\frac{\partial A}{\partial \chi_{i}}\Delta \chi_{i}.
\end{equation}
Here the slopes ${\partial A}/{\partial \chi_{i}}$ are the measured detector sensitivities of the asymmetry $A_{\rm raw}$ (defined in Eq.~\ref{eqn:Ayield}) to changes in the beam parameters $\chi_{i}$ at the helicity quartet level, and \(\Delta \chi_{i}\) is the \ac{acro-hc} difference of each beam parameter $\chi_{i}$ measured at the quartet level.  The five \ac{acro-bpm}s described in Sec.~\ref{sec:Beam:BPM} were used to continuously measure the \ac{acro-hc} beam position and angle differences at the target. The measurement of the \ac{acro-hc} energy difference relied on \ac{acro-bpm}3C12, as described in Eq.~\ref{eq:energy_equation} of Sec.~\ref{sec:Beam:Energy}.

The natural jitter of the beam can be, and was used to determine the detector sensitivities \(  {\partial A}/{\partial \chi_{i}}\).
However, better decoupling of the five sensitivities was achieved  by varying the beam parameters in a controlled manner using a beam modulation system built specifically for this purpose.  
 Decoupled position and angle motions were separately produced by varying the current in pairs of air-core magnets placed along the beamline; two pairs in $X$ and two pairs in $Y$ approximately 82 and 93 m upstream of the target. 
Optics simulations~\cite{OPTIM} were used to determine the optimum placement of the coil pairs along the beamline which produced  the offsets in position and angle desired at the target.
 Changes in energy were produced by varying the power input to a cavitity in the accelerator's South Linac, and monitored using the response of \ac{acro-bpm}3C12 at the point of highest dispersion in the Hall C arc. 
The beam was driven at $\sim$125 Hz with the modulation system for 20~s every 320~s for the duration of the experiment. 

Typical beam modulation amplitudes at the target, as well as typical monthly results measured for the \ac{acro-hc} beam properties $\Delta \chi_i$ and detector sensitivities ${\partial A}/{\partial \chi_{i}}$ can be found in Table~\ref{tab:bmod}.
The \ac{acro-hcba}s for $X$ \& $X^\prime$ are anti-correlated and largely cancel. The same is true for $Y$ \& $Y^\prime$. The uncertainties associated with the monthly \ac{acro-hc} position (angle) differences $\Delta \chi_i$ are 0.07 nm (0.01 nrad) based on the quartet level BPM resolution discussed in Sec.~\ref{sec:Beam:BPM} of 1 $\mu$m (0.2 $\mu$rad) over the $2 \times 10^8$ quartets in the monthly period shown in Table~\ref{tab:bmod}. 
\begin{table*}[t]
\centering
\begin{tabular}{ c  r  r  r }
Beam	   &  Modulation         & Msrd  $\Delta \chi_{i}$ & Msrd  $ {\partial A}/{\partial \chi_{i}}$ \\
Parameter  & Amplitude &  (monthly)  & (monthly)	\\ \hline \hline
$X$  & $\pm$ 125 $\mu$m & $-3.3 $ nm &  $-2.11 $ ppm/$\mu$m    \\ 
$Y$  & $\pm$ 125 $\mu$m  & $2.5 $ nm & $0.24 $ ppm/$\mu$m     \\
$X^\prime$ & $\pm$   5 $\mu$rad  & $-0.7 $ nrad & $100.2$ ppm/$\mu$rad  \\ 
$Y^\prime$  & $\pm$   5 $\mu$rad  & $0.02 $ nrad & $-0.0 $ ppm/$\mu$rad     \\ 
Energy     & $\pm$  61 ppm, & $0.1 $ nm  & $-1.56 $ ppm/$\mu$m \\ 
           & ($\sim$ 70 keV) &                 &    \\ 
\end{tabular}
\caption{Typical amplitudes used for driven beam modulation (column 2). 
Columns 3 and 4 provide  typical average monthly results measured during Run 2  for the \ac{acro-hc} beam parameter differences $\Delta \chi_{i}$ and detector sensitivities  $ {\partial A}/{\partial \chi_{i}}$ 
for the beam parameters $i$ listed in the first column. The total \ac{acro-hcba} for this example is only 0.4 ppb. The uncertainties associated with $\Delta \chi_{i}$ and $ {\partial A}/{\partial \chi_{i}}$
are discussed in the text.}
\label{tab:bmod}
\end{table*}

\subsection{Beam Halo Monitors}
\label{sec:Beam:Halo}

Several \ac{acro-pmt} monitors straddled the beamline between 1~m and 5~m upstream of the LH$_2$ target to monitor beam halo,  providing crucial feedback used to tune the beam. Four monitors had lucite blocks coupled to their 5.1 cm diameter \ac{acro-pmt}s, and two used small scintillator blocks.  
All six monitors used 12-stage Photonis XP2262B \ac{acro-pmt}s read out in event (pulse-counting) mode. 
Each halo monitor pair was shielded with lead and  pointed upstream at a retractable halo ``target'' 6 m upstream of the LH$_2$ target. The halo target consisted of a 2.8 cm $\times$ 5.1 cm aluminum  frame 1~mm thick with a 13 mm diameter circular hole and an 8 mm $\times$ 8 mm square hole cut out of it. The target could be positioned with a linear actuator such that either hole (or the frame) could be positioned in the beam, or it could be retracted completely out of the beam pipe. 

An absolute measure of the beam halo was obtained by calibrating the halo monitors with beam passing through the 1 mm thick halo frame. 
The most useful monitors for absolute determination of the beam halo fraction were two of the lucite monitors (one with a 2 cm thick lead block in front to suppress low-energy particles). These were well shielded on  five sides with lead, and located 16.5 cm from the beam centerline on opposite sides of the beampipe 75 cm downstream of the halo target.  The mean scattering angle of these monitors relative to the halo target was $\sim$$12.4^\circ$.  
Background from upstream of the halo target was accounted for with the halo target out. 
With this correction, the absolute halo fraction
was determined to a precision of $\sim$2$ \times 10^{-8}$ at a beam current of 180 $\mu$A.  In addition to these dedicated measurements of the halo fraction,
the 13 mm hole was in place about half the time during the experiment to provide a continuous monitor of the beam halo. Typical measured beam halo was between 0.1--1 ppm.

\section{Beam Polarization}
\label{sec:Components:Polarization}

Measurement of the beam polarization was expected to be the largest systematic uncertainty in the experiment. An existing M\o ller polarimeter has routinely provided precise beam polarization measurements 
at $\leq$1.5\% 
in Hall C for many years. However, these measurements can only be performed at beam currents  much lower than those employed in the experiment (typically $\lesssim$2~$\mu$A, although beam currents up to 20 $\mu$A 
have been employed). The measurements   are invasive, and therefore performed infrequently. 

Therefore, the  M\o ller polarimeter was augmented with a new, non-invasive Compton polarimeter which provided continuous polarization measurements at the full 180 $\mu$A of the \Qweak  experiment. A statistical precision better than 1\% per hour was achieved. The absolute polarization determined independently from the two polarimeters was cross-checked (once) with Compton polarimeter measurements at 4.5 $\mu$A bracketed with M\o ller polarimeter measurements at the same beam current.

\subsection{The M\o ller Polarimeter}
\label{sec:Components:Polarization:Moller}

The beam polarization was  measured using the existing Hall C M\o ller polarimeter~\cite{Hauger2001a} 2--3 times per week. 
Extensive studies were done for this experiment to characterize the uncertainties and ensure sub-percent precision.

\begin{figure*}[tbhp]
\centering
  \includegraphics[width=\textwidth]{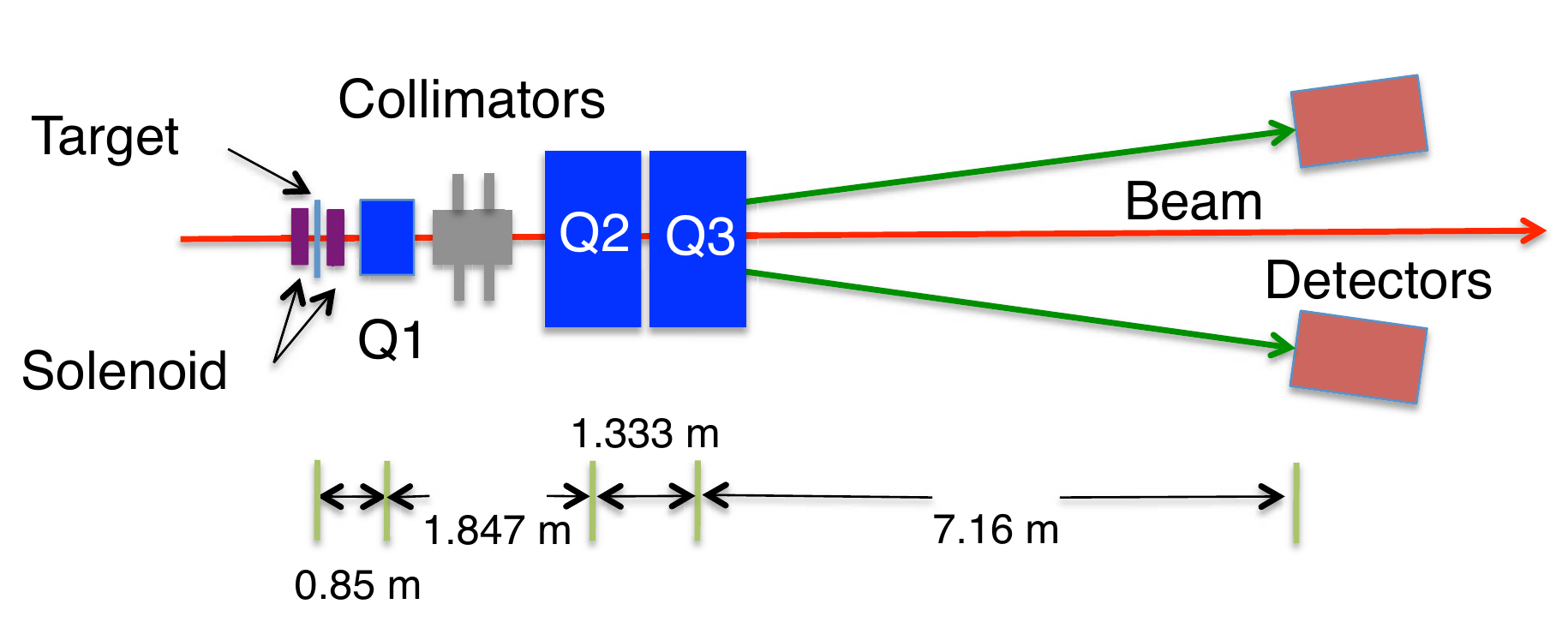}
  \caption{The Hall C \moller{} polarimeter. Only quadrupole magnets 1 and 3 were used during $Q_{\rm weak}$; quadrupole 2 was installed in preparation for the \ac{acro-jlab} 12 GeV program.}
  \label{fig:moller_setup}
\end{figure*}

The \moller{} polarimeter measured the parity-conserving $\vec{e} \, \vec{e}$ cross section asymmetry $A_{zz}$, for which the analyzing power is precisely known. The Hall C M\o ller used a split  
superconducting solenoid to brute-force polarize a 1~$\mu$m thick pure iron target foil 
along the beam direction. The 3.5~T solenoid field was sufficiently above the 2.2~T saturation point of iron to fully saturate the foil. Since only the valence electrons contribute to the magnetization, the total target polarization was only about 8\% averaged over all the electrons in the atom.
Scattered and recoil electrons were detected in coincidence using a near-symmetric apparatus, with one electron detector aperture slightly smaller than the other to cleanly define the acceptance. Use of a  narrow timing window minimized accidentals and reduced the signal from Mott scattering from the iron nucleus, the dominant background~\cite{Loppacher1996a}. Figure~\ref{fig:moller_setup} shows a schematic of the device.

Table \ref{tab:moller_systematics} summarizes the  
 uncertainties. The largest  comes from scattering off the unpolarized  inner electron shells (the Levchuk effect)~\cite{Levchuk1994a}. Since the \moller{} measurements were invasive and limited to low-current ($\sim$2 $ \mu$A), a conservative
  uncertainty  was included to account for potential effects due to extrapolation to the higher beam current used in the experiment. This concern was also addressed by comparison with the results of the Compton polarimeter
  discussed in Sec.~\ref{sec:Components:Polarization:Compton}.

During Run 1, 
an intermittent short in one of the coils of quadrupole 3 (see Fig.~\ref{fig:moller_setup}) 
affected the  acceptance and therefore the analyzing power of the polarimeter at the few-percent level.
To account for this, the \moller{} simulation used to provide the polarimeter acceptance was modified to include a correspondingly altered quadrupole field map using a POISSON magneto-static field generator~\cite{poisson}. Hall probes in the quad were used to compare 
to simulations of the polarimeter response
with and without the short.
 An  uncertainty of 0.89\% was added to the  Run 1 commissioning M\o ller polarization measurements~\cite{QweakPRL} to account for this effect, which was absent in Run 2.

\begin{table}[ttbh]
  \centering
    \begin{tabular}{|l|c|c|} \hline
                           & Uncer- & $dA/A$ \\ 
      Source & tainty & (\%) \\ \hline
      Beam position $X$     & 0.5 mm      & 0.17     \\
      Beam position $Y$     & 0.5 mm      & 0.28     \\
      Beam direction $X^\prime$    & 0.5 mrad    & 0.1      \\
      Beam direction $Y^\prime$    & 0.5 mrad    & 0.1      \\
      Q1 current                 & 2\%         & 0.07     \\
      Q3 current                 & 3\%         & 0.05     \\
      Q3 position                & 1 mm        & 0.01     \\
      Multiple scattering        & 10\%        & 0.01     \\
      Levchuk effect             & 10\%        & 0.33     \\
      Collimator position        & 0.5 mm      & 0.03     \\
      Target temperature         & 100\%       & 0.14     \\
      B-field direction          & 2$^\circ$   & 0.14     \\
      B-field strength           & 5\%         & 0.03     \\
      Spin depolarization        & --~         & 0.25     \\
      Electronic dead time       & 100\%       & 0.05     \\
      Solenoid focusing          & 100\%       & 0.21     \\
      Solenoid position($X,Y$)     & 0.5 mm       & 0.23     \\
      High current extrap. & --~         & 0.5      \\
      Monte Carlo statistics     & --~         & 0.14     \\ \hline
      ~                          & Total       & 0.83     \\ \hline
    \end{tabular}
  \caption{{The systematic uncertainties of the Hall C \moller{} polarimeter for Run 2 of the experiment. An additional 
   uncertainty was present during Run 1 due to an intermittent short in one of the quadrupoles of the polarimeter (see text). }}
  \label{tab:moller_systematics}
\end{table}

\subsection{The Compton Polarimeter}
\label{sec:Components:Polarization:Compton}

\begin{figure*}[htb]
\begin{center}
\includegraphics[width =\textwidth,angle=0]{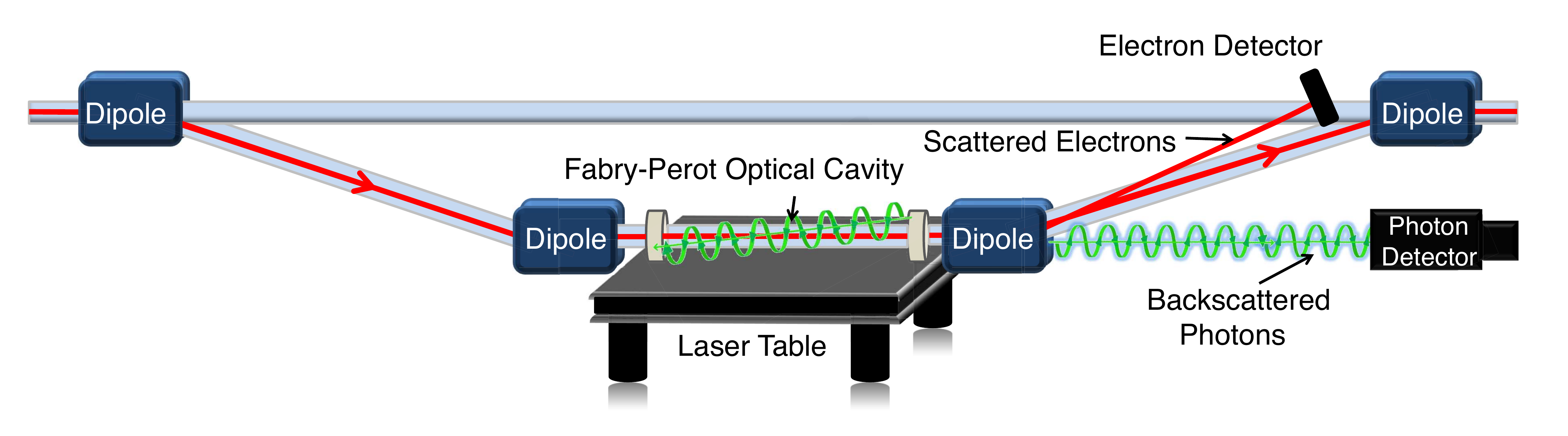}
\end{center}
\vspace*{-0.5cm}
\caption[det plane]{\label{fig:ComptonCartoon} A sketch showing the layout of the Compton polarimeter. The total length of the dipole chicane was 11.1~m, and the laser interaction region was 57~cm below the undeflected beamline. The electron beam trajectory 
is from left to right in the figure.}
\end{figure*}

A layout of the Compton polarimeter based on $\vec{\gamma} \vec{e} \longrightarrow \gamma e$ which was built for the experiment is shown in Fig.~\ref{fig:ComptonCartoon}. The electron beam was deflected vertically by two dipole magnets to where it could interact with photons in a moderate-gain laser cavity. The unscattered electron beam was deflected back to the nominal beamline with a second pair of dipole magnets. The third of the 4 chicane magnets also served  to spatially separate electrons that had undergone Compton scattering from the rest of the beam. 
These Compton recoil electrons were detected in a  multi-plane diamond strip detector.
Compton scattered photons passed through the third magnet and were detected in an array of PbWO$_4$ crystals. 
The absolute beam polarization was continuously measured to an accuracy of better than 1\% per hour  with the Compton electron detector. 

\subsubsection{Compton Laser System}
\label{sec:Components:Laser}

The photon target for the Compton electron beam polarimeter was composed of a  Coherent Verdi 10 laser~\cite{Claser1} 
with an output of 10~W at 532~nm and locked to an external Fabry-Perot optical cavity with a gain of approximately 200. The optical elements used to produce the photon target were located on an optics table 57~cm below the electron beam.
The $\sim$80 cm long optical cavity crossed the electron beam at $ 1.3 ^\circ$. 

 A variety of  optics were utilized on the optics table to control the shape, intensity, helicity, and polarization of the laser. The 100\% linearly polarized laser beam was changed to 99.9\% circular polarization in the cavity by means of a linear polarizer and a quarter-wave plate. The laser polarization was continuously measured to $\pm$0.2\% using reflected light at the entrance mirror of the optical cavity~\cite{Claser2,Claser3}.

The optical cavity was locked using the Pound-Drever-Hall  locking technique~\cite{PDH} feeding back on the laser wavelength via \ac{acro-pzt}-actuated mirrors, internal to the Verdi laser cavity.
The laser frequency was modulated using an electro-optical modulator.  
The 
modulation signal, when mixed with the signal from a photodiode monitoring the reflected
light from the Fabry-Perot cavity, provided an error signal.
 The error signal was fed into a \ac{acro-pid} feedback circuit which maintained the optical cavity lock by appropriately adjusting the laser wavelength via the internal \ac{acro-pzt} actuators.

The transmitted beam was split into multiple  beams using a holographic beam sampler 
to simultaneously monitor its polarization, power, position, and image.

\subsubsection{Compton Photon Detector}
\label{sec:Components:Photon}

Photons which were Compton back-scattered from the electron beam passed straight through the third dipole of the chicane and entered a calorimeter array composed of four 20~cm long stacked PbWO$_4$  crystals each with a cross section of 3$\times$3~cm$^2$. 
A  single 7.6~cm diameter Hamamatsu R4885~\cite{R4885}  \ac{acro-pmt} with a gain of $5 \times 10^6$ was attached with optical grease to the back face of the calorimeter. Both were inside a thermally isolated box cooled to $\sim$$14^\circ$~C which increased the light yield of the crystals by $\sim$20\% compared to room temperature. 
The photon signals were digitized with a 12-bit, 4~V range, flash \ac{acro-adc} (Struck SIS3320~\cite{SIS3320}) sampling at 250~MHz. 
An energy-weighted integral over all photon energies was performed and read out at the helicity flip rate of  960/s. 
During each helicity period, at least one set of 256 contiguous 
samples was read out to 
monitor the health 
and help determine the linearity of the detector.
The linearity was studied with a system modeled after~\cite{NIM:MFriend}, 
composed of two \ac{acro-led}s pulsing for $\sim$60~ns with one \ac{acro-led} serving as a reference signal and the other \ac{acro-led} with intensity spanning the response range of the detector.

Due to the challenging nature of the linearity measurements, absolute polarizations have not  been extracted from the Run 2 photon detector data as of this writing. The analysis is currently focusing on relative comparisons with the electron detector results. 
The quasi-independent absolute beam polarization measurements provided by the electron detector are discussed next.

\subsubsection{Compton Electron Detector}
\label{sec:Components:Compton:EDet}

The recoil electrons from the Compton scattering process were momentum analyzed in the third dipole 
magnet and detected by a set of micro-strip detectors located just upstream of the fourth 
dipole magnet.  The micro-strip electron detectors were made from 21$\times$21$\times$0.5~mm$^3$ 
plates of chemical vapor deposited 
 diamond~\cite{e6corp}. Each diamond plate 
had 96 metalized horizontal strips with a pitch of 200~$\mu$m (including a 20~$\mu$m gap) 
on one side (front) and a single metalized electrode 100~$\mu$m~\cite{ddl} thick covering the entire diamond surface
 on the opposite  side. 
Each diamond plate was epoxied to a 
60~mm $\times$ 80~mm alumina substrate.
Each of the 96 strips 
was wire bonded to gold traces
 on the alumina substrate which terminated on two 50-pin high-density connectors~\cite{conn} placed on either side of the detector  plate. 

The four detector planes were spaced $\sim$1~cm apart and inclined 10.2$^\circ$ to align them perpendicular to the electron beam exiting the third dipole in the chicane. 
The detector stack 
was attached to a vertical linear feedthrough with 30.5 cm of travel inside a vacuum can. 
Under normal operating conditions the detectors were 
lowered to a vertical distance of $\sim$7~mm from the main electron beam. 
When not in use the detectors were retracted into a section of the vacuum chamber well separated from the electron beam.
At the bias voltage of $-400$~V maintained across each plane, the raw charge signal was $\sim$9000 e$^-$ per hit.
Custom-built low-noise 
\ac{acro-qwad}s~\cite{qwads} were used with a typical gain of 100~mV/fC.

The digital signals from the \ac{acro-qwad}s were carried  via 60~m of cable to four 
\ac{acro-fpga} based general purpose logic boards~\cite{v1495}. These provided the trigger and reconditioned the signals for the independent Compton data acquisition  system~\cite{coda}. 
             
\begin{figure}[!htb]
\begin{center}
\hspace*{-0.3cm}
\includegraphics[width =0.48\textwidth,angle=0]{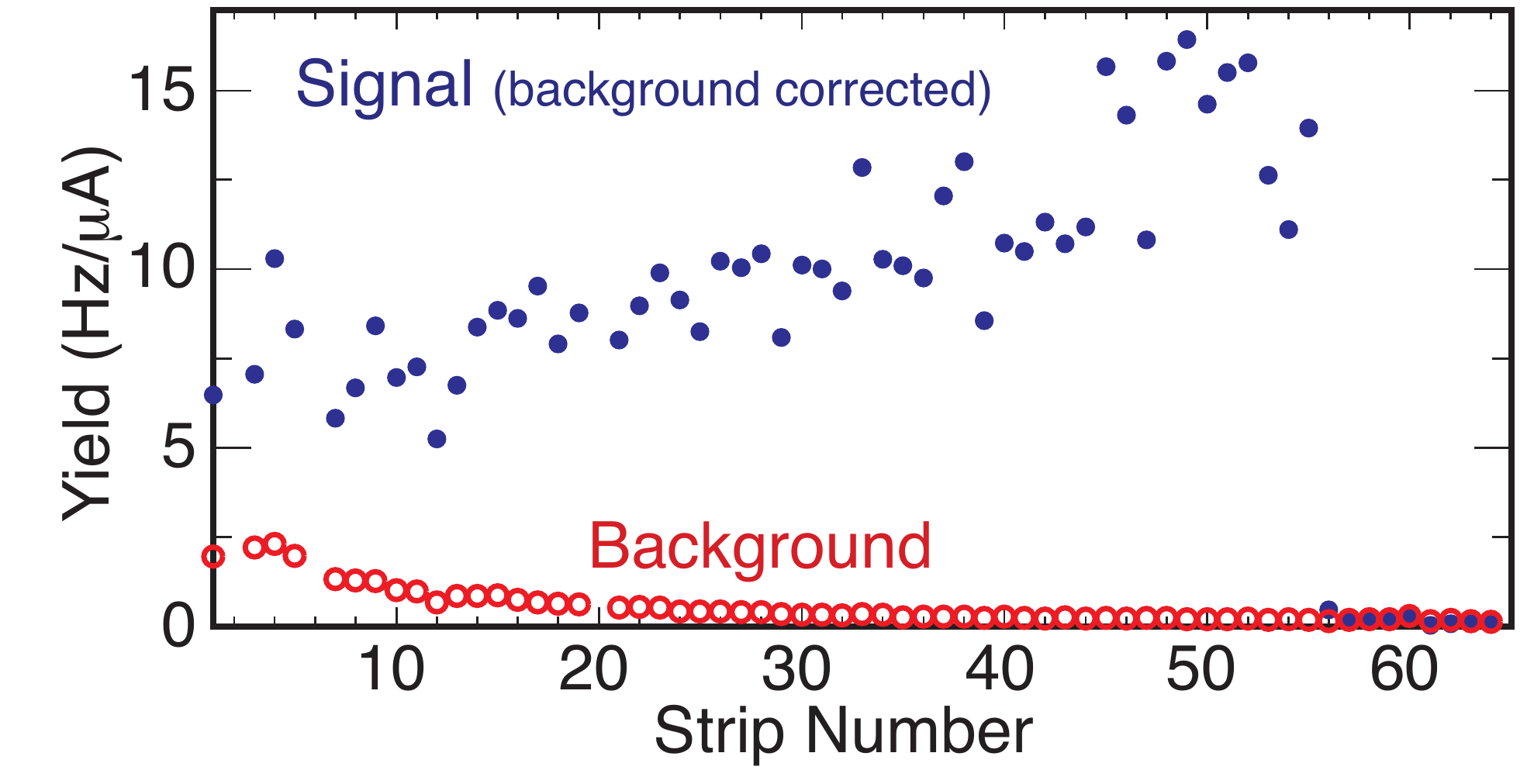}
\end{center}
\caption[det plane]{\label{fig:striphit} A spectrum of normalized yield for each strip of a single detector plane. The three empty strips correspond to strips that were too noisy and had to be masked. Strip 1 is closest (7~mm) to the beam; the Compton edge is at strip 56. }
\end{figure}

\begin{figure}[!htb]
\begin{center}
\includegraphics[width =0.47\textwidth,angle=0]{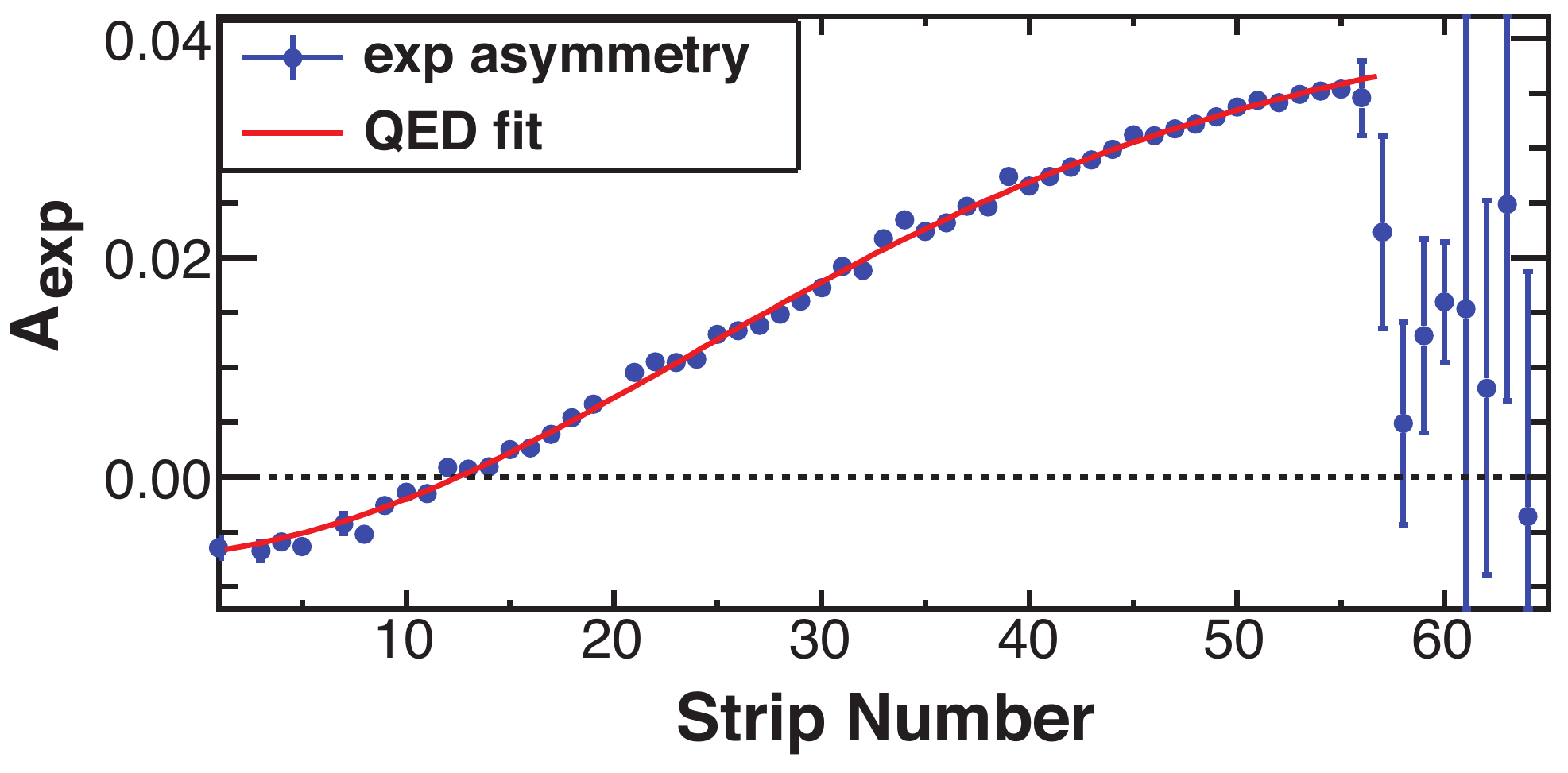}
\end{center}
\caption[det plane]{\label{fig:asym} A spectrum of the measured asymmetry for each strip along with the fit to a \ac{acro-qed} calculation shown by the solid line. The three missing strips, out of the $>$ 50 active strips, have a negligible impact on the quality of the fit and the extracted electron polarization.}
\end{figure}

The data were collected in $\sim$1~hr long runs which were later decoded and used to fill histograms of hits on each detector strip for each electron beam helicity.  Laser-off data were used to build background spectra. 
Only 3 out of the 4 detector planes were operational during the experiment. 
A typical strip hit spectrum is shown in Fig.~\ref{fig:striphit}. 
Using the background corrected strip hit spectra for each electron helicity state, the  asymmetry can be determined as a function of electron momentum. These asymmetry spectra were compared with a  
\ac{acro-qed} calculation~\cite{Prescott:1973ek}  to obtain the electron beam polarization. A typical asymmetry spectrum along with the \ac{acro-qed} calculation is shown in Fig.~\ref{fig:asym}. The electron beam polarization was continuously  monitored throughout the \qweak experiment using the Compton polarimeter electron detectors described in this section.  The beam polarization obtained using the electron detector was  consistent with the \moller{} polarimeter measurements performed at low beam currents.

Table \ref{tab:compton_systematics} summarizes the  
 systematic uncertainties associated with the Compton polarimeter. The largest two uncertainties arose from a timing issue which resulted in occasional loss of information from a plane. The effect depended on rate. The preliminary correction for this effect contributed an average of 0.7\% to the overall polarization uncertainty, with an additional 0.35\% point-to-point variation observed over the course of the experiment.

\begin{table}[ttbh]
  \centering
    \begin{tabular}{|l|c|c|} \hline
                            & Uncer- & $\Delta P/P$ \\ 
      Source 				& tainty & (\%) \\ \hline
      Laser polarization   		& 0.18 mm      & 0.18     \\
      3$^{rd}$ Dipole field		& 0.0011 T    & 0.13      \\
      Beam energy    			& 1 MeV   & 0.08     \\
      Detector $Z$ position      & 1 mm         & 0.03     \\
      Trigger multiplicity       & 1-3 plane         & 0.19     \\
      Trigger clustering         & 1-8 strips  & 0.01     \\
      Detector tilt ($X$)        & 1$^\circ$   & 0.03     \\
      Detector tilt ($Y$)        & 1$^\circ$   & 0.02     \\
      Detector tilt ($Z$)        & 1$^\circ$   & 0.04     \\
      Strip eff. variation        & 0.0 - 100\%   & 0.1     \\
      Detector Noise	         & $\leq$20\% of rate  & 0.1     \\
      Fringe Field		         & 100\%       & 0.05     \\
      Radiative corrections      & 20\%		   & 0.05     \\
      DAQ ineff. correction & 100\% (prelim) & 0.7     \\
      DAQ ineff. pt-to-pt	 & (prelim)		& 0.35     \\ \hline
      Total                          &        & 0.85     \\ \hline
    \end{tabular}
  \caption{The Hall C Compton polarimeter systematic uncertainties determined for Run 2 of the experiment. The last two rows in the table are preliminary estimates, and are expected to be smaller upon completion of their analysis. }
  \label{tab:compton_systematics}
\end{table}

\subsection{Performance}

The beam polarization was monitored by both the {\moller} and Compton polarimeters during the {\qweak} experiment.
In general the Compton polarimeter ran continuously and concurrently with data--taking for the experiment, achieving statistical errors ranging from a little more than 1\% per 1--hour run (during the latter half of Run 1) to less than 0.5\% per 1--hour run (Run 2). Each invasive {\moller} measurement took 4--6 hours, and therefore, as previously stated, was used only 2--3 times per week.

\begin{figure}[htb]
 \begin{center}
   \includegraphics[width=0.475\textwidth]{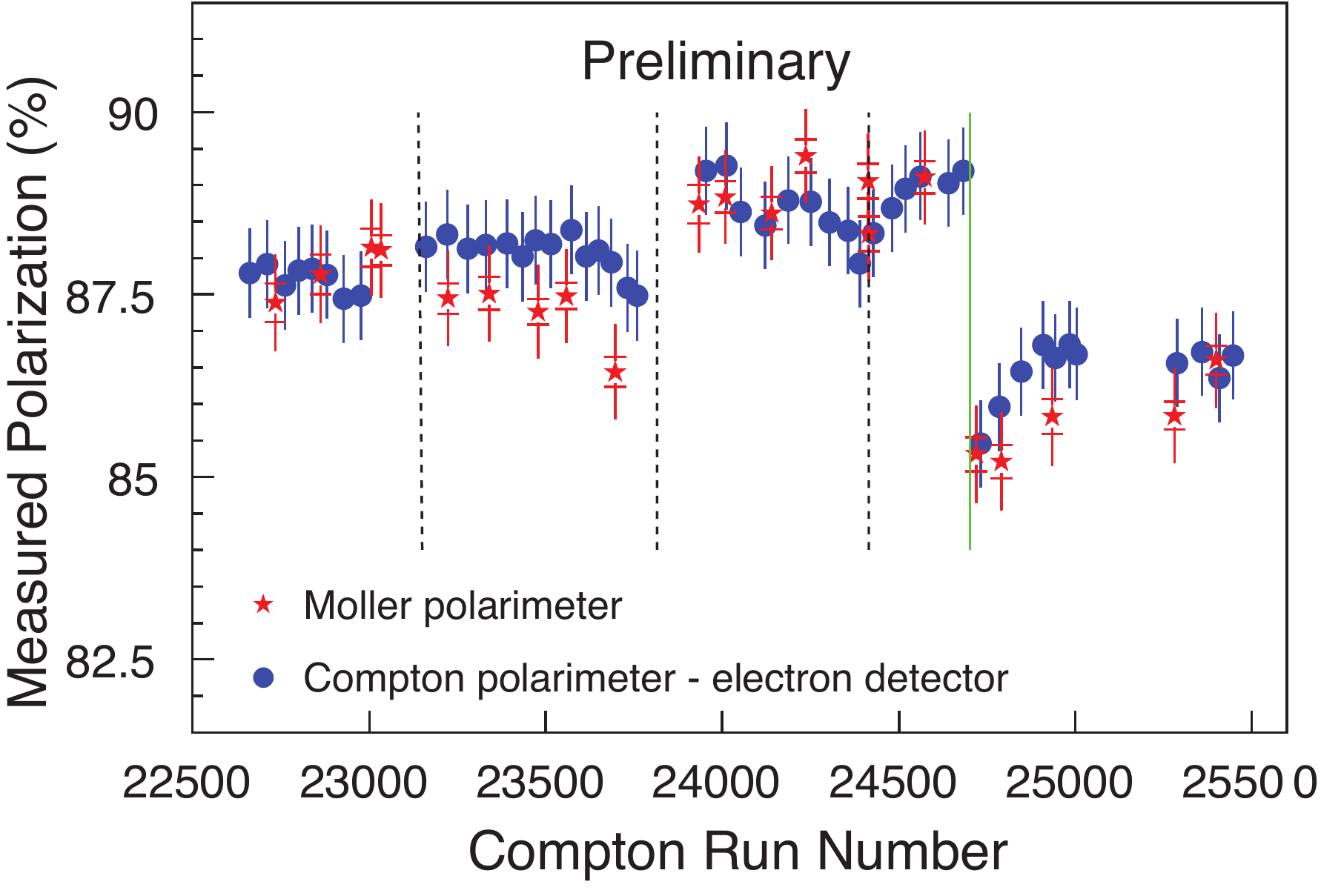}
   \caption{\label{fig:run2_pol_compare} Beam polarization measurements for the second {\qweak} run period plotted vs. Compton polarimeter run number (spanning a roughly 6-month period) 
   The 
   solid stars show the {\moller} measurements with statistical uncertainties (inner) and statistical $+$ point-to-point systematic uncertainties added in quadrature (outer) (see Sec.~\ref{sec:Components:Polarization:Moller}) - the overall normalization uncertainty of 0.64\% is not shown. The 
   solid circles show the preliminary results from the Compton polarimeter electron detector - each point represents an average over about 30 hours of data. The uncertainties shown represent statistical and estimated systematic 
   errors added in quadrature. Vertical dashed lines denote changes in the position of the electron source laser position on the photocathode, while the solid vertical
   line marks a heating and subsequent re--activation of the photocathode.}
 \end{center}
\end{figure}

During Run 2, both the {\moller} and Compton polarimeters were functioning correctly and with good efficiency. Results from both devices contributed to the extracted values of the beam polarization for that period. Fig.~\ref{fig:run2_pol_compare} compares results from both polarimeters where polarization measurements taken under similar beam conditions are plotted. The overall agreement is good. The stability of the beam polarization measured by the Compton polarimeter justifies the time interval chosen for the more infrequent {\moller}  polarimeter measurements. However, {\moller} measurements were always made immediately  before and after changes in the polarized source that were known to affect the beam polarization: notably laser spot changes or reactivation of the injector photocathode. 

The commissioning of the Compton polarimeter was completed near the middle of Run 1, so
 polarization results for the first part of Run 1  
come exclusively from the {\moller} polarimeter. The quad problem noted earlier in Sec.~\ref{sec:Components:Polarization:Moller}  significantly complicated the analysis of {\moller} data from that period and resulted in greater uncertainty for the affected Run 1 {\moller} data. 

The availability of two independent polarimeters was extremely useful. During the first run period, results from the Compton polarimeter first brought attention to potential issues with the {\moller} polarimeter, which later resulted in the discovery of the broken quadrupole. Cross checks of the two polarimeters were performed by making measurements at the same beam current (4.5~$\mu$A) - normally the {\moller} took data at $\approx$1~$\mu$A while the Compton operated at the nominal beam current of the experiment, 180~$\mu$A. An important by--product of this measurement was confirmation that the beam polarization measured at low beam currents is identical to that measured at high beam currents.

\section{The Liquid Hydrogen Target}
\label{sec:Target}

\begin{figure}[!htb]
\vspace*{-1.8cm}
\begin{center}
\hspace*{-1.8cm}
\includegraphics[width =0.95\textwidth,angle=-90]{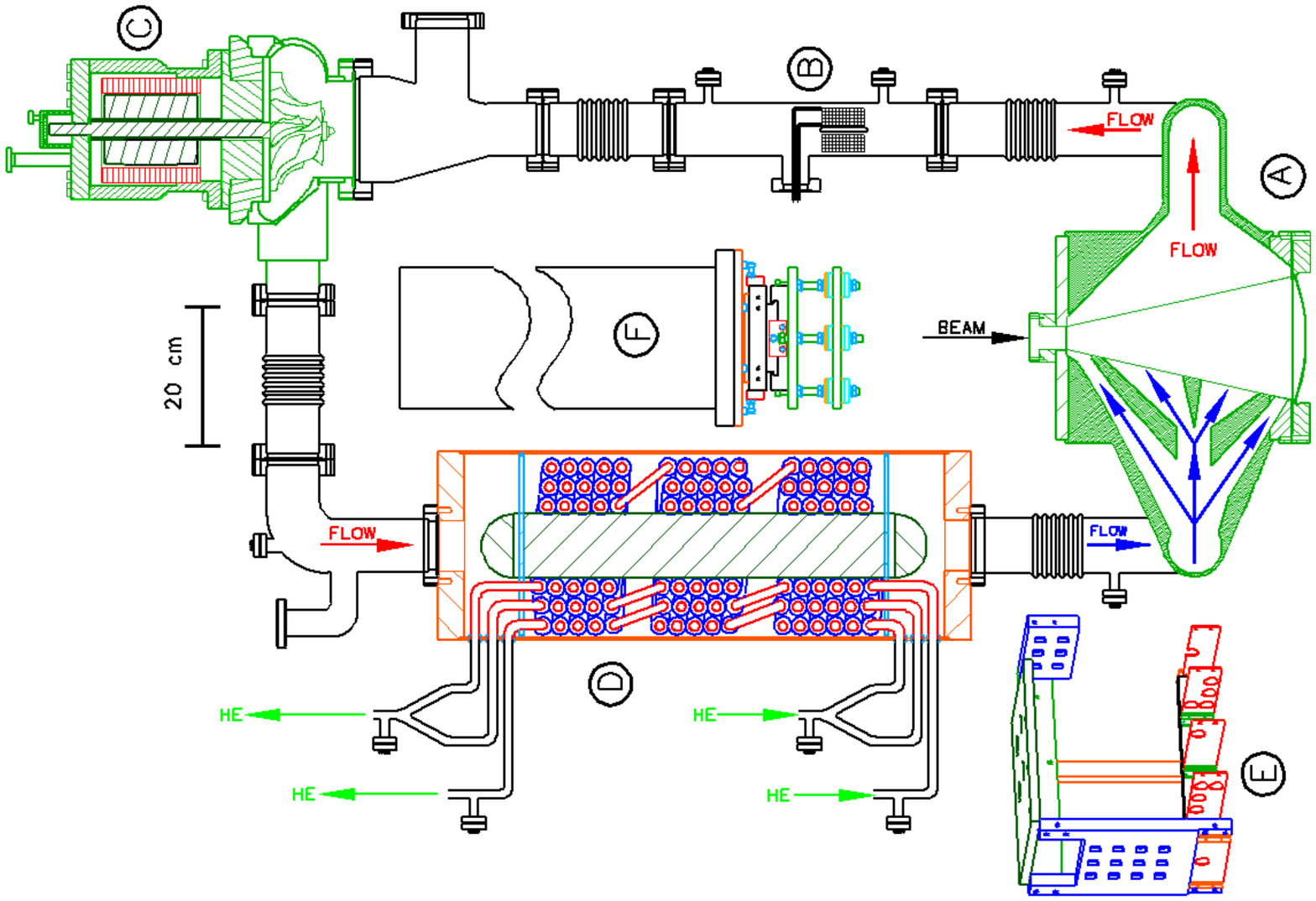}
\end{center}
\vspace*{-1.5cm}
\caption[tgt-layout]{\label{fig:tgt-layout} A schematic showing the components of the \Qweak target. A: The beam interaction cell (pitched 90$^\circ$ in this figure in order to illustrate the flow pattern), B: the heater, C: the centrifugal re-circulation pump, D: the hybrid heat exchanger, E: the solid target ladder, which was mounted directly below the cell, and F: the long thin stainless steel  pipe which mechanically supported the loop, as well as the manual cell adjustment mechanism at its lower end. }
\end{figure}

The \Qweak liquid hydrogen (LH$_2$) target (see Fig.~\ref{fig:tgt-layout}) 
 consisted of a closed hydrogen loop whose main components were a pump to circulate the H$_2$,
a \ac{acro-hx}  to liquify the H$_2$ and remove the heat deposited by the $e^-$ beam,
a cell with thin windows where the beam interacted with the H$_2$, and
a heater to replace the beam power when the beam was off and to regulate the temperature of the H$_2$. 
The target was designed~\cite{LH2Tgt_1, LH2Tgt_2} to operate at 20 
K and 207-228 kPa.
It was connected at all times to 
storage/ballast tanks with a total volume of 23,000 \ac{acro-stp} liters.
The volume of LH$_2$  was $\sim$ 58 liters.

The ionization energy loss of the
1.16 GeV electron beam traversing the 34.4 cm of LH$_2$ was
 2.1 kW. A further 0.7 kW of cooling power was provided for viscous heating (180~W), pump heat (150~W), conductive and radiative heat load (150~W), as well as reserve power for the heater (250~W).

The target loop was affixed to  a 1.6~m long 
stainless steel  
 pipe (see Fig.~\ref{fig:tgt-layout} F) in an evacuated scattering chamber connected to the beam line. 
A small fast-acting gate valve isolated the chamber from the upstream beamline. The downstream end of the chamber was provided with a custom~\cite{GNBValves}  40.6~cm diameter, extended-stroke gate valve which isolated the chamber from the downstream beamline as well as the thin downstream vacuum window on the chamber.  
The gate retracted into a  lead box when the valve was open and the beam was on to protect the ethylene propylene diene monomer (M-class) rubber (EPDM) seals on the gate from radiation damage.
A spoked aluminum 2024-T4 vacuum window with eight 0.89 mm thick windows was attached to the downstream flange of the gate valve. The scattered electrons passed through the open gate valve, through these windows, and into the collimation system on their way to the experiment's detectors.

\subsection{Target Components}
\label{sec:Target:Components}

The target cell's central LH$_2$ volume was a conical section oriented along the beam axis such that all electrons scattered less than 14$^\circ$ passed through the larger diameter exit window. This comfortably included scattered electrons in the experiment's 
$5.8^\circ$$ <$$ \theta$$ < $$11.6^\circ $ acceptance. 
The Al 2219 cell contained strategically segmented inlet and outlet manifolds (see Fig.~\ref{fig:tgt-layout} A) which directed the flow of LH$_2$ transversely across the beam axis (at $\sim$3 m/s)
and toward the center of both windows (at $\sim$7 m/s).
The precise geometry of the cell and its manifolds was arrived at iteratively using 
\ac{acro-cfd} simulations.

The 22.2~mm $\phi$ entrance window of the cell was 0.097~mm thick Al 7075-T6. The Al 7075-T6 exit window of the hydrogen cell was a 0.64 mm thick machined surface 305 mm in diameter with a 254 mm radius of curvature. The unscattered beam passed through a thin  spot 15 mm in diameter and 0.125 mm thick at the center of the exit window.
The LH$2$ thickness seen by the beam between the entrance and exit windows was 343.6 mm (after correction for thermal contraction and pressure expansion), or 3.9\% expressed in radiation lengths.

In order to provide the nearly 3 kW of cooling power required, a hybrid counterflow \ac{acro-hx} was built (see Fig.~\ref{fig:tgt-layout} D) that made use of $\sim$14~K helium coolant from the \ac{acro-esr} as well as $\sim$5~K helium coolant from the \ac{acro-chl}.
Typical target coolant mass flows were $\sim$14~g/s (5~K source) and $\sim$40~g/s (14~K source). 
The unusually high 14~K mass flow was achieved by recovering the unused enthalpy of the returning 5~K coolant to pre-cool the 14~K helium supply in the \ac{acro-esr}.

The \ac{acro-hx}~\cite{Meyer} was composed of 12.7 mm $\phi$ copper fin tube with 6.3 fins/cm. The fin tube 
was wound in three 15-turn layers 
contained in a 27.3 cm $\phi$ stainless steel shell 70.6 cm long.
A 9.2 cm $\phi$ solid Al mandrel ran the length of the central axis of the cylindrical \ac{acro-hx} to divert the H$_2$ flow across the fin tubes.

The 3 kW capacity heater (see Fig.~\ref{fig:tgt-layout} B)
consisted of
1.83 mm $\phi$ nichrome wire  wrapped in four 23-turn layers  through perforated G10 boards.    
The total resistance (cold) was 1.3 $\Omega$.
A 60~V, 50~A Sorenson \ac{acro-dc} power supply~\cite{sorensen} 
was used to energize the heater.
The power sent to the heater was determined by a \ac{acro-pid} feedback loop looking at the hydrogen temperature as well as the e$^-$ beam current.

The LH$_2$ was circulated around the target loop with a
homemade centrifugal pump rotating at typically 29.4 Hz (see Fig.~\ref{fig:tgt-layout} C). The pump provided
a differential pressure (head) of 7.6~kPa (11~m), 
and a LH$_2$ mass flow of 1.2 kg/s (17.4$\pm$3.8 liters/s) determined from measurement of the temperature~\cite{Cernox} difference across the heater.
The 220 kPa system pressure was well above the  parahydrogen vapor pressure (94 kPa at 20~K) to mitigate cavitation.

The pump was made by adapting a commercial aluminum automotive turbocharger impeller and volute
 to an \ac{acro-ac} induction 
 motor~\cite{Baldor}. Several turns of copper pipe carrying returning 20~K helium coolant were wrapped around a custom motor housing to help remove heat from the motor. 
Initially, bearings employing ceramic balls and race with a teflon retainer failed. 
They were  replaced with bearings using ceramic balls, a stainless steel race, and graphite impregnated vespel retainers. The pump was further modified  to promote a small flow of LH$_2$ across the bearings. One end of the motor shaft spun the 142~mm $\phi$ impeller, the other spun a small tachometer magnet.

\subsection{Solid Targets}
\label{sec:Target:SolidTgts}

A remotely controlled 2-axis motion system with 600 mm of vertical travel and 86 mm of horizontal travel was used to position the LH$_2$ target or any of 24 solid targets on the beam axis. The
solid targets were distributed across  three arrays in an aluminum target ladder assembly 
(see Fig.~\ref{fig:tgt-layout} E)
in good thermal contact with the bottom of the LH$_2$ target cell. Each target in the upper two arrays was 2.5 cm square.
The lower array was composed of various combinations of foils in 2 rows and 3 columns at five ($Z$) positions along the beam axis between the upstream (entrance) and downstream (exit) LH$_2$ cell windows. The combinations of ``optics targets" in this  array were used to aid the development of vertex reconstruction algorithms at $\sim$100~pA beam currents.

A second array of 12 targets arranged in 4 rows and 3 columns was situated at the same ($Z$) plane along the beam axis as the upstream window of the target cell. Likewise, a downstream array of six targets arranged in 2 rows and 3 columns was located at the $Z$ of the exit window of the LH$_2$ cell. These  two arrays  were used for separate background subtraction of the upstream and downstream aluminum cell windows of the LH$_2$ target. Different thickness aluminum background targets were provided in both the upstream and downstream matrices to benchmark radiative corrections~\cite{Myers}. Targets of pure aluminum, thick and thin carbon targets, and beryllium were also provided. Other targets in these arrays were used to measure the relative location of the beam and the target system using a BeO viewer in conjunction with a TV camera looking at the targets, as well as thin aluminum targets with various size holes in their centers. 

\begin{figure}[ht]
  \centerline{\includegraphics[width=0.5\textwidth,angle=0]{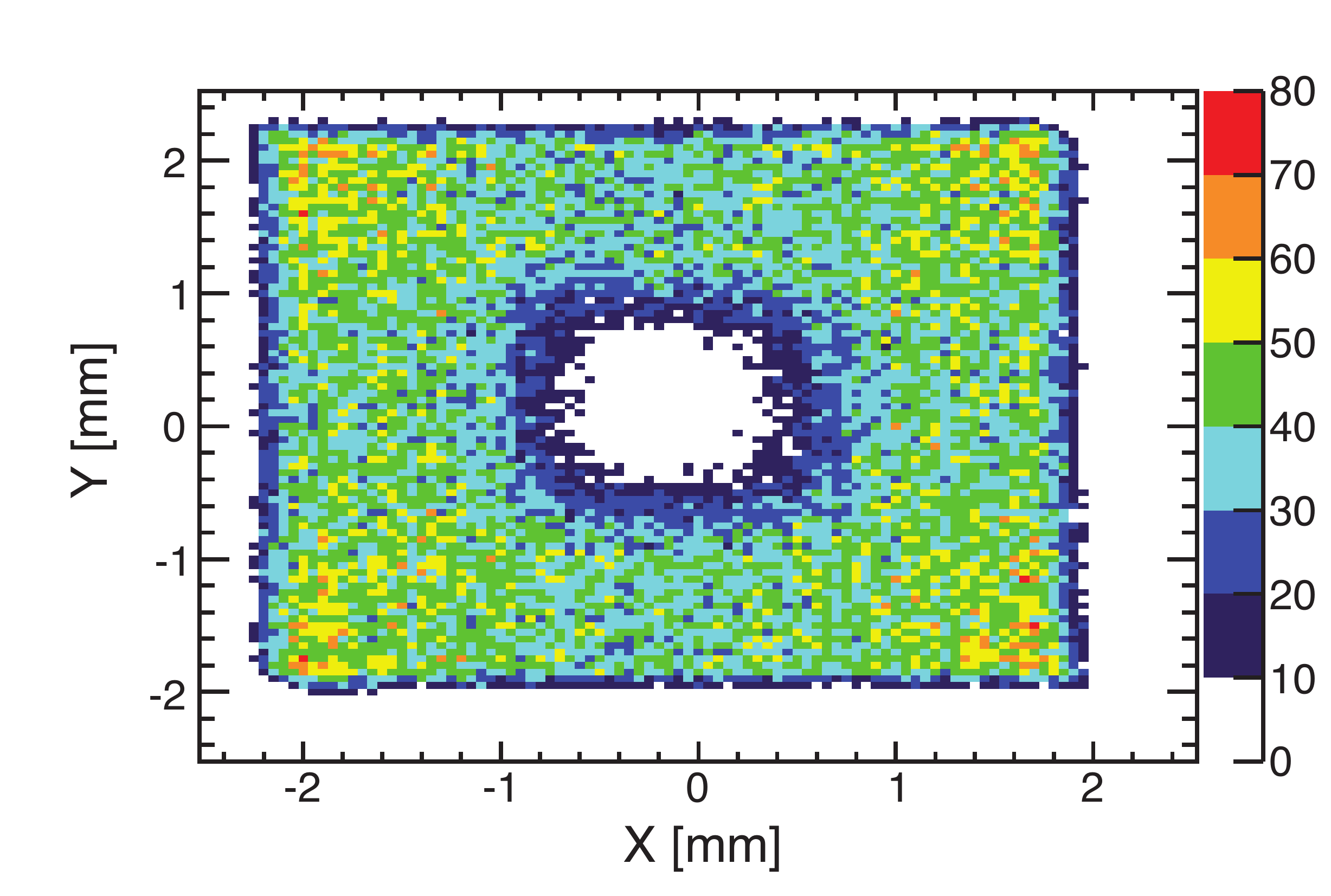}}
\caption{\label{fig:holetgt} Profile of the beam position on the hole target. 
The central area devoid of events represents the 2 mm $\times$ 2 mm hole in the target illuminated by a 4 mm $\times$ 4 mm dithered beam.}
\end{figure}

These latter hole targets were especially useful to position the  target system with respect to the beam. One mm thick aluminum ``hole targets" with two mm  square holes punched out of their centers were moved into the beam. 
A 2-dimensional profile of the beam position at the target was generated using the dithering/raster magnets  (described in Sec.~\ref{sec:accelerator}). Only beam which missed the hole, and could thus scatter into the detectors generated a trigger (see Fig.~\ref{fig:holetgt}).
 By measuring the hole profiles at both the upstream and downstream $Z$ locations, the $X, Y$, pitch, roll, and yaw of the extended target could be accurately determined. Offsets in $X$ and $Y$ could be corrected in real time using the 2-axis motion system. Pitch, roll, and yaw offsets were corrected with a manual cell adjustment mechanism when the target was warm.
The success of the target positioning achieved using the hole targets was confirmed after the experiment by inspection of spots left by the beam on the target cell windows as well as the solid targets. In all cases, the spots were well within 1 mm of the center of each target. 

\subsection{Target Performance}
\label{sec:Target:Performance}

Except for the early failure of the LH$_2$ pump bearings mentioned above, 
the target met expectations. 
About eight hours were required to condense the hydrogen in the target. 
Warming the target to room temperature (by stopping the coolant flow) took about two days.

The temperature \ac{acro-pid} feedback loop  on the heater kept the target temperature at 20.00 $\pm$ 0.02 K while the beam was on. Damped oscillations of 100 mK were observed for 2--3 minutes when the 180 $\mu$A beam was interrupted.  To improve temperature stability upon restoration of beam, the beam  was ramped back to full current at the rate of 5 $\mu$A/s.  

At high beam currents, the small bulk reduction of the nominal 71.3 kg/m$^3$ density of the LH$_2$ target was obscured by percent-scale nonlinearities in the main detector signal chain and the \ac{acro-bcm}s used to normalize the signals.
A bulk LH$_2$ density reduction of 0.8\% $\pm$ 0.8\% at 180 $\mu$A was estimated by comparing changes in the detector yield as a function of beam current for the LH$_2$ and solid targets.

The target was also operated at a beam current of 2 $\mu$A with various pressures of cold H$_2$ gas, as well as with the target loop evacuated, in order to characterize the background from the cell windows. 

The primary metric of target performance was its contribution to the main detector asymmetry  width $\sigma_A$ (measured over quartets), as discussed in Sec.~\ref{sec:Asymmetry}. 
This contribution arises from target noise near the helicity reversal frequency and includes density fluctuations from all sources. Because of the high beam current employed in the experiment, fast helicity reversal 
was essential in reducing the ratio of target noise to counting statistics to a nearly negligible level.

\begin{figure}[hhbt]
\hspace*{0.75cm}
\centerline{\includegraphics[width=0.6\textwidth,angle=0]{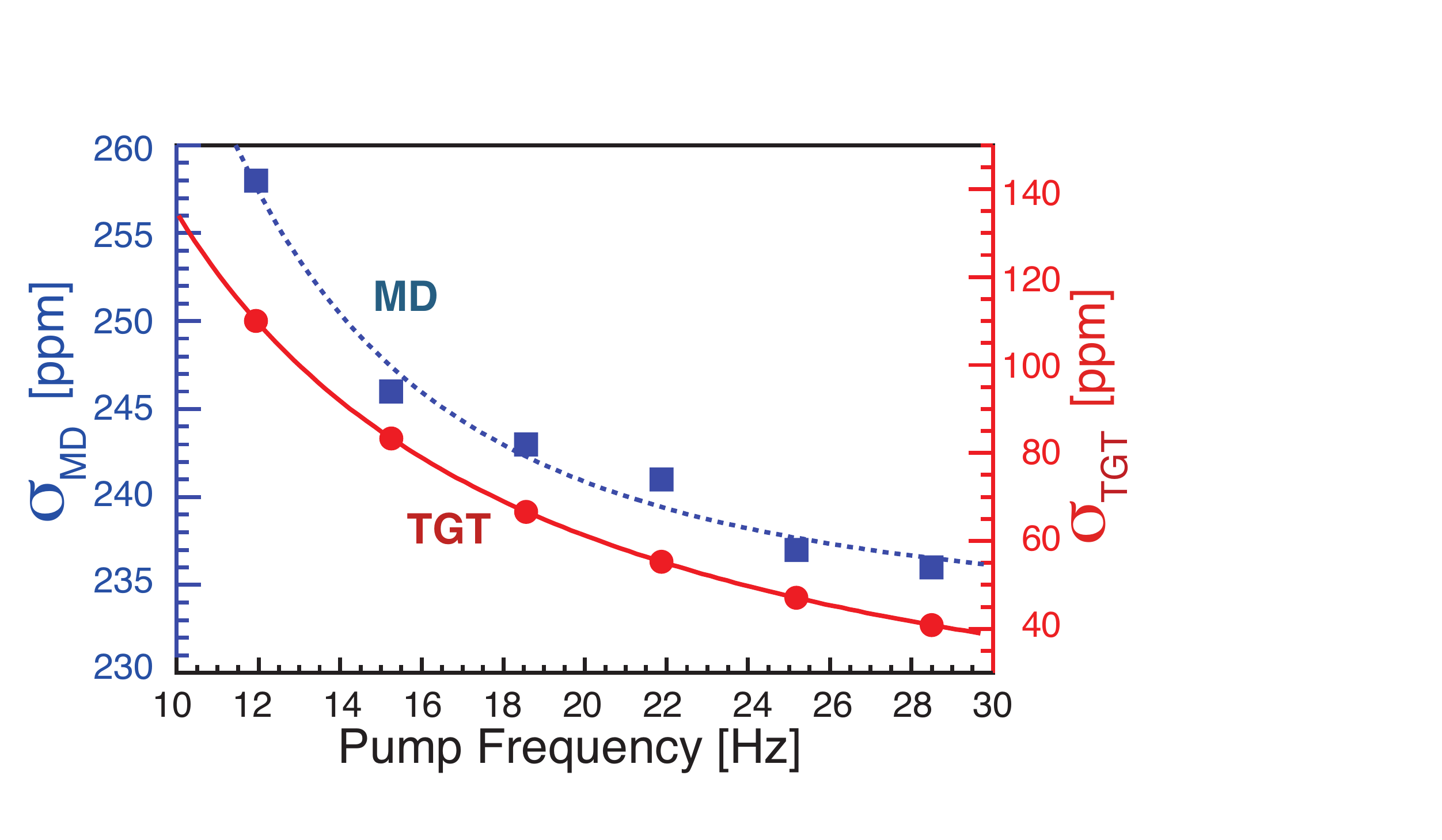}}
\caption{\label{fig:boiling} The measured quartet-level main detector asymmetry width measured at varying rotational speeds of the LH$_2$ re-circulation pump (solid square points, dotted fit). The fit function is $\sigma_{MD}^2 = \sigma_A^2 = 233^2 + {1889^2}/{f^{2.288}}$, where f is the pump frequency in Hz. Conditions were 169~$\mu$A with a 4x4~mm$^2$ raster. 
The solid circles and fit curve are the corresponding target noise results  (right hand vertical scale) deduced assuming the variation in the main detector asymmetry width is due exclusively to varying target noise.}
\end{figure}

The target noise was explicitly measured~\cite{Smith:2012ur} using three independent techniques, by measurement of the asymmetry width in the main detectors as a function of either beam current, rastered beam spot size at the target, or the rotational frequency of the hydrogen re-circulation pump. The latter is the cleanest and surest method, however consistent results were obtained using all three methods. Results from one of the target noise studies are provided in Fig.~\ref{fig:boiling}. 
At the nominal conditions of the experiment (180 $\mu$A, 4$\times$4 mm$^2$ raster, 28.5 Hz pump speed), the target noise 
was 53 ppm with an estimated 5 ppm uncertainty.

\section{Collimation and Shielding}
\label{sec:collimation_and_shielding}

The experiment was carefully designed to  mitigate the extraordinarily high levels of radiation resulting from the use of a large beam current on a long target. Besides choosing radiation-hard materials 
(\textit{e.g.}, 
Spectrosil 2000 synthetic quartz~\cite{S2000} for the detectors), and materials with relatively short half-lives when activated 
(\textit{e.g.}, 
aluminum beamline components instead of stainless steel),
two collimation systems and heavy use of shielding around the target area and the main detectors were employed. Portions of the collimator region shielding and the detector shielding hut are shown in Fig.~\ref{fig:QweakApparatus}.

\subsection{Triple Collimator}
\label{sec:Components:Collimator}

The main collimation system (see Fig.~\ref{fig:QweakApparatus}) consisted of a triplet of lead antimony (95.5\% Pb, 4.5\% Sb) collimators each  
with eight sculpted apertures that passed  scattered electrons into each of the experiment's eight octants. The first was a cleanup collimator 15.2 cm thick centered just 74 cm downstream of the target center to provide initial cleanup.  Its apertures were sculpted with 14 sides to allow electrons scattered from a  hypothetical $9\times9$~mm$^2$ beam envelope  anywhere along the target length to pass through the defining aperture of the second collimator. 
A pair of retractable 5 cm thick tungsten blocks could be positioned behind two opposing apertures of the first collimator 
to block scattered electrons for dedicated, intermittent background studies.

The downstream face of the   second  collimator  defined the acceptance for scattered electrons. It was centered 2.72~m downstream of the target center, and was 15.0 cm thick.  The electrons passed through eight six-sided openings, each approximately 400~cm$^2$ in area, defining an angular acceptance from the upstream end of the target of $\theta = 5.8^{\circ}-10.2^{\circ}$, and $\theta = 6.6^{\circ} - 11.6^{\circ}$ from the downstream end of the target. 

A third  cleanup collimator 11.2 cm thick was located 3.82 m downstream of the target center, at the entrance to the magnet. It was sandwiched between aluminum plates for support. It provided several centimeters  of clearance to the  elastic electron profile.

\subsection{Lintels}
\label{sec:Lintels}

Lead lintels  
were installed between the coils of the magnet to shield the detectors from  line-of-sight neutrals generated at the inner apertures of the defining collimator. The  lintels were located 70~cm upstream of the magnet's center, with a size of 26.2~cm radially, 70~cm long between adjacent magnet coils, and 10~cm thick with a forward pitch of 20.85$^{\circ}$.
They provided 2 cm of clearance to the elastic electron envelope, and are discussed further in Sec.~\ref{sec:bkgsims}.

\subsection{Beam Collimator}
\label{sec:WPlug}

The experiment was designed to minimize line-of-sight between the target and the aluminum beampipe in order to reduce backgrounds in the main detectors. Simulations showed that this could be almost completely achieved with a water-cooled tungsten-copper beam collimator 21 cm long  fit snugly in the central aperture of the most upstream collimator.
The upstream face of this 7.9 cm diameter beam collimator was attached to the central hub of the scattering chamber vacuum window only 47 cm downstream of the target cell's exit window. The beam passed through an evacuated tapered conical section machined out of the center of the collimator which was 14.91 mm in diameter at the upstream end and 21.5 mm in diameter at the downstream end. From there it was flanged to the downstream beamline. 
The power deposited on the beam collimator was $\sim$1.6 kW, derived from the measured water flow and temperature difference across it. 

\begin{figure*}[hhtb]
\centering
\includegraphics[width=0.89\textwidth,angle=0]{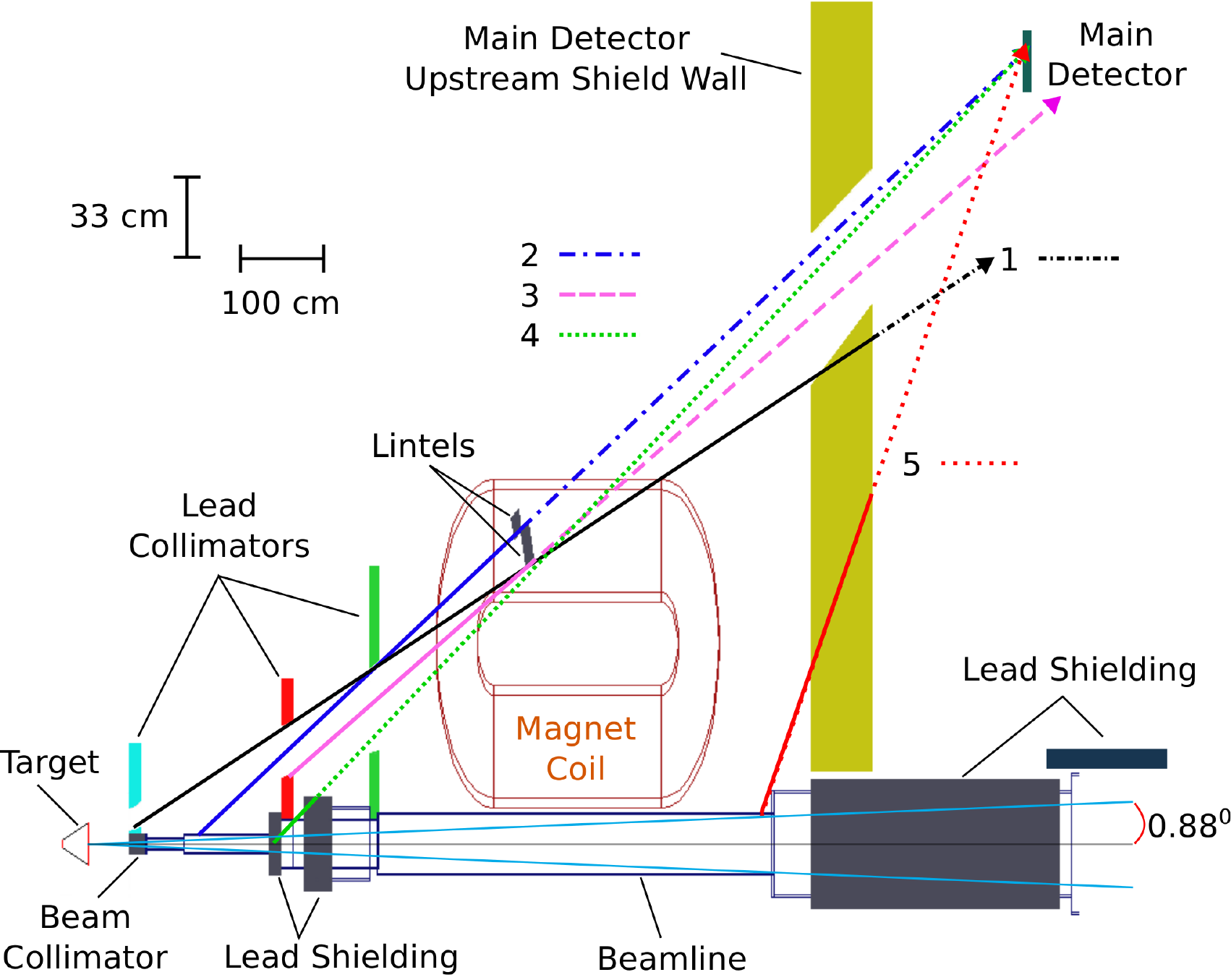}
\caption{\label{fig:BkgRays} 
A simplified cross-sectional elevation view through the beam axis illustrating the neutral background sources downstream of the target and how the main detectors were shielded from them. The vertical scale is amplified by a factor of three for clarity.  The beam passes from left to right in this figure. Only the topmost main detector is shown, along with the corresponding neutral particle trajectories that were shielded as discussed in the text. The trajectories are represented as solid lines until encountering the shielding put in place to  prevent them from reaching the main detectors, after which they are represented by different line types. The $\pm$0.88$^\circ$ cone inside the beam pipe represents the maximum angle passed by the beam collimator for events generated on the downstream face of the target, as discussed in the text.  
}
\end{figure*}

The maximum angle $\theta_{\rm max}=0.88^\circ$ passed by the beam collimator corresponds to events scattered at the downstream face of the target which intercepted the downstream aperture of the beam collimator. Including the corners of the 4x4 mm$^2$ square raster increases $\theta_{\rm max}$ to $1.11^\circ$. There were several regions along the downstream beamline which intruded on this cone, 
depicted in Fig.~\ref{fig:BkgRays}. Neutrals from the first region (ray 2 in Fig.~\ref{fig:BkgRays}) were mitigated by the lead lintels described in the previous section~\ref{sec:Lintels}. The lintels also blocked neutrals generated on the inner radius of the defining collimator apertures, represented by ray 3 in Fig.~\ref{fig:BkgRays}.  The second region along the beamline (ray 4 in Fig.~\ref{fig:BkgRays})
was discovered during the setup period of the experiment, using dosimetry and trial shielding. It was at the upstream face of the defining collimator, aggravated by the presence of one of the two stainless steel bellows along the beamline downstream of the target. 
After the setup run, an additional 5.1~cm of lead shielding was clamped along 15~cm of the beampipe upstream of the defining collimator, and after Run 1 an additional 30.5~cm of lead was added along the beampipe downstream of the defining collimator.
The third region was at the exit of the magnet (ray 5 in Fig.~\ref{fig:BkgRays}), just upstream of the detector hut shielding wall. This region was well shielded by the detector hut shielding wall, discussed next in Sec.~\ref{sec:Shielding}, as well as by surrounding the 
entire length of
beamline inside the detector shield hut with 5.1 cm of lead shielding. An additional (fourth) region along the beamline downstream of the main detectors was covered by the lead beamline shielding and did not contribute to the background. 
Finally, the main detectors were well shielded from neutral particles originating in the target by the triple-collimator system, as shown by ray 1 in Fig.~\ref{fig:BkgRays}.

\begin{figure}[ht]
\hspace*{-0.5cm}
\centering
\includegraphics[width=0.5\textwidth,angle=0]{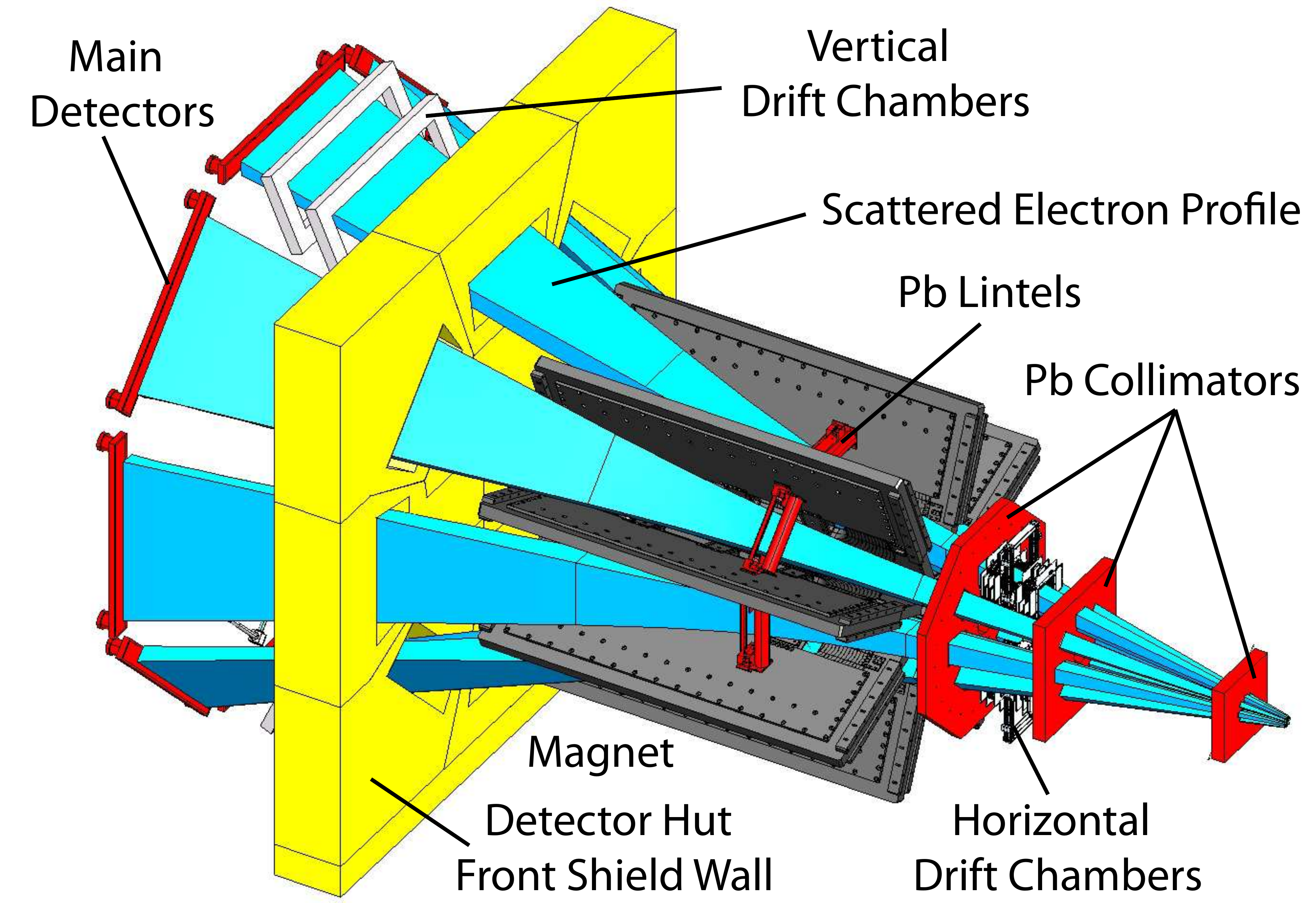}
\includegraphics[width=0.5\textwidth,angle=0]{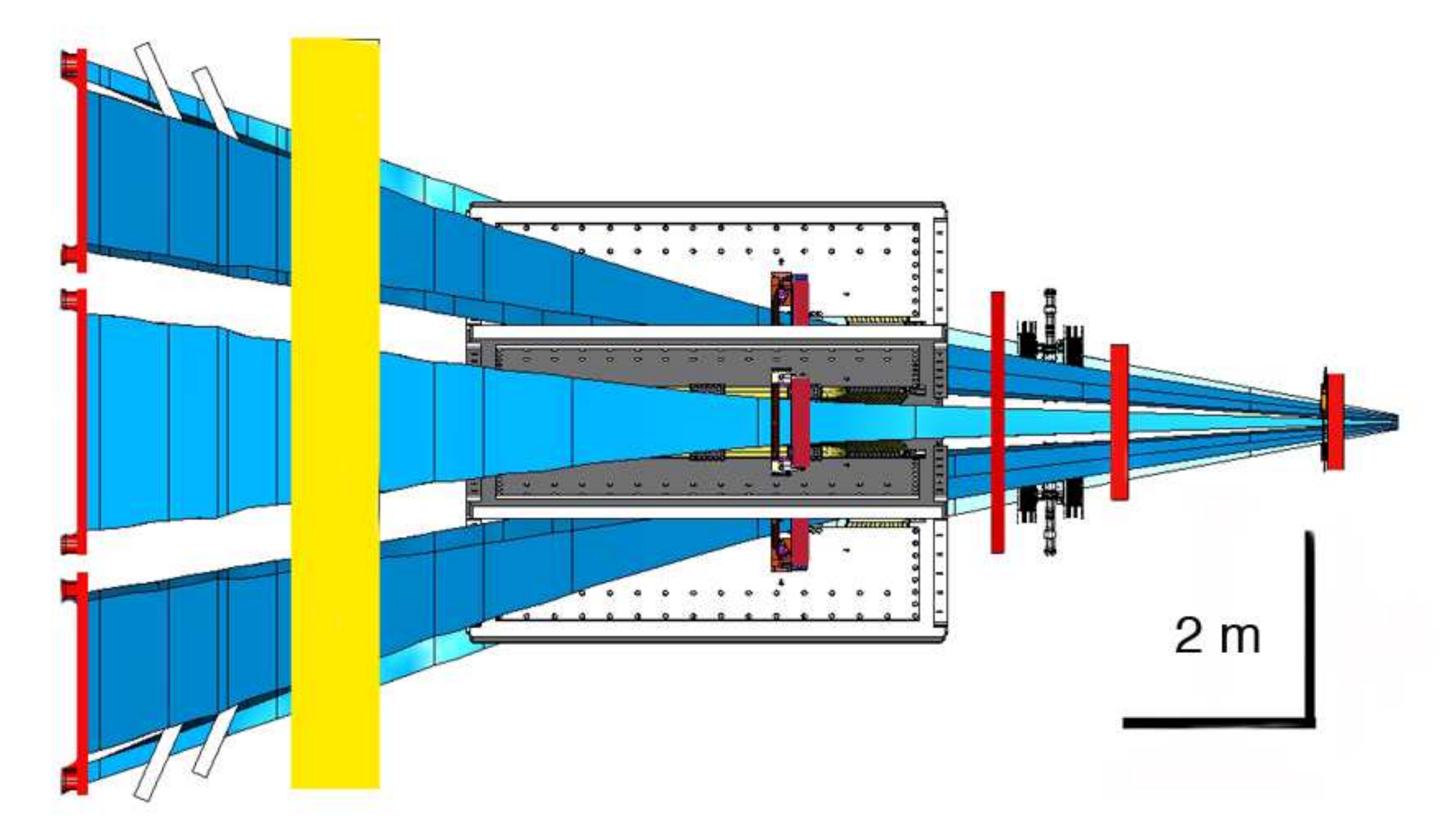}
\caption{\label{fig:envelopes}  \ac{acro-cad} drawings showing the eight scattered electron profiles  defined by the collimator triplet  passing through the apertures in the detector hut upstream shielding wall.
The upper figure is from a perspective similar to \protect{Fig.~\ref{fig:QweakApparatus}}. The bottom figure is a cross-sectional view, with scales provided. In both figures the beam goes from right to left. 
The scattered electron profiles were obtained from Monte-Carlo simulations and then overlaid into the CAD model. They originate at the LH$2$ target and terminate on the 8 synthetic quartz main detectors.
The lintels between the magnet coils  are also shown. The tracking chambers are  depicted between the second and third collimators as well as downstream of the detector hut.  
}
\end{figure}

\subsection{Shielding}
\label{sec:Shielding}
The region immediately downstream of the target scattering chamber between the first and second collimators was completely enclosed in concrete shielding 61 cm thick. Further downstream, the main detectors were  enclosed in a separate shielding hut  made of 122 cm thick concrete shielding with shielded entrances. The 80 cm thick upstream wall of this hut was formed from 10 tightly fitting interlocking sections. The sections consisted of high-density (2700 kg/m$^3$) barite loaded concrete (Ba$_2$SO$_4$). Stainless steel rebar (and stainless steel lifting fixtures) were used due to the proximity to the magnet. The apertures in this front wall provided several 
cm clearance for the elastic electron envelopes determined by the defining collimator, as shown in Fig.~\ref{fig:envelopes}. The area around the beam-pipe penetration was filled with lead.  The shielding hut downstream of the main detectors was made using 122 cm thick iron shielding blocks.

\section{The Spectrometer}
\label{sec:Components:Spectrometer}

The  
\ac{acro-qtor} magnetic spectrometer focused elastically scattered electrons within the acceptance profile defined by the triple collimator system onto  eight rectangular fused silica detectors. Its design was loosely based on the  \ac{acro-blast} magnet~\cite{blast}, and provided a large acceptance for $ep$ elastics and a high degree of azimuthal symmetry in an iron-free magnet to minimize parity-conserving $A_{zz}$
backgrounds. The \ac{acro-qtor} spectrometer spatially separated elastic and inelastic events at the focal plane. In conjunction with the triple collimator system, the spectrometer separated elastic events from line-of-sight trajectories (photons and neutrons) originating in the target. It also swept away low-energy electrons from the copious M\o ller interactions in the target.

The \ac{acro-qtor} spectrometer consisted of eight identical resistive coils electrically connected  in series and arrayed azimuthally around the beamline, centered 6.5 m downstream of the target. Each coil was composed~\cite{wang2007magnetic} of a double pancake of 13 turns of copper conductor. Each racetrack-shaped pancake had  straight sections  2.20 m long, and semi-circular curved sections of inner (outer) radius 0.235 m (0.75 m). The oxygen free, high-conductivity conductor was formed from long copper bars brazed together, of cross section  5.84 cm $\times$ 3.81 cm with a central hole of diameter 2.03 cm for cooling water supplied 
at 13.3 liters/s. The nominal resistance of each 3900~kg coil was 1.76 m$\Omega$ at 20$^\circ$~C. 
The design current density was 500 A/cm$^2$.
Each of the eight coils was mounted in an aluminum coil holder. The coil holders were in turn mounted in a large aluminum frame assembled with silicon-bronze fasteners to minimize magnetic material (see Fig.~\ref{fig:Installation}).

Due to the iron-free nature of the magnet, it did not have to be cycled through a hysteresis curve to obtain a reproducible field. The field was determined from a \ac{acro-dcct}
at the output of a 2 MVA, $\pm$10 ppm current-regulating power supply~\cite{alpha}. A Hall probe was installed as a cross-check on the stability of the \ac{acro-qtor} power supply current. This probe helped identify intermittent periods when radiation damage affected the  stability of the \ac{acro-dcct}.

\begin{figure}[!htb]
\begin{center}
\includegraphics[width=0.48\textwidth]{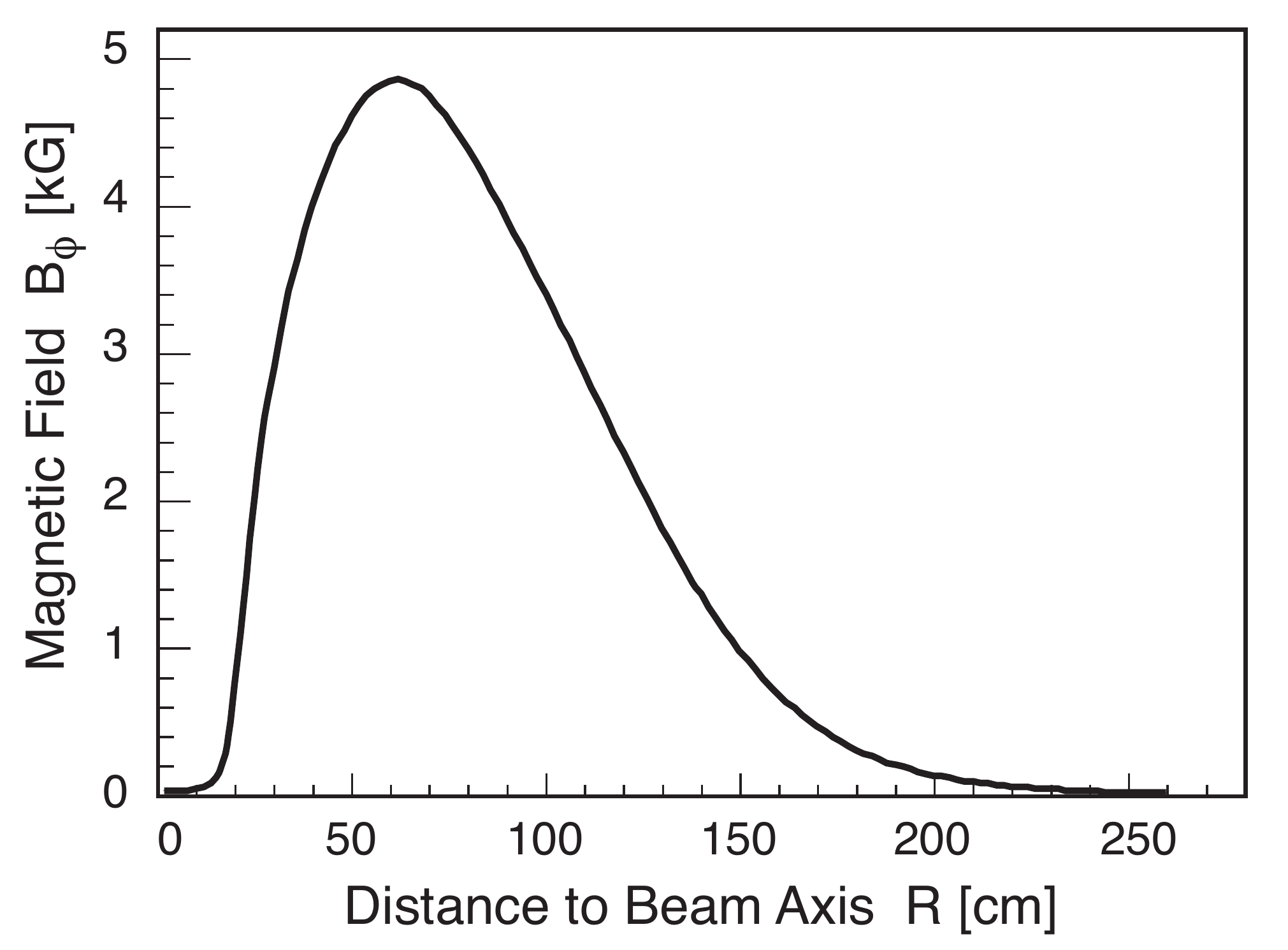}
\end{center}
\caption[The shape of QTOR magnetic field]
{\label{fig:qtor_field_shape} Radial distribution of $B_\phi$ of the \ac{acro-qtor} magnetic field magnitude at $z=0$ (the center of the magnet) and $\phi=0$ (halfway between the left two coils when looking downstream). The field is dominated by the azimuthal component $B_{\phi}$. Its radial and axial components $B_R$ and $B_Z$ are small, however defocussing in the azimuthal direction was nevertheless large enough to double the required length of an individual detector.}
\end{figure}

The shape of the magnetic field is depicted in Fig.~\ref{fig:qtor_field_shape}. Electrons were deflected radially outwards by the magnet. 
At the mean scattered electron angle of 7.9$^\circ$, the $\int B \, dl$ was about 0.9 T-m.
The collimated elastic electrons in each octant were focused into an envelope which was roughly 10~cm tall in the dispersive direction, but almost 2~m wide in the non-dispersive direction at the position of the main detector array 5.78 m downstream of the magnet center. Due to  $\phi$-dependent aberrations, curvature of the elastic event envelope resulted in a mustache-shaped image on the focal plane, as discussed in Sec.~\ref{sec:Components:Current:FocalPlaneScanner}.

The \ac{acro-qtor} field was carefully simulated and techniques were developed to analyze the results of field mapping~\cite{wang2007magnetic} carried out initially 
at \ac{acro-bates} 
using a 3-axis mapper. 
The mapper measured positions to $\pm$0.3~mm and magnetic fields to $\pm$0.2~G. 
It employed a 3-axis gantry that moved a probe over a 4~m $\times$ 4~m $\times$ 2~m range. The probe consisted of two high-precision 3-axis Hall effect transducers, temperature sensors, clinometers and photodiodes.
Zero-crossing measurements of certain fringe field  components as well as direct field measurements in the envelope of the scattered electron trajectories were performed. 
Simulations of the effects of coil misalignments (of  ideal coils) indicated that they had to be positioned within $\leq$3~mm radially, and $\leq$0.1$^\circ$ azimuthally of their ideal positions. The \ac{acro-qtor} magnet center had to be within $\leq$3~mm of the beam axis, and the eight field integrals along the electron trajectories had to be matched to within 0.4\%. 

\begin{figure}[!htb]
\vspace*{-0.5cm}
\begin{center}
\hspace*{-0.5cm}
\includegraphics[width=0.525\textwidth,angle=0]{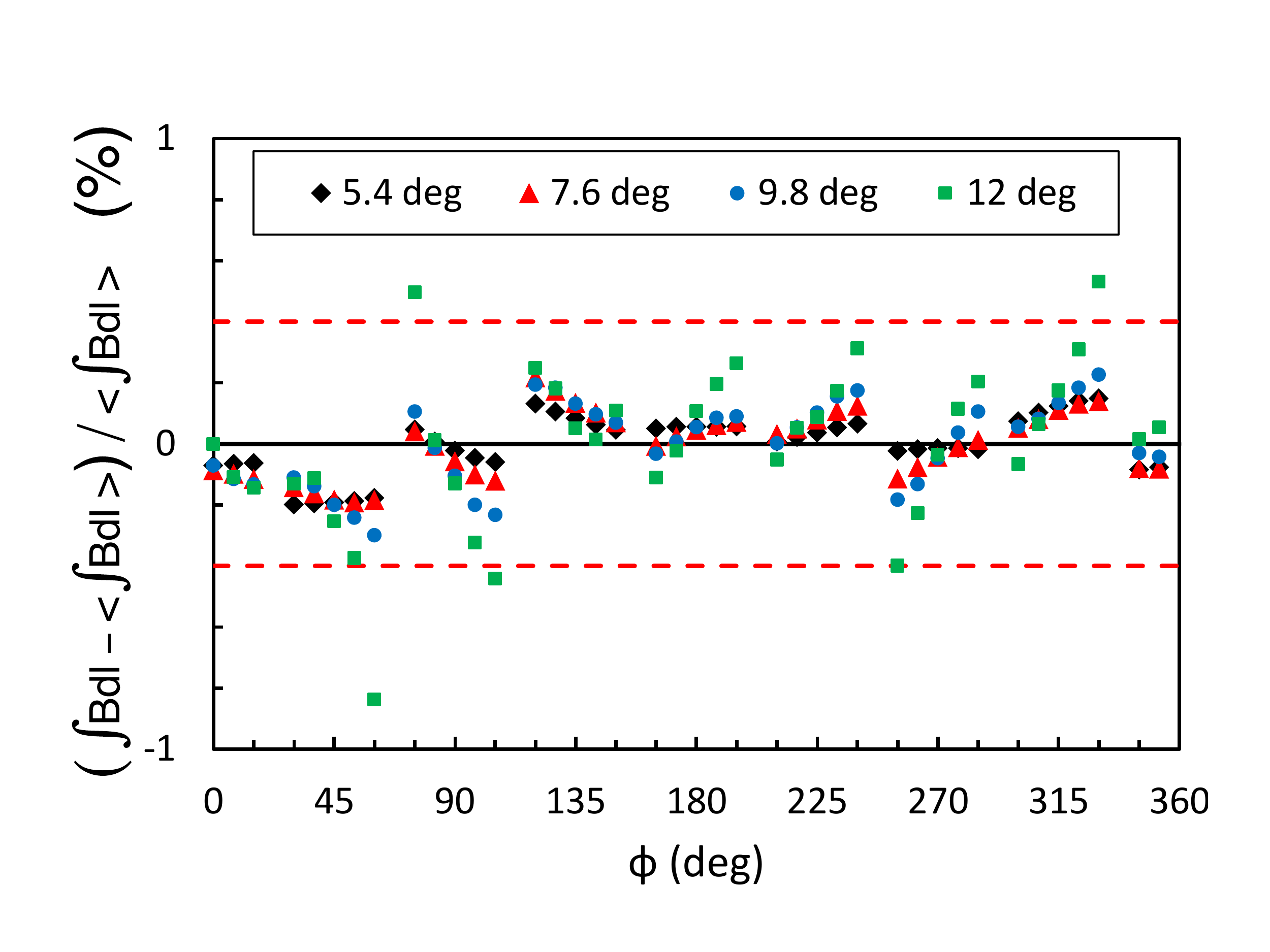}
\end{center}
\vspace*{-0.5cm}
\caption[The shape of QTOR magnetic field]
{\label{fig-bdl} 
Relative difference $\left( \int{ B  dl} - \left< \int{ B  dl} \right> \right) / \left< \int{ B  dl} \right>$ between the eight-octant average $\left< \int{ B  dl} \right>$ and the calculated $\int{ B  dl}$ for each octant based on the actual coil positions as determined from the zero-crossing field mapping measurements. The results for each octant are plotted as a function of the azimuthal angle $\phi$ for the four different polar angles $\theta$ indicated in the legend. $\phi$=0 corresponds to the center of the top octant, and increases going clockwise around the beam axis. The dashed lines indicate the specification on this parameter. Note that the smallest and largest polar angles plotted are just outside the scattered electron acceptance defined by the collimation system. The same is true for the azimuthal range.
}
\end{figure}

The mapping indicated that the coil positions were well within the desired 3~mm of their ideal positions in $X$, $Y$, and $Z$ of the magnet's local coordinate system, except for two coils that had $-$3.1~mm and 3.8~mm displacements in $X$, the radial outward direction. The measured coil average of the angular displacements in $X$, $Y$, and $Z$ were 0.04$^\circ$, 0.07$^\circ$, and 0.14$^\circ$, respectively. $Z$ is along the axis of the spectrometer, and $Y$ is perpendicular to the coil measured from its center.
Measured coil-to-coil variations in $\int{ B  dl}$ were $\leq$0.3\% except at the outermost radii of some of the coils, corresponding to the largest scattering angles and lowest scattered electron rates, where  variations up to about 0.5\% were found. 
Fig.~\ref{fig-bdl} shows the difference between the average $\int{ B  dl}$ from all eight octants, and the calculated $\int{ B  dl}$ based on the actual coil positions determined from the zero-crossing measurements. This difference is plotted as a function of the azimuthal angle $\phi$ for several polar angles $\theta$ in each octant to illustrate the high degree of azimuthal symmetry achieved across the spectrometer. 
The \ac{acro-qtor} magnet was fiducialized at \ac{acro-bates}, transported to \ac{acro-jlab} and installed in  experimental Hall C. After installation, the upstream half of the magnet was remapped and calibrated to verify the alignment and magnet performance. 

\subsection{Spectrometer Performance}
\label{sec:Components:Spectrometer:Performance}

During the \Qweak experiment, \ac{acro-qtor} was operated routinely at a current of 8900 A \ac{acro-dc} (123 V). The cooling water temperature rose $\sim$20~C across the magnet. 
Some problems were experienced early in the experiment due to cooling water restrictions, which led to burst cooling hoses and even a burst water cooled lead which damaged the load sensing resistors in the power supply. A phase failure in the power grid damaged the \ac{acro-dcct}, which was repaired, recalibrated, and checked using the Hall probe. A blown silicon-controlled rectifier in the power supply and a bad breaker in the 2 MVA service feed also caused  significant down time. Radiation damage to the power supply was mitigated with the addition of steel shielding. 
The intense flux of low-energy electrons from Moller scattering was deflected radially outward and away from the magnet by the magnetic field. Furthermore, the magnet coils themselves were in the shadow of the collimation system, and thus received a radiation dose far smaller than the 1 Mrad seen by the main quartz detectors.

A combination of glass witness plates (which recorded the location of the beam envelope through radiation damage induced darkening) and the tracking chambers verified that the eight beam envelopes were  radially symmetric within $\pm$0.5~cm.
However, the nominally field-free region along the central symmetry axis (see Fig.~\ref{fig:qtor_field_shape}) 
contained small contributions amounting to $\sim$5 kG-cm,
which arose from minor misalignments of the magnet coils.
Although this had no effect on the scattered electrons of interest in the experiment, it did  steer low-energy electrons in the beam pipe towards the 2 o'clock position looking downstream, breaking the azimuthal symmetry in the downstream luminosity monitors described in Sec.~\ref{sec:dslumis}.  A more serious consequence was the resulting $\sim$1 kW beam power deposited on a flange near the end of the beamline ($\sim$14~m downstream of the magnet center) which led to vacuum leaks, and impacted the experiment's efficiency in Run 1. This section of downstream beamline was redesigned for Run 2, cooling was added, and the vacuum problems were eliminated. 

\section{Integrating Mode Electron Detectors}
\label{sec:Current_mode_dets}

In addition to the eight main detectors which were used to measure the experiment's main quantity of interest, the elastic asymmetry defined in Eq.~\ref{eqn:Ayield}, several ancillary detection systems were also employed.  Dedicated background detectors 
helped monitor and quantify backgrounds and their asymmetries. A focal plane scanner was used to map out the profile of events in 1 cm$^2$ pixels over the face of one of the main detectors at the full current used in the experiment. Upstream and downstream arrays of luminosity monitors were also used. 

\subsection{The Main \v{C}erenkov Detectors}
\label{sec:maindetectors}

\begin{figure}[!htb]
\begin{center}
\includegraphics[angle=0,width=0.475\textwidth]{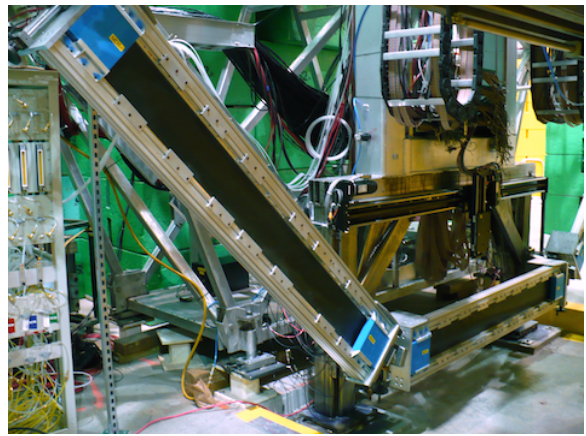}
\end{center}
\caption[maindets]{\label{fig:maindets} Two of the eight main detectors, before installation of the lead pre-radiators. The 
squares standing out on each end of the detectors are lead shielding covering the light guides and PMTs. The quartz sampling scanner is also shown above the lower detector in the figure.}
\end{figure}

The challenges associated with the  main detector 
were to detect elastically scattered electrons in integrating mode with low noise, low background, high linearity, and excellent azimuthal symmetry  and radiation-hardness over a focal plane area totaling several square meters.
The main detector system~\cite{wang2011cerenkov} employed a set of eight \v{C}erenkov detectors 
made of non-scintillating, low-luminescent synthetic quartz bars (Spectrosil  
2000 fused silica~\cite{S2000})  
which were extremely radiation-hard and insensitive to neutral backgrounds. Each detector consisted of two 100~cm $\times$ 18~cm $\times$ 1.25~cm radiators (made by Saint-Gobain Quartz~\cite{SaintGobain}) and two 18~cm $\times$ 18~cm $\times$ 1.25~cm long light guides (made by Scionix~\cite{Scionix}). All surfaces  were finished with 25 \AA  ~(\ac{acro-rms}) surface polishing. The systematic point-to-point variations in thickness were $\pm 250~\mu$m. All edges were beveled to a width of 0.5~mm to reduce chipping. Pairs of bars and lightguides were glued together end-to-end, forming  240~cm long bars with three main glue joints: 
light guide-to-radiator, radiator-to-radiator, and radiator-to-light guide 
(see Fig.~\ref{fig:maindets}).
The index of refraction of the quartz was n=1.482 at a wavelength of 280 nm, corresponding to a threshold $\beta=0.67$ and a \v{C}erenkov cone angle of 47.6$^\circ$.

The detectors were instrumented with 130 mm diameter \ac{acro-pmt}s on each end of each detector bar.  
Each \ac{acro-pmt} (Electron Tubes 9312WKB~\cite{ET9312}) had a multi-alkali S20 (Na$_2$KSb:Cs) photocathode, UV transmitting glass window and \ac{acro-dc} coupling with an electrostatic shield at cathode potential.  The PMTs had  10 stages of high-gain, high-stability SbCs dynodes with a linear focus design for good linearity and timing. They were sensitive in the wavelength range from 200 to 900 nm, with a peak quantum efficiency of about 23\% at 260 nm,  well matched to the \v{C}erenkov light wavelengths in quartz. Each \ac{acro-pmt} was magnetically shielded using a double-layer mu-metal case.  

The \ac{acro-pmt}s were  glued onto the downstream faces of the light guides. The optical glue used to attach the quartz pieces and the \ac{acro-pmt}s was SES406 by Shin-Etsu~\cite{SES406}. This glue was chosen for its mechanical strength and stable light transmission under high radiation doses. Both the quartz and the glue were tested for radiation damage at doses up to 1 MRad at a $^{60}$Co facility~\cite{NCSU}, approximately the dose they were expected to receive over the course of the experiment. 

To suppress soft neutral backgrounds, 
a 2 cm thick lead pre-radiator was installed in front of each quartz bar. This increased the light yield by a factor of seven and improved the  signal-to-background  ratio  by $\sim$20, although shower fluctuations in the pre-radiator also introduced an additional excess noise ($\sim$10\%) to the total asymmetry width. 

The eight quartz \v{C}erenkov detectors were arranged symmetrically about the beam axis (3.44~m from the beam axis to the outer edge of each bar) at the focal plane of the spectrometer (5.78~m downstream of the center of the \ac{acro-qtor} magnet). 
The mounting and support structure consisted of a housing for each quartz and \ac{acro-pmt} assembly, an exoskeleton around each housing, and a general support structure for all eight detectors, referred to as the Ferris wheel. Each light-tight housing consisted of an aluminum frame which supported the quartz bars and the \ac{acro-pmt}s, and thin PORON\textsuperscript{\textregistered} covers~\cite{PORON}. 
The housing was mounted inside the exoskeletons, which  reinforced the mechanical strength of the main detector housings and provided mounting structures for the pre-radiators and \ac{acro-pmt} shielding.
Lead plates 5~cm thick were mounted on the exoskeleton just in front of the \ac{acro-pmt}s and 
at their inner radius side. This provided 10 radiation lengths attenuation for ${\cal O}(1)$ GeV particles (electrons) coming from the upstream direction and from the beamline. The exoskeletons also provided the mounting interface to the overall support structure (the Ferris wheel). The Ferris wheel located the main detectors at  the desired radius in the focal plane. The attachment of the exoskeletons to the Ferris wheel incorporated manual radial motion capabilities with a range of about 15 cm. 

The Ferris wheel also supported cable trays and access platforms with ladders. 
All the mechanical structures were built with aluminum in order to provide a ``low-Z'' 
and iron-free 
environment around the detectors.  
The entire detector system was placed in a shielded hut (see Sec.~\ref{sec:collimation_and_shielding}) to reduce the background.

\v{C}erenkov light generated by scattered electrons traveled along the quartz bar via total internal reflection and was collected at each end by the \ac{acro-pmt}s. An average of 98 \ac{acro-pe}s 
were generated for each incident electron. The \v{C}erenkov signals were read out with two types of custom made \ac{acro-pmt} bases: one for high-gain ($2 \times 10^6$) event-mode calibration running at nA level beam currents and one for low-gain  ($\sim$440) integrating-mode production running for asymmetry measurements at beam currents up to 180~$\mu$A. The low-gain operation in integrating mode reduced non-linearity effects from the \ac{acro-pmt}. This was achieved by using only the first seven dynode stages and keeping the remaining stages and anode at the same bias voltage as dynode 7, as well as operating the \ac{acro-pmt} at a relatively low bias of around -1~kV. 

\subsection{Integrating Mode Detector Performance}

\begin{figure}[thhttt]
  \centering
  \includegraphics[width=0.5\textwidth]{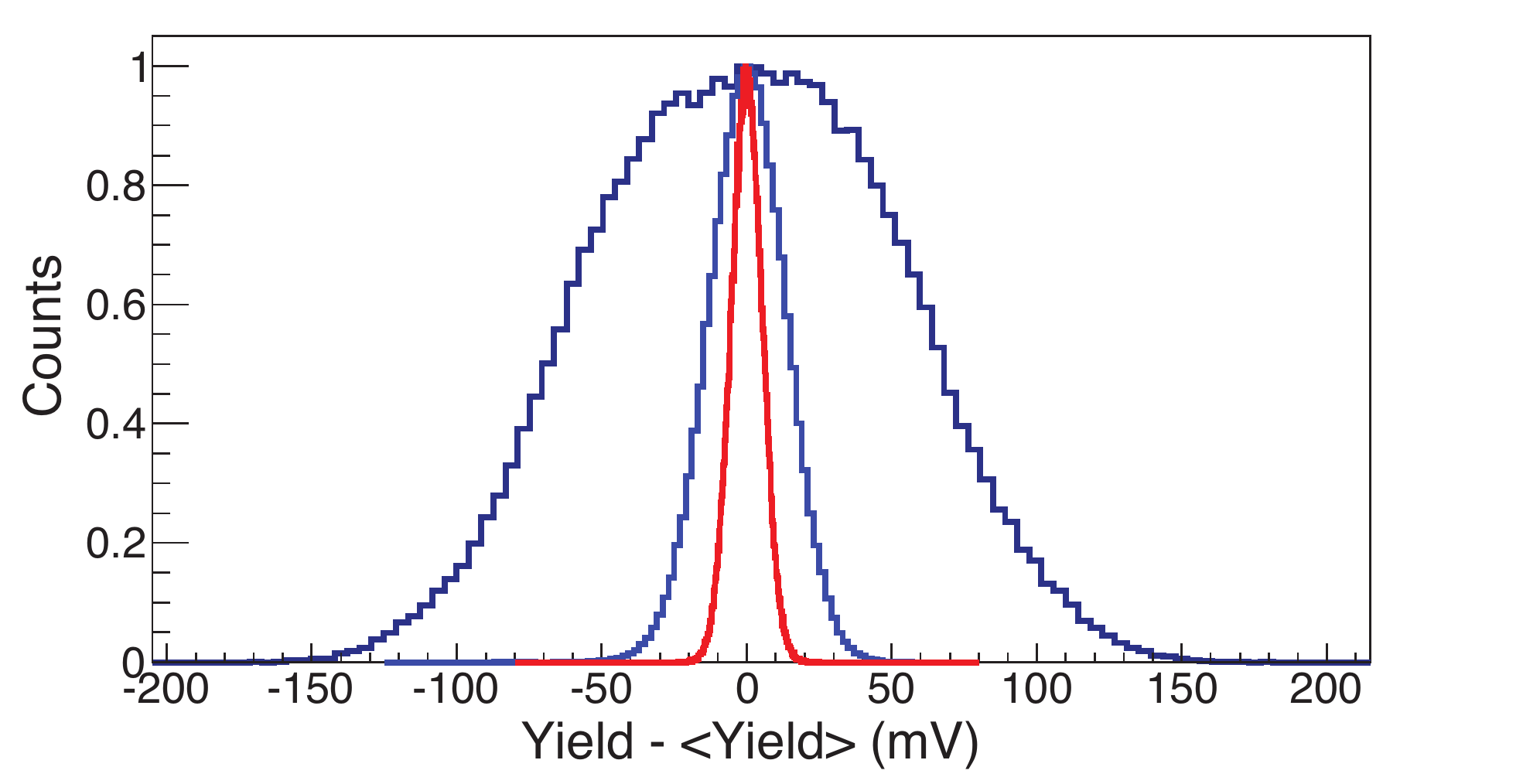}
  \caption{\label{fig:qweak_md_yield1} A typical 6 minute main detector \ac{acro-pmt} yield distribution after subtraction of the mean value of the yield. The outer ($\sigma$=50.3 mV) distribution corresponds to the raw signal and the middle ($\sigma$=13.4 mV) peak are the same data after normalization to the beam current. The 
  innermost peak is a Gaussian distribution with a width corresponding to a calculated estimate of the shot (statistical) noise, which experiences no gain drifts due to factors such as temperature changes like the former two yield curves do.  
  }
\end{figure}

The rate of scattered electrons incident on each of the eight main detectors was over 850 MHz, precluding the counting of individual events. Instead, the raw current of each of the 16 \ac{acro-pmt}s was read out using the electronics described in Sec.s~\ref{sec:Components:Preamp} and  \ref{sec:Components:DAQ:CurrentMode} and saved for later analysis, which included pedestal subtraction and beam charge normalization (see Fig.~\ref{fig:qweak_md_yield1}). 
Pedestal data were acquired 
every eight hours by taking 1--5 minutes of data with no beam in the experimental hall. 
The \ac{acro-rms} width of each tube's raw pedestal distribution over a typical 5 minute period varied between \(0.20-0.25 ~\mbox{mV}\). More interesting was the width of the \ac{acro-hc} pedestal differences measured over helicity quartets, which was only 30 $\mu$V. That may be used to estimate the electronic noise contribution to the main detector asymmetry width as 5 ppm  (relative to the nominal 6~V signal magnitude). This is negligible compared to the overall main detector asymmetry width of 230 ppm, which was dominated by the statistical width as described in Sec.~\ref{sec:Helicity}. Similar results were obtained by replacing the detector inputs to the \ac{acro-adc}s with a 9~V battery. 

These pedestal data also provided an opportunity to search for a false asymmetry in the main detector signal chain due to electronic pickup. 
Helicity-dependent changes in the mean value of the pedestals were searched for and excluded with a typical sensitivity of $\pm$10 nV  per monthly Wien setting. Given the signal magnitude of $\sim$6~V, this meant that any false asymmetry from \ac{acro-hc} pedestals was $<$ 2 ppb per Wien setting. Because this method regularly tested the actual detectors, electronics, and current monitors (albeit necessarily at low duty factor), it was complementary to measurements by the continuously monitored battery signals which provided a 1 ppb limit on electronic pickup in the \ac{acro-pmt}s or \ac{acro-pmt} gating every 8 hours.

\begin{figure}[ttbhh]
  \centering \hspace*{0.1cm}
   \includegraphics[width=0.5\textwidth, angle=0]{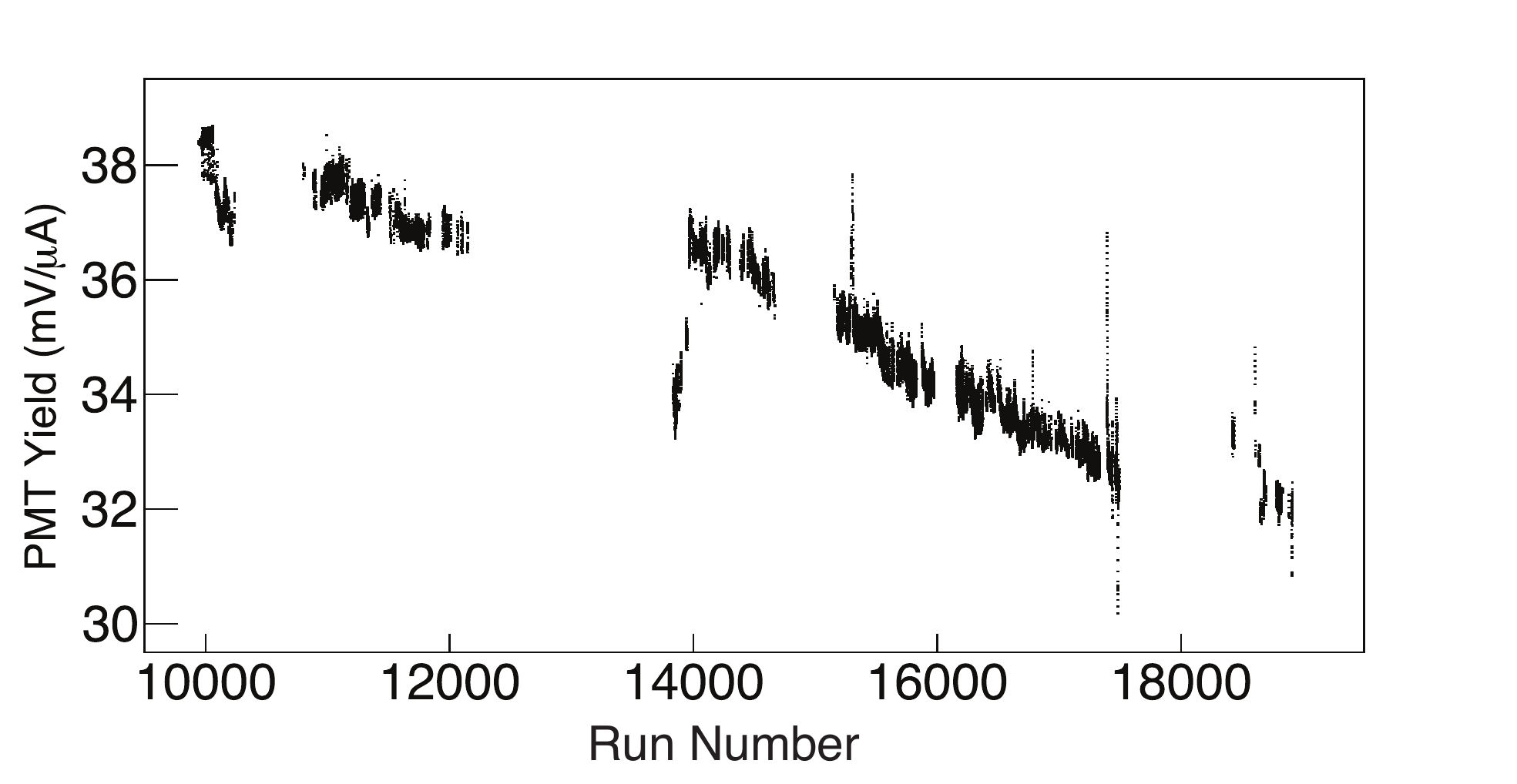}
   \includegraphics[width=0.47\textwidth, angle=0]{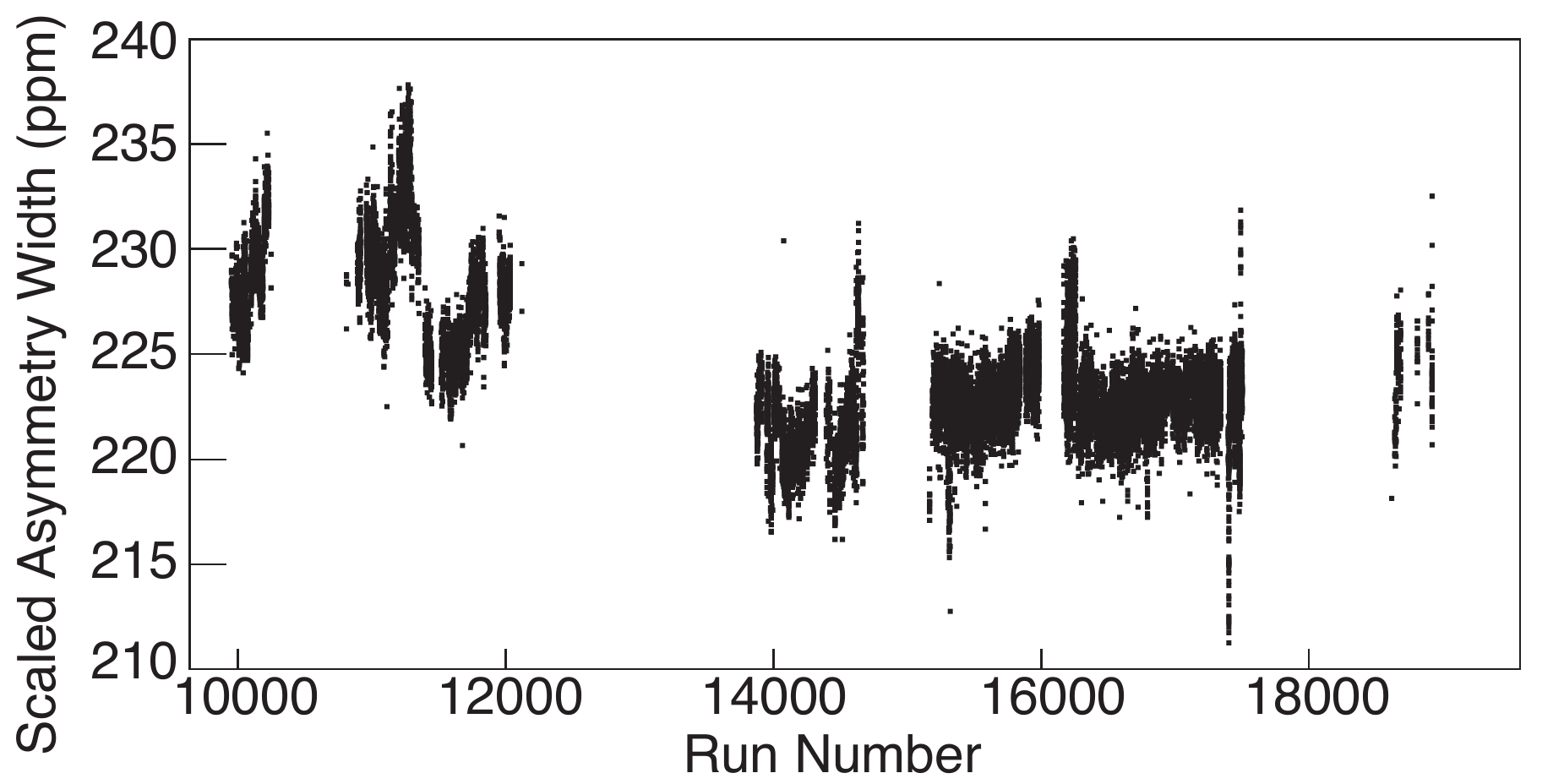}
     \caption{\label{fig:qweak_md_yield} Top: Typical charge normalized main detector phototube yields over the course of the experiment. Run 2 started around run 14000. The $\sim$10\% decrease in yield over time is attributed to gain degradation. The change at run 14000 corresponds to a period when the beam current was raised as Run 2 got underway. Spikes near runs 15500 and 17500 are the results of temporary issues with heat dissipation in the \ac{acro-bcm} and \ac{acro-daq} electronics, respectively. 
     Bottom: The main detectors' asymmetry width over the whole experiment, scaled by $\sqrt{(I/180)}$, where $I$ is the beam current in $\mu$A, As described in the text, this metric combines detector, target, and \ac{acro-bcm} performance as well as regression to correct for  \ac{acro-hcba}. 
      }
\end{figure}

The health of the 16 individual \ac{acro-pmt}s was tracked continuously so that damaged hardware could be  identified  and replaced. 
 Figure~\ref{fig:qweak_md_yield}  displays how a typical \ac{acro-pmt} yield changed over the course of the entire \Qweak running period. Drifts in the detector yield over time scales of hours were typically 1\% due to \ac{acro-pmt} gain drifts arising from temperature variations in the hall, while a 10\% gain degradation was observed over the course of all of Run 2. 

The primary metric used to assess detector performance 
was the width of the asymmetry distribution (see Sec.~\ref{sec:Asymmetry}). The general health of the experiment as a whole could be monitored by ensuring that widths in excess of counting statistics could be attributed to known sources such as the \ac{acro-bcm} resolution and target density fluctuations. 
For a \(175 \,\mu\mbox{A}\) run, the average single-\ac{acro-pmt} asymmetry width (averaged over helicity quartets) was \(\sim 680~\mbox{ppm}\). The full
16-tube (8 radiator) combination  width was \(\sim 230 ~\mbox{ppm}\). 
Fig.~\ref{fig:qweak_md_yield} (bottom) shows how the asymmetry width of the average of all 16 \ac{acro-pmt}s varied over the whole experiment.  

Potential additional 
sources for increased asymmetry widths are non-linearities either in the detectors (hardware and electronics) or  the \ac{acro-bcm}s. 
The inherent non-linearity of the main detector \ac{acro-pmt}s and associated electronics was  studied in detail prior to the experiment with \ac{acro-led}s, 
with a non-linearity of $\sim$0.8\%  at the signal levels corresponding to those experienced during the experiment.

\subsection{Event Mode Electron Detectors}

The low-gain bases of the main detectors were swapped out with high-gain bases in order to use them at low beam currents (50 pA - 200 nA) in an event-by-event counting mode. Further amplification ($\times 20$) was provided locally before each signal was sent to the counting room electronics outside the experimental hall.  This configuration  
provided pulse height, timing, and coincidence information to be used in conjunction with  drift chamber information to reconstruct kinematic quantities such as \(Q^2\) and scattering angle. Data taken in this configuration also proved useful in calibrating  
the performance of individual phototubes. Comparisons of data acquired using \ac{acro-led}s with data obtained from scattered electrons in event mode 
were used to determine the number of photoelectrons generated per event. 
These numbers varied by 8$-$18~\ac{acro-pe}/\ac{acro-pmt} from the average value of 98 \ac{acro-pe}s per event. The light collected from an event near the middle of the bar was about 2/3 of the light collected from an event near the end of the bar.

Timing information from the main detectors was also used both in track reconstruction as well as for systematic measurements of dilution factors for various background processes. By requiring a coincidence for events recorded by tubes on opposite ends of a single quartz bar, accidental events could be suppressed.

\subsection{Luminosity Monitors}
\label{sec:uslumis}

The luminosity monitors (\ac{acro-lumis}) were auxiliary detectors located where the scattered flux was much higher than at the main detectors, while the expected physics asymmetry of the contributing processes was much smaller than at the main detectors. Two sets of detectors - the upstream and downstream luminosity monitor arrays - were used during the experiment. The desire for a high scattered flux required the detectors to be in areas where they received a large radiation dose. The extra efforts to make these detectors radiation-hard worked very well.
Both the upstream and downstream luminosity monitors made use of rad-hard Spectrosil 2000 quartz (fused silica) radiators~\cite{S2000}. Special signal connectors (Kings teflon~\cite{kings}) on the \ac{acro-pmt}s were used to avoid the especially radiation sensitive insulators on standard BNC connectors. Long light guides helped distance the \ac{acro-pmt}s from the radiation field. 
The light guides were formed from single sheets of highly reflective aluminum (Alanod Miro-Silver 27~\cite{Alanod}) and continuously flushed with N$_2$ gas to eliminate the corrosive effects of moist air and to reduce backgrounds.
Pre-amplifiers  located close to the detectors were heavily shielded with lead. Careful alignment was performed early in the experiment before activation made personnel access problematic. 
 
\subsubsection{Upstream Luminosity Monitors}

\begin{figure*}[tthbt]
  \centering

  \includegraphics[width=1.\textwidth]{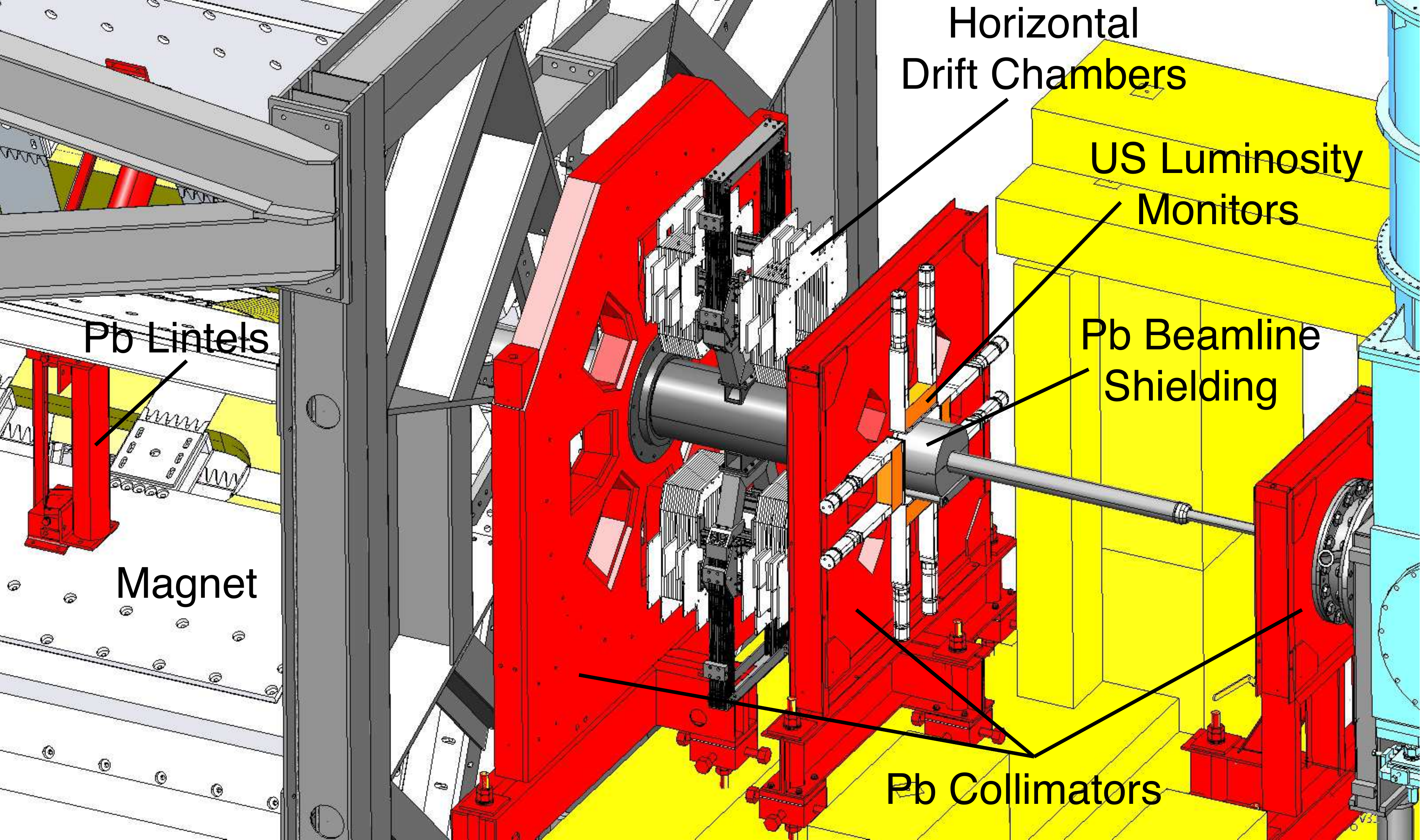}
    \caption{\label{fig:HDCs} A \ac{acro-cad} drawing of the triple collimation region. The target is in the scattering chamber on the right side of the drawing; the beam direction is to the left. The lead collimators and the lead lintels between the magnet coils are shown in this view. The horizontal drift chambers  are shown  just upstream of the third collimator. The four upstream luminosity monitors are shown mounted to the upstream face of the middle collimator, each with two light guides and \ac{acro-pmt}s. The active elements span the light guide pairs in the horizontal and vertical octants. }
\end{figure*}

The upstream  \ac{acro-lumis} 
(see Fig.~\ref{fig:HDCs})
were intended to provide a way to measure target noise if other methods failed to do so. As it turned out, the target noise was  small and  well measured by three other techniques (see Sec.~\ref{sec:Target:Performance}).  
However, the upstream \ac{acro-lumis} were a crucial tool to link together beamline background asymmetries observed with differing configurations of other background detectors  in the main detector shielding hut (described in Sec.~\ref{sec:Components:Current:Background}). 

The four upstream \ac{acro-lumis}~\cite{Leacock} were Spectrosil 2000 quartz radiators~\cite{S2000} measuring 7 cm $\times$ 27 cm $\times$ 2 cm. They were situated 2.67 m downstream of the target on the upstream face of the defining collimator at a scattering angle of about 5$^\circ$ where they were expected to be primarily sensitive to M\o ller electrons. The detectors each saw a rate of about 115 GHz (scaled from event mode to a beam current of 180 $\mu$A), about half of which came from sources other than the target. They were read out with 5.1 cm diameter Hamamatsu R375 quartz window phototubes~\cite{R375} operated in vacuum photodiode mode (unity gain voltage divider) at each end. The light from the detectors was transmitted to the \ac{acro-pmt}s   through 35~cm long N$_2$ filled light guides.  
The remainder of the integrating-mode electronics chain is described in Sec.~\ref{sec:Components:DAQ:CurrentMode}. The unity-gain bases were swapped with modest-gain ($\sim$$ 10^{6}$) bases so they could be used as relative beam current monitors in event (pulse counting) mode during the low beam current running.

\subsubsection{Downstream Luminosity Monitors}
\label{sec:dslumis}

The downstream luminosity monitors  were situated 17 m downstream of the target at an angle of 0.5$^\circ$, sensitive  to similar rates of scattered electrons from M\o ller ($e-e$) and Mott ($e-p$) interactions in the target. 
Based on pure counting statistics considerations, these detectors were anticipated to have a smaller asymmetry width than the main detectors.  
In practice, their asymmetry width was only slightly smaller ($\sim$200 ppm typically) than the main detectors.  
Target (53 ppm) and BCM (60 ppm) noise contributed in quadrature to the anticipated statistical width (14 ppm).  
Excess noise from a variety of sources, including the beamline and beam monitor resolution, also contributed to the observed 200 ppm asymmetry width. 
However, they provided sensitivity to false asymmetries at this level, because the expected asymmetry from the contributing physics processes was much smaller than that of the main detector.  The relation of false asymmetries measured in these detectors to any potential false asymmetries in the main detectors is being studied.
They also proved useful as beam current and relative position monitors during the extremely low current running used for the event mode of the experiment.

\begin{figure}[tthbt]
\centering  \includegraphics[width=0.5\textwidth]{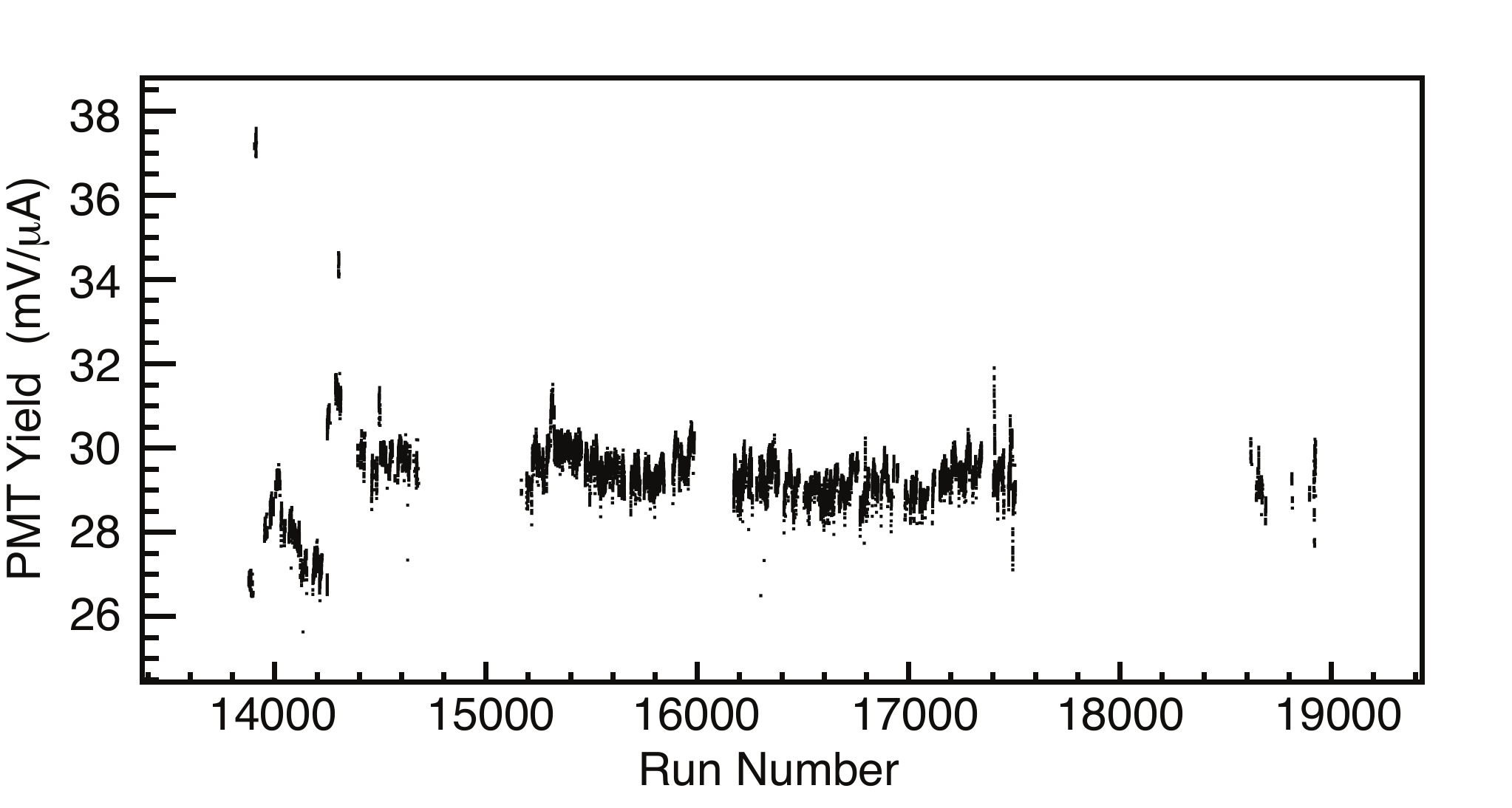}
\caption{\label{fig:dslumi_y} 
Typical charge normalized downstream luminosity monitor yields over the course of Run 2 of the experiment.  The behavior was typically very stable with no long term gain degradation observed.  Discrete jumps near the beginning of the running period were the result of deliberate beam position changes of $\sim$0.5~mm during studies to determine the experiment's neutral axis. The relatively large size of the jumps reflects the expected large beam position/angle sensitivity of these very forward angle monitors. 
}
\end{figure}

Each of the eight downstream \ac{acro-lumis}~\cite{Leacock} consisted of a  piece of Spectrosil 2000 quartz~\cite{S2000} measuring 4 cm $\times$ 3 cm $\times$ 1.3 cm with a $45^{\circ}$ taper at one edge. Each of the quartz radiators had a 2 cm thick lead pre-radiator in front of it to suppress low energy backgrounds. They were inserted into flanged cups which penetrated the 61 cm diameter beampipe to within 13 cm of the nominal beam axis. Each quartz piece was read out with the same type of phototube and light guide described in Sec.~\ref{sec:uslumis} for  the upstream \ac{acro-lumis}, using the integrating mode electronics described in Sec.~\ref{sec:Components:DAQ}. 
Fig.~\ref{fig:dslumi_y} displays how a typical downstream luminosity monitor phototube yield behaved over the course of the Run 2 period. 
As with the upstream \ac{acro-lumis}, the unity gain bases were swapped with modest gain ($\sim$$ 10^{6}$) bases 
 during the low beam currents used in the event mode of the experiment. 
 The downstream \ac{acro-lumis}  each saw about 150 GHz of scattered electrons (scaled up from event mode measurements) and  withstood a dose of about 2 Grad over the life of the experiment. 

\subsection{The Focal Plane Scanner}
\label{sec:Components:Current:FocalPlaneScanner}

In order to measure the profile of events reaching the main detectors over the whole 4$-$6 orders of magnitude of beam current used in the experiment, a small scanning \v{C}erenkov detector was built~\cite{Jie}. It provided confirmation early in the experiment that the scattered electron envelope on the main detectors agreed with simulations.
It consisted of two overlapping  $\mathrm{1~cm^3}$ cubes of synthetic quartz 
 (Spectrosil 2000) fused silica \v{C}erenkov radiators~\cite{S2000}.  
The geometrical overlap of the two sensitive elements formed a small fiducial area of $1\times1$~$\rm cm^2$ to cope with the maximum electron flux of about 1 MHz/cm$^2$, allowing operation in event mode at most beam currents. Each quartz radiator was 
optically coupled via a 50 cm long air-core light guide to a Photonis XP2268 5~cm \ac{acro-pmt}. 
The light pipes were lined with highly specular Alanod Miro4 reflector~\cite{Miro4}, providing a light transport efficiency better than 93\%. 
The two \ac{acro-pmt}s were read out in coincidence. Accidental coincidences 
were reduced by configuring the light guides in a non-overlapping V-shape. Despite configuring the  air light guides in the super-elastic region, they were exposed to a large photon flux from the upstream beamline which contributed to a larger than expected accidental rate.

\begin{figure}[hhbbt]
\begin{center}
\includegraphics[width=0.475\textwidth]{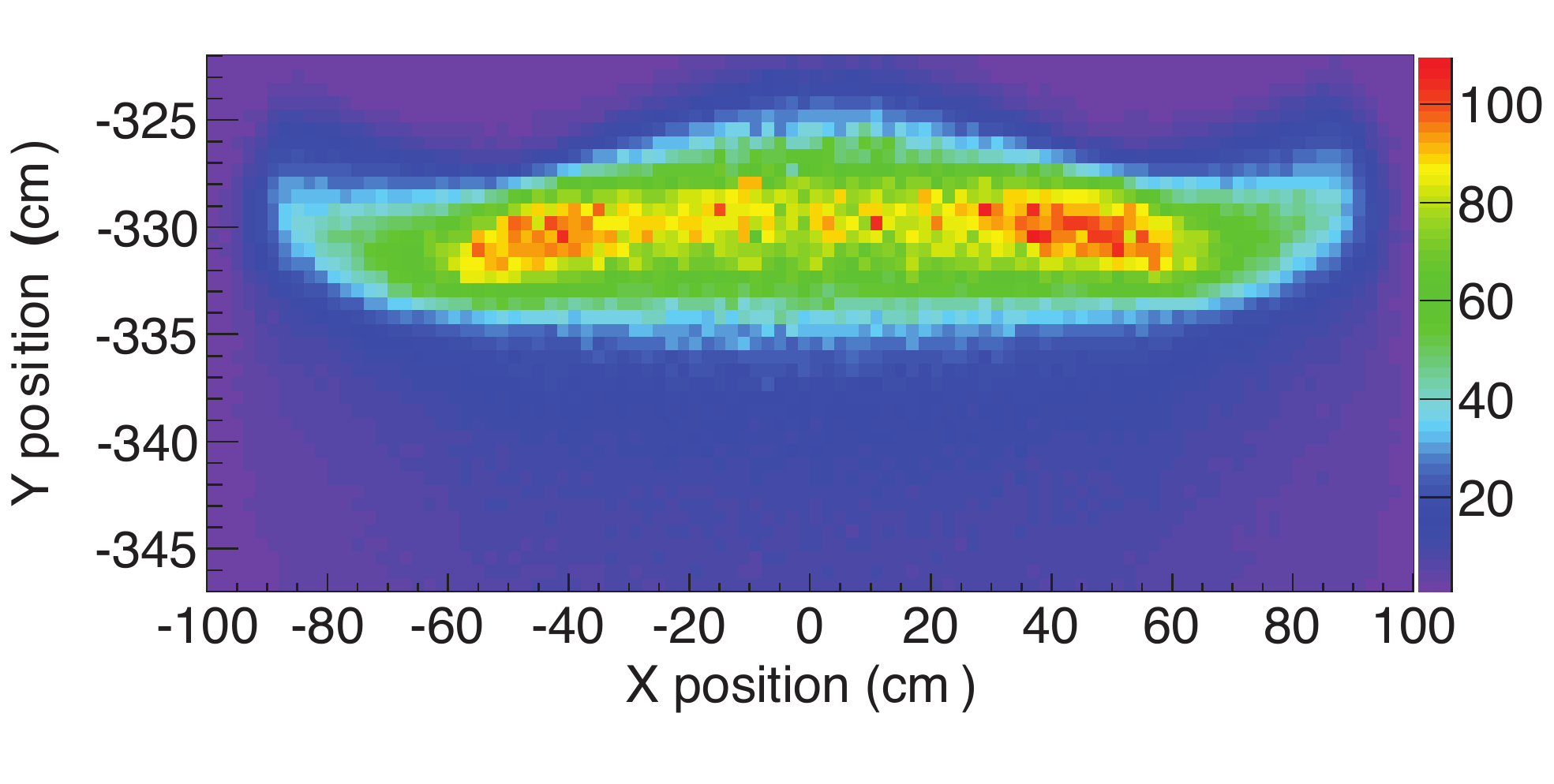}
\caption{ \label{fig:scanner_ratemap} Scattered electron flux distribution in the bottom octant obtained by the scanner. The x-axis is in horizontal direction, the y-axis is in vertical direction. The 
scale indicates the relative electron flux.}
\end{center}
\end{figure}

A 2D linear motion system~\cite{Lintech90}  was employed to move the scanner detector  along a predefined path while measuring position-dependent rates. The 2D linear motion system consisted of two stainless steel 
 ball-screw driven tables with a range of 200~cm~$\times$~26~cm and a position resolution
of 100~$\mu$m. 
The moving tables were driven by servo-motors controlled by a custom-built control box housing two Danaher S300 servo-amplifiers~\cite{DanaherS300} and a Galil DMC-4020 motion controller~\cite{Galil}. Two linear displacement draw wire potentiometer position sensors~\cite{ATech} 
measured instantaneous detector positions. 
The $\sim$~500 mR/h radiation field inside the shielded main detector hut required the controller to be shielded with lead.
A full scan (see Fig.~\ref{fig:scanner_ratemap}) could be completed in half an hour when the detector was moved  $\sim$6~cm/s. 

\subsection{Background Detectors}
\label{sec:Components:Current:Background}

To continuously monitor the asymmetry of the diffuse background and search for potential leakage of the helicity reversal signal, 
a set of background detectors were constructed and placed at specific locations in the detector hut. 
The background detectors included one complete main detector assembly (identical to the other eight main detectors) placed 
in the super-elastic region, just downstream of the nominal focal plane as a beamline background monitor, and three smaller dark boxes numbered from 1 to 3.

Background detectors 1 and 2 consisted of a dark box containing a bare \ac{acro-pmt} of the same type as the main detector \ac{acro-pmt}s, shielded with double layers of mu-metal, an integrating-mode \ac{acro-pmt} base and an \ac{acro-led} light source. Background detector 1 was placed in a well-shielded fixed location. Its  
\ac{acro-led} light, \ac{acro-pmt}, and low-gain base delivered a low-noise signal ($\sim$10 ppb over 8 hours) to provide a noise floor reference for the main detector electronics chain. It was used to search for 
leakage of the helicity reversal signal, and thus needed to have similar cabling to the main detectors while being well-shielded from beam induced backgrounds. 
 
Background detector 2 
was moved in the super-elastic region next to different main detector \ac{acro-pmt}s to  characterize the ``\ac{acro-pmt} background'' at each \ac{acro-pmt}. 
   Background detector 3 was identical to 1 and 2, 
  but had a piece of fused quartz (identical to the light guide extension used on the main detectors) glued to it in the same configuration as the \ac{acro-pmt}s were glued to the light guides in the main detectors. Background detector 3 was also 
  moved in the super-elastic region next to the \ac{acro-pmt} of any of the main detectors to serve as a ``\ac{acro-pmt}+lightguide background'' detector. The \ac{acro-led}s in background detectors 2 and 3 were used only for checkout. Background detectors 2 \& 3 both measured the asymmetry of the diffuse background, and needed to be close to a main detector radiator to measure the relevant background, ideally while minimizing cross-talk due to showering.

\section{Event Mode Detectors}
\label{sec:Tracking}

The tracking system was used for calibration measurements performed with low beam currents (50 pA $-$ 200 nA) to
extract the acceptance-weighted average of $Q^2$ ($<$$Q^2$$>$) of the asymmetry measurement.  $<$$Q^2$$>$ was formed from 
the measured scattering angle and scattered momentum
distributions, the analog response of the main \v{C}erenkov
detectors, the known incident beam energy, and a detailed GEANT 4~\cite{GEANT4a,GEANT4b} simulation including all radiative effects. 
The tracking
system was also used to characterize various backgrounds and monitor
the performance of the main \v{C}erenkov detectors.

To accomplish this, 
horizontal drift chambers (\ac{acro-hdc}s) were used on the upstream side of the \ac{acro-qtor} magnet, and vertical drift chambers (\ac{acro-vdc}s) on the downstream side. 
\ac{acro-hdc} and \ac{acro-vdc} designs are distinguished by the dominant drift direction
relative to the wire planes. In an \ac{acro-hdc} the ionization electrons
drift parallel to the wire plane, and in a \ac{acro-vdc} they dominantly drift
perpendicular to the wire plane.
Scintillation counters were placed between the \ac{acro-vdc}s and the main detectors. During integrating mode, all of the tracking  detectors were retracted from the experiment's acceptance. During event-mode running, two octants at a time (180$^\circ$ apart) were instrumented with tracking detectors. Separate rotation systems were employed for the \ac{acro-hdc}s and the \ac{acro-vdc}s to
  rotate the chamber systems around the beam axis to cover each of the four
  octant pairs.  One octant pair could be covered with
  either of the two sets of chambers for redundancy.
  The tracking system consisted of eight chambers and two scintillators: two \ac{acro-vdc} chambers, two \ac{acro-hdc} chambers, and one scintillator in each of two octants.

 During the tracking measurements, the \ac{acro-hdc}s were used in conjunction with the  \ac{acro-vdc}s at beam currents of 50 pA on the 35 cm LH$_2$ target. The $\sim$100 kHz rate observed in the \ac{acro-hdc}s was dominated by 
  $\sim$50 MeV M{\o}ller electrons which were swept away by \ac{acro-qtor} before they reached the \ac{acro-vdc}s or the main detectors.  The elastically scattered electrons of interest typically had a rate of $\sim$200 Hz under these conditions.  
  Data were also obtained with various thin solid targets under similar conditions to study backgrounds and to tune the track reconstruction algorithms.
  
\subsection{The Horizontal Drift Chambers}
\label{sec:HDCs}

For elastic scattering, ignoring radiation and energy losses, the
  four-momentum transfer squared $Q^2$ is given by 
  \begin{equation} { Q}^2   = 2E^2 \frac{(1 - \cos \theta)}{1 + \frac{E}{M}(1 - \cos \theta)}
  \end{equation}
  where $M$ is the proton mass, $E$ is the known incident beam energy,
  and $\theta$ is the lab scattering angle.  The  (\ac{acro-hdc}s) established the scattered electron trajectory before
  the magnet in the event mode of the experiment. They tracked back to
  the target to establish the interaction vertex and scattering angle
  $\theta$ required for determining the $Q^2$ distribution.

\begin{figure}[hhbbt]
\begin{center}
\includegraphics[width=0.475\textwidth]{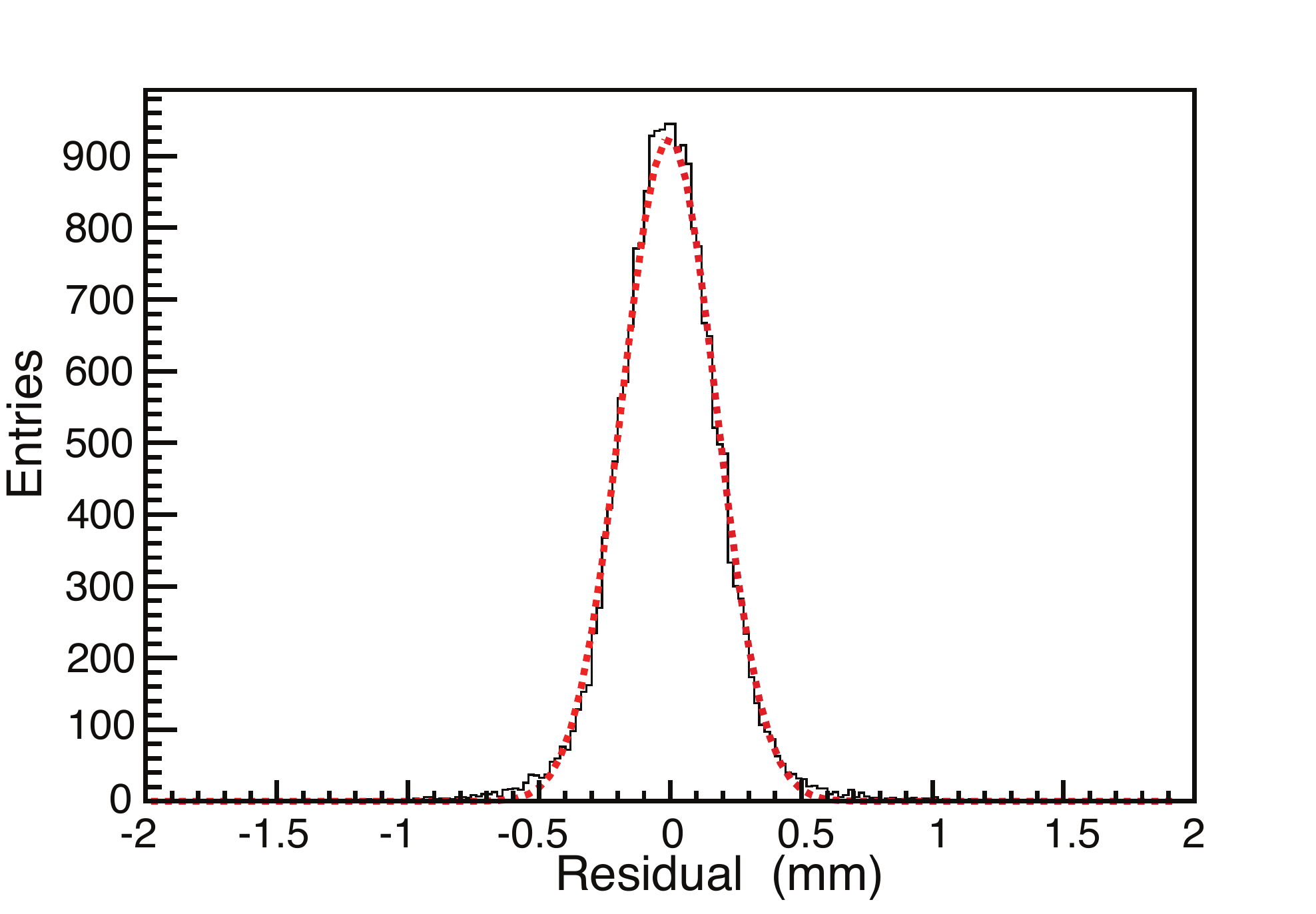}
\caption{ \label{fig:HDCresiduals} 
\ac{acro-hdc} track fit residuals under nominal running conditions (see text).  Displayed are the track fit residuals for a single plane of a 12-plane track fit through two  \ac{acro-hdc}s.  The dashed line is a Gaussian with RMS=178~$\mu$m, which is a measure of the characteristic chamber resolution.
  } 
\end{center}
\end{figure}

A total of five \ac{acro-hdc}s were constructed, with the fifth one serving as a spare.  
Each chamber consisted of six wire planes with 32 sense wires (20 $\mu$m diameter gold-plated tungsten wires strung at a nominal tension of 20~g) and 33 field wires (75
$\mu$m gold-plated beyllium-copper wires strung at a nominal tension of 30~g) per plane. The wire pitch was 5.84 mm, and the spacing between wire planes was 19.0 mm. The frame material was Ertalyte~\cite{Ertalyte}.  The six wire planes 
were in a $XUVX^\prime U^\prime V^\prime$ configuration, with the $U,V$ wires at angles of $\pm 53.1^\circ$ relative to the $X$ wires.   In the installed orientation, the typical electron track made an angle of $\sim$$7^\circ$  relative to the normal of the wire planes, so there was no need to offset identically strung planes by a half drift cell to resolve the left-right ambiguity.  
An automated scanning system with a digital camera and the standard vibrating wire technique~\cite{vibwire} was used
during chamber construction to measure wire positions and tension, with typical standard deviations of $\sim$50~$\mu$m for deviations of wire positions from expected and $\sim$3~g for wire tension about nominal.  
The wire planes were separated by cathode planes made of double-sided aluminized mylar foil.  The active area of each chamber was 28 cm $\times$ 38 cm.  The operating gas mix
was 65\% argon-35\% ethane.  The cathode planes and field wires were held at a potential of $-$2150 V, while the sense wires were held at ground.  Each completed chamber was tested with cosmic rays,
with measured position resolutions of $150 - 200~ \mu$m, and single plane efficiencies of \textgreater 99\%.
The single-wire position resolution was determined from the difference of a given wire's drift distance to the distance expected from a straight-line fit to $\sim$12 wires in a typical two-chamber \ac{acro-hdc} track. A typical  residual obtained during running under nominal conditions ($\sim$50 pA beam current on the LH$_2$ target with $\sim$100 kHz chamber rate) is shown in Fig.~\ref{fig:HDCresiduals}.

\subsubsection{HDC Positioning System}
\label{sec:HDC:Positioning}

The \ac{acro-hdc}s were situated between the defining (second) and third lead collimators just upstream of the \ac{acro-qtor} magnet. Each of two octants was covered by a pair of chambers which constituted an “arm”.
The two chambers which formed an arm for a given octant were separated by 42 cm; the upstream (downstream) chamber center was 3.15 m (3.57 m) from the center of the target. 

Each arm could be positioned radially inward to cover a given octant during tracking measurements, or outward to clear the scattered electron acceptance during production running. Furthermore, the arms were attached to a central hub which rotated about the beam pipe. The two arms were fixed 180$^\circ$ apart. Each arm could cover one of five different octants, which means that together both arms covered all eight octants, and that two octants could be covered by both arms (at different times). The  \ac{acro-hdc}s were positioned radially and rotated about the beam axis manually. 

\subsubsection{HDC Electronics}
\label{sec:HDC:Electronics}

On-board Nanometric N-277 preamplifier/discriminator cards~\cite{nanometric} sent the signals from each sense wire $\sim$27~m  over twisted pair ribbon cable to  \ac{acro-tdc}s located in a hermetically shielded   electronics hut in the experimental hall. 
The \ac{acro-tdc}s used were the \ac{acro-jlab} F1TDCs~\cite{F1TDC}, which are high-resolution multi-hit \ac{acro-tdc}s in 64-channel \ac{acro-vme}  modules.

\subsection{The Vertical Drift Chambers}
\label{sec:VDCs}

The \ac{acro-vdc}s were used in conjunction with the tracks from the \ac{acro-hdc}s and the known magnetic field of \ac{acro-qtor} to determine the scattered momentum and thereby identify elastically scattered electrons.

A total of five \ac{acro-vdc}s were constructed, with the fifth serving as a spare~\cite{Leckey}. The chambers were patterned on an earlier design used in \ac{acro-jlab}'s Hall A for many years~\cite{Fissum_et_al}, but with an
increase in overall size as well as modifications to some materials and other details which made them more cost effective and gas tight.

Each chamber consisted of two anode wire planes held at ground potential. Each plane included 279 sense wires. The sense wires were 25 $\mu$m diameter gold-plated tungsten wires strung at a nominal tension of 60 g. The wire pitch was 4.97 mm.  The two wire planes per chamber were strung in a $UV$ configuration, with the $U$ and $V$ wires oriented at angles of $\pm 26.56^{\circ}$ from the long axis of the chambers.

High voltage (\ac{acro-hv}) cathode planes operated at $-$3800 V were situated 12.7 mm above and below each wire plane. The outer HV planes were 12.7 $\mu$m thick Mylar foils, aluminized on one side. The HV plane located between the two wire planes was the same material but was aluminized on both sides.

The frame material, used for both the wire planes and the \ac{acro-hv} planes, was 1.27 cm thick G10-FR4~\cite{G10}, a mesh of compressed glass fibers and epoxy/resin. The frame pieces were sand-blasted to create a smooth, uniform surface to ensure uniform spacing between the cathode planes and the sense wires. Each rectangular frame (235 cm $\times$ 84 cm) was made of four separate pieces doweled together and bonded with Araldite epoxy (AY 103 resin and HY 199 hardener). The active area for each chamber was 204.5 cm $\times$ 53.3 cm. The entire 10.2 cm thick stack of frames was held together by two 1.9 cm thick  aluminum tooling plate frames with central cut-outs.  12.7 $\mu$m thick aluminized Mylar foils were stretched across the aluminum frames to contain the 50\% argon and 50\% ethane gas mixture.

An automated scanning system with a \ac{acro-ccd} camera attached to a stepper motor and linear encoder, and the usual vibrating wire technique~\cite{vibwire} were used during chamber fabrication to measure wire positions and tensions. With respect to the nominal values (4.97 mm pitch, 60 g tension), typical standard deviations were 78 $\mu$m and 5.6 g, respectively.

\begin{figure}[hhbbt]
\begin{center}
\includegraphics[width=0.475\textwidth]{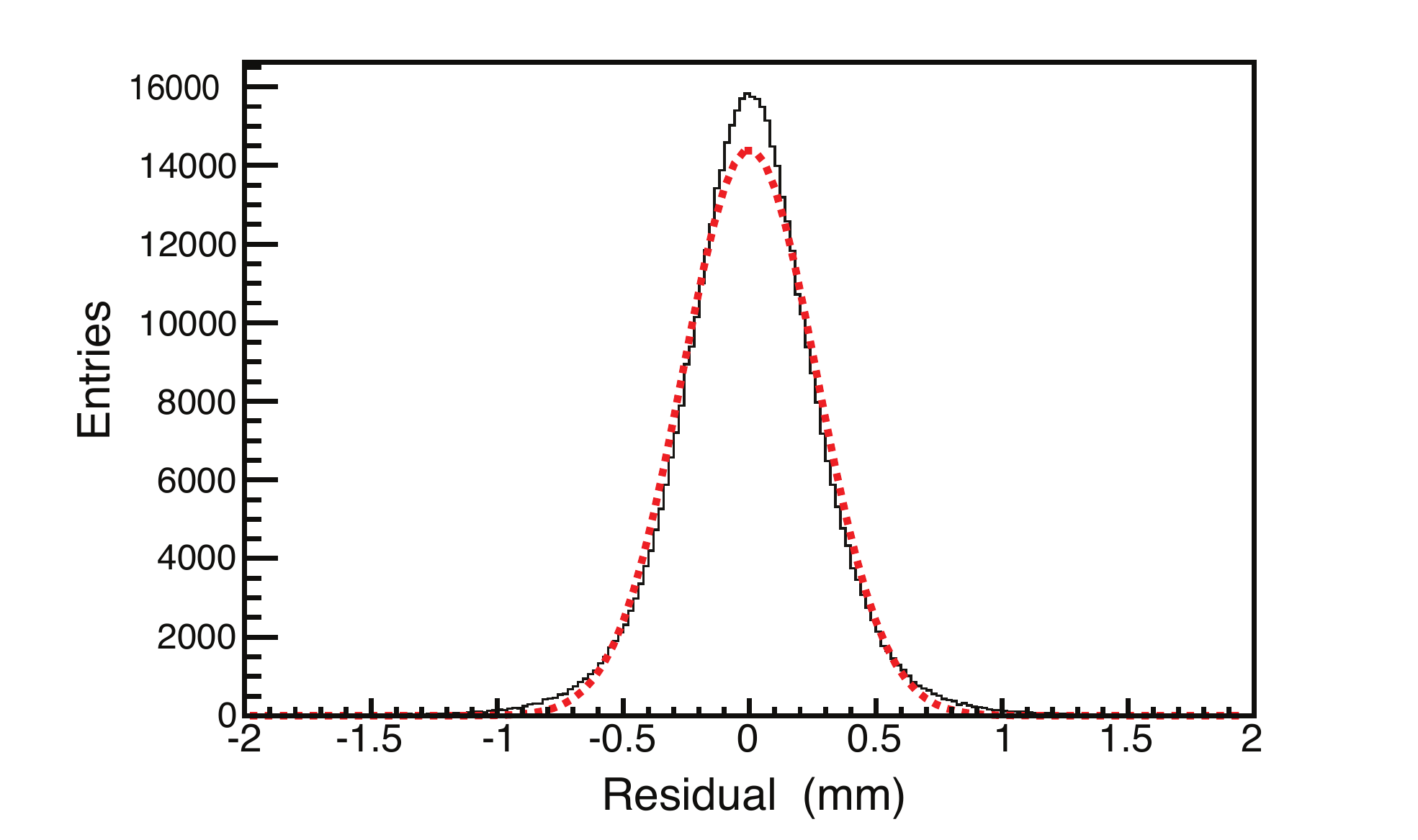}
\caption{ \label{fig:residuals} 
 Solid curve: A typical single-wire position residual formed from $\sim$24-hit tracks in the VDC (typically 6 wires firing in each of two planes in each of two chambers). 
Dashed curve: a Gaussian fit to the data,
with standard deviation of 264~$\mu$m. The RMS width of the data is 295~$\mu$m.
} 
\end{center}
\end{figure}

Each completed chamber was tested with sources and cosmic rays.  The dark current (current from cosmic ray flux alone) was $\sim$100~pA. The gas gain was determined to be $\sim$$ 2 \times 10^5$ at the operating voltage ($-$3800 V). The single-wire efficiency was $>$98.8\%.
At the $\sim$50~pA beam current used for the tracking measurements on the LH$_2$ target, the rate on the \ac{acro-vdc} was only $\sim$300 Hz. This was more than two orders of magnitude lower than the corresponding rate on the \ac{acro-hdc} chambers, which experienced high rates from M\o ller scattering on the LH$_2$ target. These M\o ller electrons were swept away by the \ac{acro-qtor} spectrometer and did not contribute to the rate observed in the \ac{acro-vdc}s. 
A typical residual showing the \ac{acro-vdc} position resolution achieved during these nominal running conditions is shown in Fig.~\ref{fig:residuals}.

\subsubsection{VDC Positioning System}
\label{sec:VDC_rotator}

The \ac{acro-vdc}s were located downstream of the \ac{acro-qtor} magnet, and upstream of the main detectors. Each of two octants was covered by a pair of chambers which constituted an arm.

In a vertical drift chamber, for which an ionizing track should typically fire about six wires per wire plane, the optimal incident track angle with respect to the wire planes is about 45$^{\circ}$. Given the 22.5$^\circ$ average lab angle of the elastic scattered electrons after the \ac{acro-qtor} magnet, the \ac{acro-vdc} pairs needed to be held at  24.4$^{\circ}$  from the vertical (optimized via GEANT~\cite{GEANT4a} simulation).
Each pair of \ac{acro-vdc}s in an arm was held in place at this angle by two 2.54 cm thick stainless steel side plates. Each \ac{acro-vdc} pair was separated by 53 cm (center-to-center) in the beam direction, and the center of the pair was located 107 cm upstream of the main detectors.

The two arms, held 180$^{\circ}$ apart, were mounted on rails which were in turn mounted to a central rotating hub.  The positioning system allowed the 1100 kg \ac{acro-vdc} assemblies on each arm to be retracted from the acceptance (linear motion) during integrating-mode running, and rotated to cover different pairs of octants during event-mode running. Each arm could cover one of five different octants, meaning that all eight octants could be covered, and that one octant pair could be covered by either arm (at different times) for systematic studies.

Pins locked the chambers in their retracted or extended positions. The linear and rotation motions were both automated. An  
electric cylinder~\cite{IDCEC3} with an S6961 controller~\cite{S6961} was used for the linear motion. The rotation was done using a Sumitomo Drive Technology 
motor~\cite{Sumito} with a HF-320 $\alpha$ controller~\cite{HF320} which drove a chain surrounding the central hub. Positioning reproducibility was better than 3 mm (azimuthal) and $< 1$ mm (radial).

\subsubsection{VDC Electronics}
\label{sec:VDC_Electronics}

Custom-made preamplifier/discriminator cards employing the CERN MAD chip ~\cite{mad1,mad2} were mounted directly to the \ac{acro-vdc}s. The  \ac{acro-lvds}  
output of these cards was carried through 16-channel ribbon cables 15~m long which absorbed the rotation and translation of the \ac{acro-vdc}s. Thirty~m of twisted pair fixed cables then carried the 558 signals from each \ac{acro-vdc} to a level translator in a shielded electronics hut in the experimental hall. 

The level translator converted the \ac{acro-lvds} signals to 
\ac{acro-ecl} signals which were split into two signals. These formed the inputs to a novel digital delay multiplexing system~\cite{MUX}.
Delayed signals from every 8$^{\rm th}$ wire in a group of 141 wires were ganged together, enabling a significant reduction (factor of $\sim$9) in the number of readout channels required. 
The arrival time of the multiplexed \ac{acro-vdc} signals~\cite{MUX} 
with respect to the trigger scintillators (see Sec.~\ref{sec:TriggerScints}) was measured and digitized with one F1TDC module~\cite{F1TDC} per \ac{acro-vdc}.

\subsection{The Trigger Scintillators}
\label{sec:TriggerScints}

Plastic scintillators were used to provide the fast timing trigger to the electronics for event mode readout.  The scintillators were also used in the  analysis of event mode data to study neutral backgrounds in the detectors.  

The trigger scintillators were 218.45~cm $\times$ 30.48~cm $\times$ 1.00~cm Bicron BC-408 plastic~\cite{BC408}  manufactured by Saint Gobain~\cite{SaintGobain}.  One scintillator was positioned on each arm of the \ac{acro-vdc} rotator directly downstream of the \ac{acro-vdc}s. 
Light guides on each end of the scintillator were formed from strips of UV-transparent lucite coupled to UV-transparent disks.
Photonis XP-4312B 7.6~cm diameter \ac{acro-pmt}s were coupled to the 7.6~cm diameter UV-transparent disks, and a Photonis VD123K transistorized voltage divider was used with a 3~M$\Omega$ resistance.  The \ac{acro-pmt} high-voltage was set between $-1725$ and $-1860$~V.  

A CAEN N842 8-channel constant fraction discriminator~\cite{CaenN842} was used in conjunction with a CAEN V706 16-channel meantimer module~\cite{CaenV706} for optimum timing performance. The efficiency of the scintillators was $>$99\%, and the measured timing resolution using cosmic rays was $\sim$460~ps (\ac{acro-rms}). 

\subsection{Track Reconstruction Software}
\label{sec:Tracking_Software}

The offline track reconstruction software was used to determine the scattered electron kinematics. This was done by comparing the straight-line track segments before (from the \ac{acro-hdc}s) and after (from the \ac{acro-vdc}s) the known spectrometer toroidal magnetic field. The overall algorithm was patterned after that developed by the HERMES collaboration~\cite{Hermes}.

Standard procedures~\cite{timetodx} were used to convert the drift time measurement to drift distance for each wire hit in the tracking chambers. For the \ac{acro-vdc}s, individual wire-by-wire timing offsets were determined from the measured scattered electron tracks. A common timing offset was adequate for the \ac{acro-hdc}s.
For the \ac{acro-vdc}s, the difference in times between the two ends of the digital delay lines was used to demultiplex and identify the wire number of each wire fired.

Once hit locations were determined, a pattern recognition algorithm 
was used to separately identify valid line segments in the \ac{acro-hdc}s and in the \ac{acro-vdc}s. The algorithm used a template matching scheme, similar to the one developed for the ARGUS experiment~\cite{ARGUS}.  The templates were based on two-dimensional projections of a track segment. The pattern of wire hit locations was compared to a series of templates, each of progressively finer spatial resolution, which were generated from simulated tracks. Each template was stored as a bit pattern, with the spatial region of the hit stored as a ``1" and regions without a hit as a``0". These allowed the construction of a searchable tree of valid patterns~\cite{tree}, which could be quickly and efficiently compared to the pattern corresponding to the observed wire hits.

\begin{figure}[hhbbt]
\begin{center}
\includegraphics[width=0.475\textwidth]{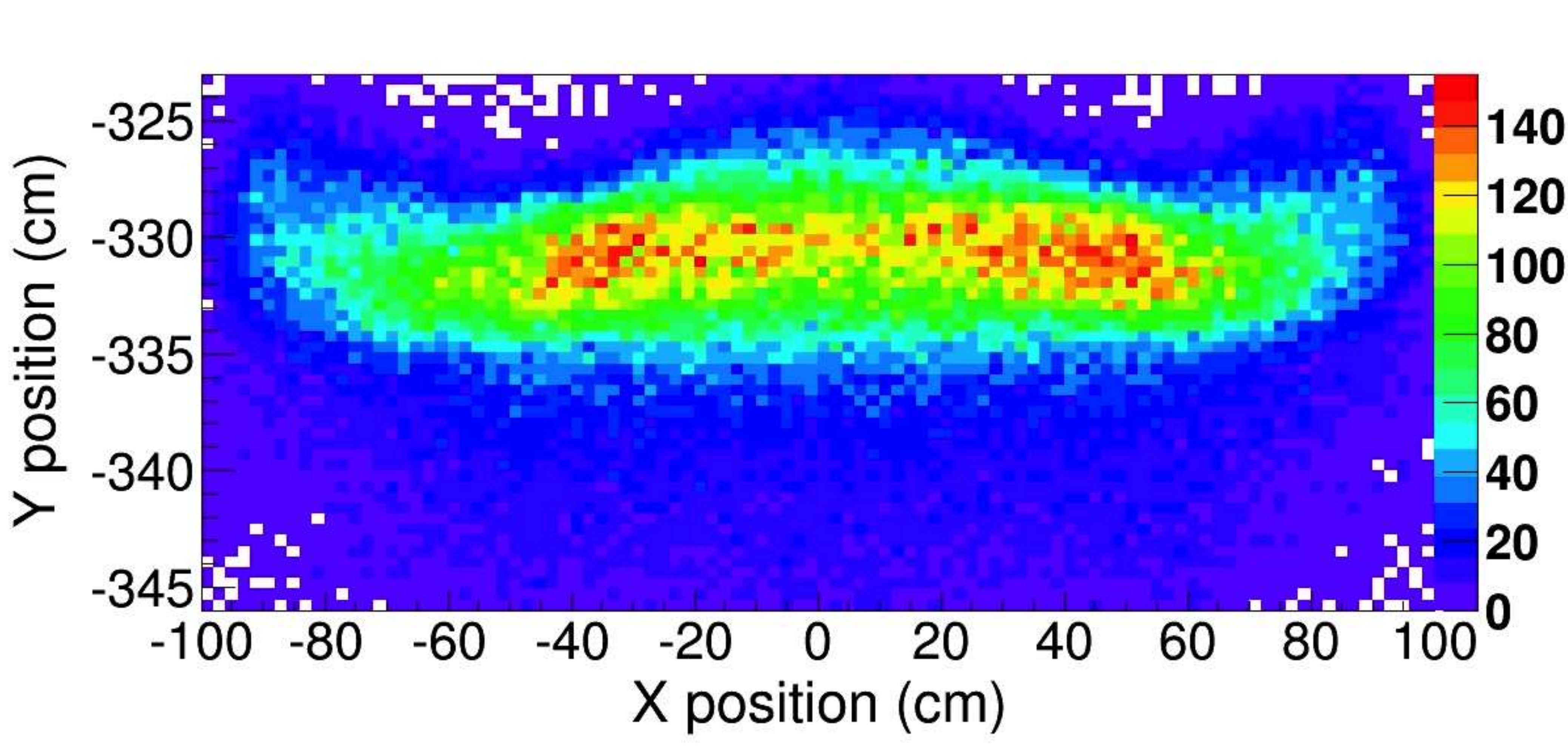}
\caption{ \label{fig:vdcprofile} 
A projection of tracks found in the \ac{acro-vdc}s to the main quartz detector in the bottom octant. The trigger used for these events did not include the 2~m wide main detector itself, which extended from -326 cm to -344 cm in $Y$. The $X$-axis is in horizontal direction, the $Y$-axis is in vertical direction. The  
scale indicates the relative electron flux. } 
\end{center}
\end{figure}

Once a valid template was found for the wire hits corresponding to a given wire plane orientation in a given set of chambers in each arm ($X,U,V$ for the \ac{acro-hdc} pair, $U,V$ for the \ac{acro-vdc} pair), the data from each orientation was combined in a least-squares fit to form a three-dimensional track segment. 
Fig.~\ref{fig:vdcprofile} shows tracks from the \ac{acro-vdc}s projected to the main detectors, which compares well to the corresponding Fig.~\ref{fig:scanner_ratemap} provided by the focal-plane scanner  in Sec.~\ref{sec:Components:Current:FocalPlaneScanner}.
%

The track segment determined before the magnet from the \ac{acro-hdc}s was then ``swum" numerically through the magnetic field using a 4th-order Runge-Kutta~\cite{RungeKutta} procedure. This requires an initial estimate of the energy of the scattered track; this estimate was derived from the known beam energy, the scattering angle from the initial track segment, and the assumption of two-body kinematics for elastic scattering.  If the resulting track after swimming through the field  matched, within crude limits, the location and angle of the actual track segment found in the \ac{acro-vdc}s, the event was accepted as valid. The scattered energy was then iteratively improved using the Newton-Rapheson method, by comparing the radial (dispersive direction) position of the \ac{acro-vdc} track with that of the swum track, and repeating the swimming process until convergence.

\section{Electronics and Data Acquisition}
\label{sec:Components:DAQ}

The \acf{acro-daq}  was built to function in one of two modes:  ``integrating mode'' or ``event mode'' (see Sec.~\ref{sec:Logistics}).
Integrating mode was used with beam currents  up to 180~$\mu$A to record the average detector and beamline
instrumentation signals in each interval of stable beam helicity.
Event mode was used with beam currents well below 1~$\mu$A to record trajectory information for individual
particles in the spectrometer, triggered by a hit in one of the detector elements.
The two operational modes were largely independent, but some instrumentation was common, such as the focal
plane scanner (see Sec.~\ref{sec:Components:Current:FocalPlaneScanner}).

\subsection{Integrating Mode Preamplifiers}
\label{sec:Components:Preamp}

Low-noise pre-amplifiers based on the OPA2604~\cite{OPA2604} and OPA2227~\cite{OPA2227} by Burr-Brown were used to convert the \v{C}erenkov main quartz detector anode currents to voltage signals. The pre-amps were designed~\cite{qwads} with two \ac{acro-i2v} channels per \ac{acro-rf}  shielded package. The transimpedance was selectable using an internal switch from 0.5 to  4.0 M$\Omega$ for the main detectors and downstream \ac{acro-lumis} and 0.5 to 50 M$\Omega$ for the upstream luminosity monitors.  The output dynamic range of $\pm$10~V was matched to the input range of the \ac{acro-adc}  (see below) and drove  130~m of RG-213 cable. A ganged output offset was internally adjustable in the range $\pm$1.2~V. The pre-amp required a +5~V supply. 
An internal \ac{acro-dc}-\ac{acro-dc} converter stepped up the supply voltage 
to $\pm$15~V while providing isolation from the external power supply.  A 26 KHz bandwidth filter was applied to the output signals using a single pole filter. Because the detector signals were normalized to beam current, a 26 kHz filter was also used on the \ac{acro-bcm}s.

Before the experiment, a prototype pre-amp was tested to 18 kRad in the mixed electron and photon field of the \ac{acro-jlab} $^{137}$Cs irradiation facility  with no increase in noise. During the experiment, the pre-amps were located within a few meters of the detectors and received an estimated dose of at least 1 kRad from mostly electromagnetic background. No noise degradation was observed. 

In integrating mode, low-capacitance RG-62 (93 $\Omega$) cable was used to carry the anode current signal from the \ac{acro-pmt} bases to the pre-amplifiers. 
In event mode, standard RG-58 (50 $\Omega$) cable was used.   Under nominal operating conditions, all main  detector pre-amps were set to 2 M$\Omega$ transimpedance, then the main detector \ac{acro-pmt} \ac{acro-hv}’s were adjusted until the average signal magnitude was 6 V. At the corresponding anode current of 3 $\mu$A, the \ac{acro-pmt}s showed a modest  gain drop of only $\sim$10\% throughout the experiment (see Fig.~\ref{fig:qweak_md_yield}). 
Further details, including a schematic diagram, can be found in~\cite{wang2011cerenkov}.

\subsection{Integrating Mode Instrumentation}
\label{sec:Components:DAQ:CurrentMode}

\begin{figure}[hhbbt]

\begin{center}
\includegraphics[width=0.5\textwidth]{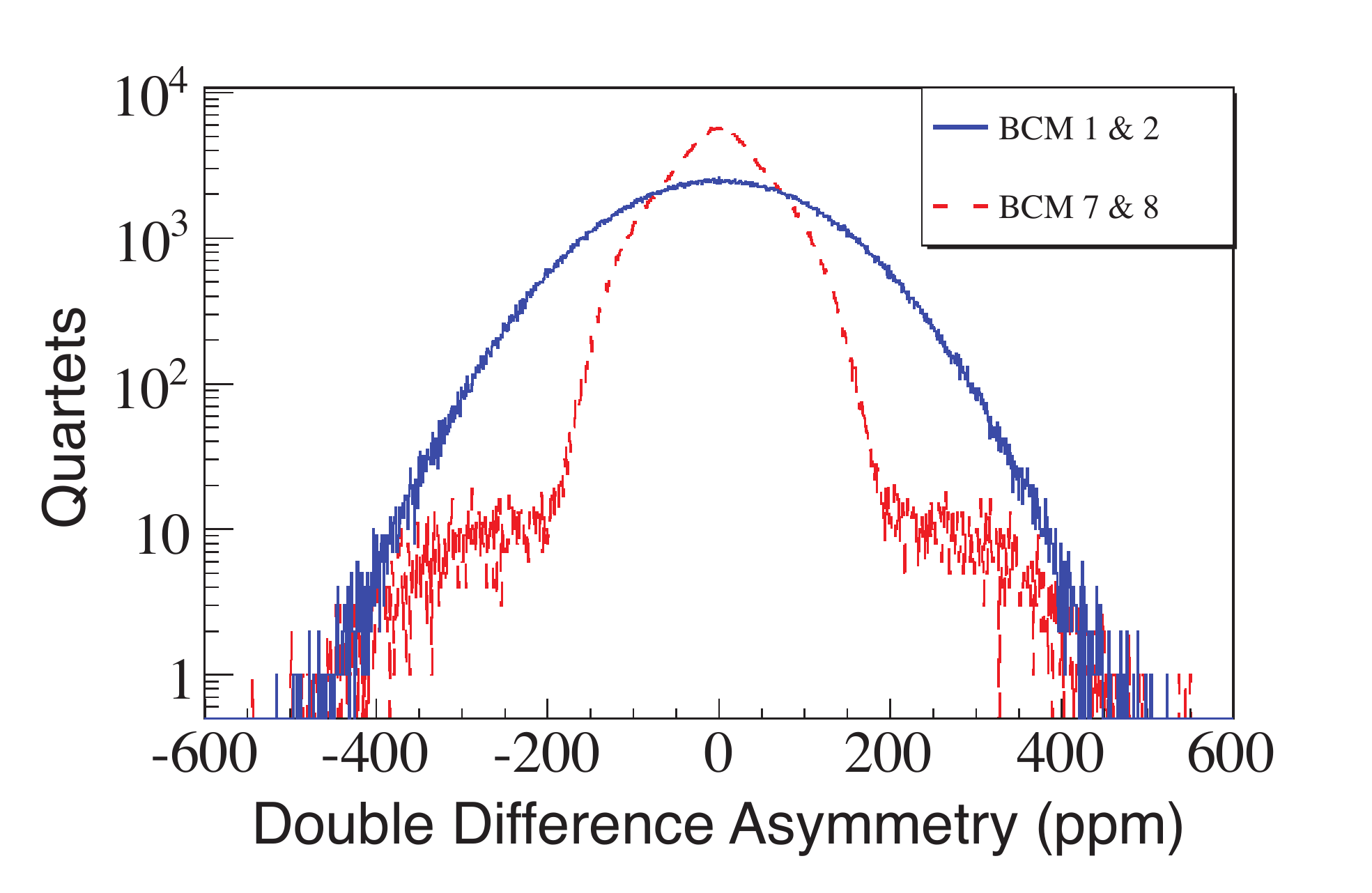}
\vspace*{-1.5cm}
\caption{ \label{fig:adc_chain} 
A simple schematic of the main detector signal chain. 
 } 
\end{center}
\end{figure}

The goal of the integrating-mode instrumentation was to record the integrated signals or yields from all the detectors
and beam monitors during each period of stable beam helicity $T_\text{Stable}$, as well as recording the beam helicity itself.
Most detector and beamline instrumentation signals were transformed into time-dependent voltages, and the average
voltages were measured during the  $T_\text{Stable}$ interval by either a sampling-integrating \ac{acro-adc} 
(the \ac{acro-vqwk} modules built by \ac{acro-triumf}~\cite{qwads}) 
or by a gated scaler counting the output of a voltage-to-frequency converter.
For a few detectors, such as the halo detectors (see Sec.~\ref{sec:Beam:Halo}), the
instrumentation was a gated scaler counting pulses from a discriminator. 
The main detector signal chain is depicted schematically in Fig.~\ref{fig:adc_chain}.

The voltage signals produced by the detector and beam monitor I-to-V pre-amplifiers, as well as voltage signals produced by the \ac{acro-bcm} receivers, 
were digitized in the  \ac{acro-vqwk}~\cite{qwads} \ac{acro-adc}s.  Each of the eight input signals of a \ac{acro-vqwk} module was sampled
at 500~kHz by an AD7674 18-bit \ac{acro-adc}~\cite{AD7674} accepting a full-scale range of $\pm 10$~V.  An \ac{acro-fpga} on the \ac{acro-vqwk}
module  synchronized and accumulated the sample readings into the
reported value.  For each gate trigger, the \ac{acro-vqwk} module accumulated a preset number
of samples to produce the reading for that trigger; in addition to the total sum, four sub-sums were
accumulated representing each quarter of the gate. 
The input stage of the \ac{acro-vqwk} 
module had a 5-pole low-pass filter with a 50~kHz cutoff to
prevent aliasing of the input signals.

As discussed in Sec.~\ref{sec:Components:Preamp}, the bandwidth of the \ac{acro-bcm}s and \ac{acro-bpm}s was matched to the 26 kHz bandwidth of the integrating-mode preamplifiers. The delay of these signals was also matched, by making use of the 60 Hz pulsed (tune) beam structure available at \ac{acro-jlab}. To allow for the fact that  pure analog and digital \ac{acro-bcm} receiver  signals  arrived out of time by $\mathcal{O}\left(10\right)$ $\mu$s, gates sent to \ac{acro-adc} modules containing late signals were programmed to begin  digitizing after a programmable delay. The flexibility  to match both bandwidth and delay between the detectors and the \acp{acro-bcm} was a crucial feature of the electronics.

\begin{figure}[hbt]
\begin{center} 
\includegraphics[width=0.47\textwidth]{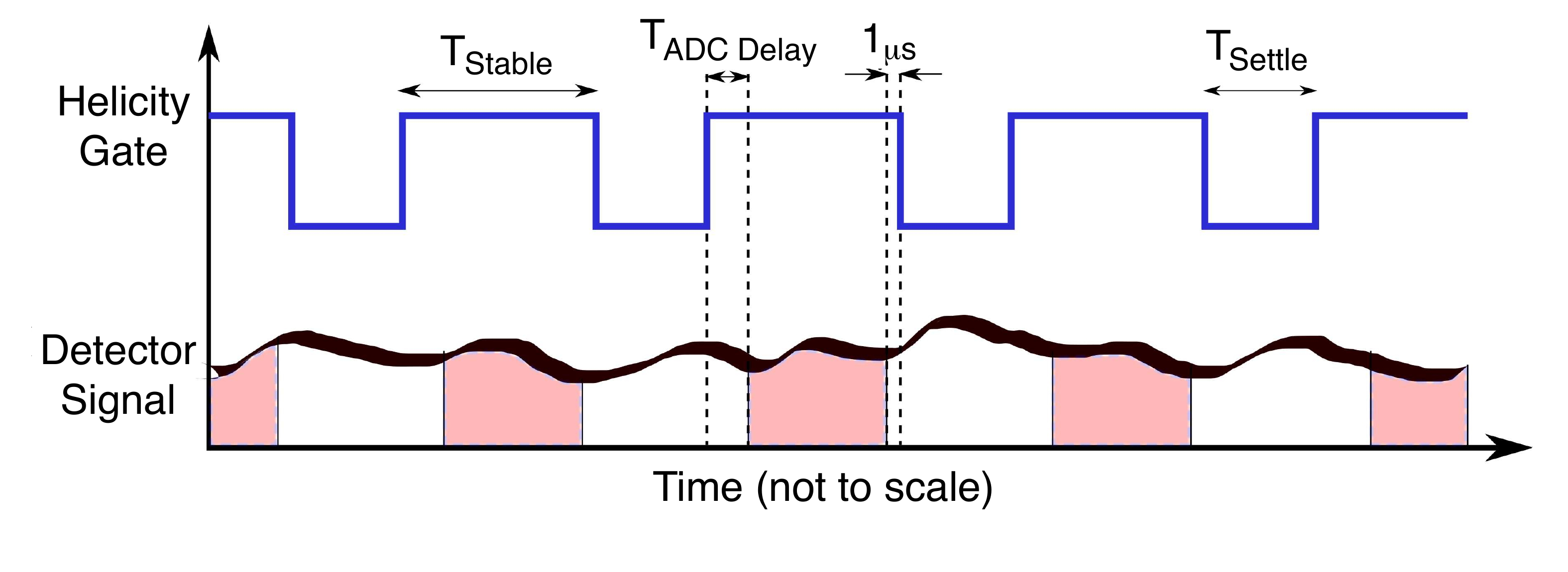}
\caption{ \label{fig:adcgate_timing} 
The readout timing diagram of the integrating-mode \ac{acro-daq}. The shaded area in the detector signal indicates the signal region digitized by the \ac{acro-adc}s.
The  horizontal axis scale is exaggerated to show details of the signal timing.  Further information on the helicity signal timing is summarized in Fig.~\ref{fig:helicity_timing}.
}
\end{center}
\end{figure}

At the 960/s reversal rate, the \ac{acro-vqwk} \ac{acro-adc}s
acquired 464 samples per Helicity Gate (see Fig.~\ref{fig:helicity_timing}),
giving a sampled average of the input voltages over a 928~$\mu$s interval. 
The accumulation of samples was started $T_{\rm ADCDelay}$=42.5~$\mu$s after 
the beginning of the $T_{\rm Stable}$=971.65~$\mu$s interval in the Helicity Gate signal.
Thus the last sample was taken about 1~$\mu$s before the
end of the $T_{\rm Stable}$ interval.
$T_{\rm Stable}$ and the time set aside for the helicity
transition to fully complete ($T_{\rm Settle}$=70~$\mu$s) were set by the polarized source helicity
board (see Sec.~\ref{sec:Source:HelicitySignal}).   
$T_{\rm ADCDelay}$ was set internal to the \ac{acro-vqwk} modules, and was used to prevent distortion of the voltage samples by the beginning of the gate signal. 
Use of the $T_{\rm ADCDelay}$ contributed an additional 4.1\% deadtime on top of the 6.7\% deadtime associated with $T_{\rm Settle}$ discussed in Sec.~\ref{sec:Source:HelicitySignal}. The readout timing diagram
of the integrating-mode \ac{acro-daq} is summarized in Fig.~\ref{fig:adcgate_timing}.

The \ac{acro-rms} width of the quartet asymmetry generated on a channel digitizing a constant voltage signal 
(from a battery) was about 3~ppm, compared to the $\sim$230 ppm \ac{acro-rms} width of the detector asymmetries.
 The random noise introduced by the electronics chain was thus negligibly  small compared to the fluctuations from counting statistics.

A set of three SIS3801~\cite{SIS3801} scalers and one STR7200~\cite{STR7200} scaler were used to provide rate measurements
for detectors and monitor channels that were not instrumented by the \ac{acro-vqwk} modules. 
During the integrating-mode running, only the Helicity Gate triggers were collected; all other
trigger types (see next section) were disabled.  There were 32 \ac{acro-vqwk} modules
in three \ac{acro-vme} crates, providing digitization of the detectors as well as beamline instrumentation in
Hall C and  the injector.  Two additional \ac{acro-vme} crates supported
the scalers and the trigger control electronics.

The total data rate was about 6.5~MB/s.  The data were written to disk as a series of
data files with a maximum size of 1.9 GB.  
These files, known as ``runlets", each represented about 5--6 minutes of data collection. A typical hour-long run would consist of about 10--12  runlets. 
The analysis 
also benefited from the segmentation
into runlets, as beam and experimental conditions were generally stable on a  5 minute timescale, but could vary more over the course of an hour-long run.

\subsection{Event Mode Instrumentation}
\label{sec:Components:DAQ:TrackingMode}

In event mode, detector 
information was collected based
on detector-based triggers, as opposed to the periodic helicity-based triggers
collected in integrating mode.
Pre-scaling and trigger selection was done by a \ac{acro-jlab} Trigger Supervisor
module~\cite{ts}, with several trigger inputs.  The helicity-based trigger was replaced by
a slower (usually 10 Hz) clock, to control readout from scalers.  The most common
trigger sources were:  the trigger scintillators, the main \v{C}erenkov
detector bars, the focal plane scanner, and one of the ``background'' detectors.  
Occasionally other trigger sources were used for tests and background measurements.

The instrumentation was changed for event mode as well.  All of the  
drift chambers were read out using \ac{acro-jlab} F1TDC modules~\cite{F1TDC}.
The main detectors and trigger scintillator signals were passively split,
with one copy being discriminated and input to  F1TDC modules, and the
other copy passing through an additional 190~ns of delay cable before
being input to a CAEN V792~\cite{V792} Charge-to-Digital Conversion (QDC) module.
All of the timing signals from the
scintillators and \v{C}erenkov detectors were also input to scalers.

\section{Software}
\label{sec:Software}

Simulations were performed based on the GEANT3~\cite{GEANT3}, GEANT4~\cite{GEANT4a,GEANT4b} and GARFIELD~\cite{GARFIELD} simulation packages.  GEANT3 simulations were used in the design phase of the experiment to optimize the acceptance, and to study background from the aluminum target cell and other sources.  GEANT4 was used to maximize the photoelectron yield from the quartz main detectors, and in the analysis of tracking data (see Sec.~\ref{sec:Tracking}).  GARFIELD was used in the design of both the horizontal and vertical drift chambers to optimize the gas mixture, field and sense wire positions, and the cathode plane spacing.

The physical processes implemented in both GEANT simulations included all electromagnetic and low-energy hadronic processes above a threshold value. In addition, generators were written for elastic $e-p$, inelastic $e-p$, and M{\o}ller scattering in the target, and for scattering from the aluminum target windows.

\subsection{Acceptance, Rate, and Momentum Transfer Simulations}\label{sec:acceptantance}

As part of the \Qweak acceptance optimization, the collimator design, the position of the main detector quartz bars, 
and the current setting of the \ac{acro-qtor} magnet were varied.  First, the maximum solid-angle acceptance that would clear the 
\ac{acro-qtor} spectrometer support structure was found.  Then, for a given collimator design, the elastic rate, mean $Q^2$, uncertainty 
on \Qwp, and the background dilution (contamination) from inelastic and target window scattering were calculated as a function of the 
main detector radial position. The final combination of magnetic field and detector position
chosen was an optimization that minimized the \Qwp  uncertainty, through a trade-off maximizing the elastic rate and acceptance while minimizing
the inelastic and target window background dilutions.

\subsection{Background Simulations}
\label{sec:bkgsims}

In addition to aiding in the design and optimization of the acceptance-defining collimator, the \Qweak GEANT3~\cite{GEANT3} simulation was
also used in the design of the upstream shielding 
wall of the detector hut, the  
``lintel" collimators, and the general study of backgrounds from the target 
windows and the beamline, as well as other background processes. 
 
An example of one of many improvements to the experiment's design that came out of these studies is the lintels alluded to 
in Sec.~\ref{sec:Lintels}. 
The simulations~\cite{Myers} showed that electrons from elastic and M\o ller scattering generated photons along the  
inner edge of the defining collimator apertures which had direct line of sight to the detectors. 
Lead ``lintel" collimators were added between the coils of the magnet to provide line-of-sight shielding 
between the \v{C}erenkov detectors and the photons from this background source.  
Simulations showed that the lintels reduced this background by about 90\%. 
Elastically scattered electrons which passed through the defining collimator apertures were 
deflected by the spectrometer field and not affected by the lintels. 
The lintels were also designed to avoid the intense ``fountain'' of low-energy electrons from M\o ller scattering 
in the target, in order to prevent them from becoming a net source of photon background.

The apertures in the upstream wall of the detector shielding hut were also carefully designed with GEANT3 simulations~\cite{Myers}. 
These apertures essentially consituted a fourth collimator, the only one downstream of the spectrometer magnet.
Backgrounds from the aluminum target windows, beamline, collimators, and shielding wall were all studied with simulations. 
The results  were used to devise methods to measure and suppress these backgrounds. 

\subsection{Detector Simulations}

The detector simulation software was developed~\cite{wang2011cerenkov} 
 based on the GEANT4~\cite{GEANT4a} framework. The basic method was to 
track particles from the target to the detectors using the detailed geometries of the detectors and shielding as well as the measured magnetic field of the spectrometer as input. 
By turning on all relevant physics processes along the particle trajectories, the detector response (light production and transport) could be simulated for various trial geometries. The simulation results were benchmarked by comparing them to independent experimental results wherever possible and the resulting deviations were then used to modify the simulation code until consistent results were obtained or a reasonable uncertainty could be assigned to the simulation results. The detector design was optimized through several such iterations.

The detector geometries were implemented in particular detail, including the quartz bars, glue joints, lightguides, \ac{acro-pmt}s, detector housing, \ac{acro-pmt} housing, \ac{acro-pmt} lead shield, pre-radiator, quartz bar holders and detector windows (covers). 
Relevant material properties were included, such as wavelength spectra, index of refraction, surface reflectivity of the \ac{acro-pmt} photocathode, and the quartz surface roughness. 
The simulation investigated the consequences of possible defects, such as a small mismatch between two quartz bars, quartz bar and/or light guide misalignment, and oversized bevels or chipped edges. Detector geometries such as detector thickness, pre-radiator thickness, and light guide shape were studied by turning on the \v{C}erenkov process to determine the overall efficiency of the detectors for various design choices. 

Extensive simulations were performed to study effects of quartz bar surface properties, to determine quartz bar and light guide shapes, quartz bar tilt angle, to study the  position dependence of the light yield, as well as to study background and the $Q^2$ distribution in the detectors. The length (200 cm) and width (18 cm) of the quartz bars were determined by the elastic beam spot size on the focal plane. 
The thickness (1.25 cm) was optimized using simulations by balancing the competing aspects of maximizing light yield and minimizing shower activity and background~\cite{wang2011cerenkov}. A similar procedure was performed to find the optimal thickness (2~cm) for the pre-radiator. 

The optimal \v{C}erenkov detector light guide geometry was determined by simulating the light yield for various configurations, such as with the \ac{acro-pmt} on the edge or on the face of the light guide for rectangular, trapezoidal, or wedge guide geometries, or without a light guide (\ac{acro-pmt} on the active quartz volume edge only). 
Although the geometry variations showed   no significant effect on the excess noise due to light yield variations, the mean photoelectron yield was largest with the \ac{acro-pmt} on the face of a rectangular light guide. 

The detector tilt angle was also optimized by determining the excess noise as a function of detector tilt angle, the light yield uniformity as a function of electron hit positions along the length of the quartz bar and tilt angle, as well as the uniformity of the $Q^2$ distribution across the quartz bar. The light yield was
smaller for a zero tilt angle, but significantly more uniform.
The optimal tilt angle chosen was zero -- perpendicular to the beam direction.

The simulation studies showed that the total light yield in the detector depended on the position of the incident electron  
-- a combined effect of the shower activities in the quartz bar, the various path lengths of the electrons in the quartz bar, and the various light transmission distances in the quartz bar. The light yield was approximately uniform in the short, radial dimension  but  nonuniform in the long (200 cm) direction of the quartz bar. This hit position dependence of the light yield affected the excess noise in the detector as well as the light-weighted $\langle Q^2\rangle$ determination. Because events near the middle of the quartz radiator had a lower light yield and the lower $Q^2$ events were focused in the center of the detector, the combination of the two effects biased the $Q^2$ upward by about 1.5\%.

\begin{figure}
  \centering
  \hspace*{-0.5cm}
     \includegraphics[width=0.55\textwidth, angle=0]{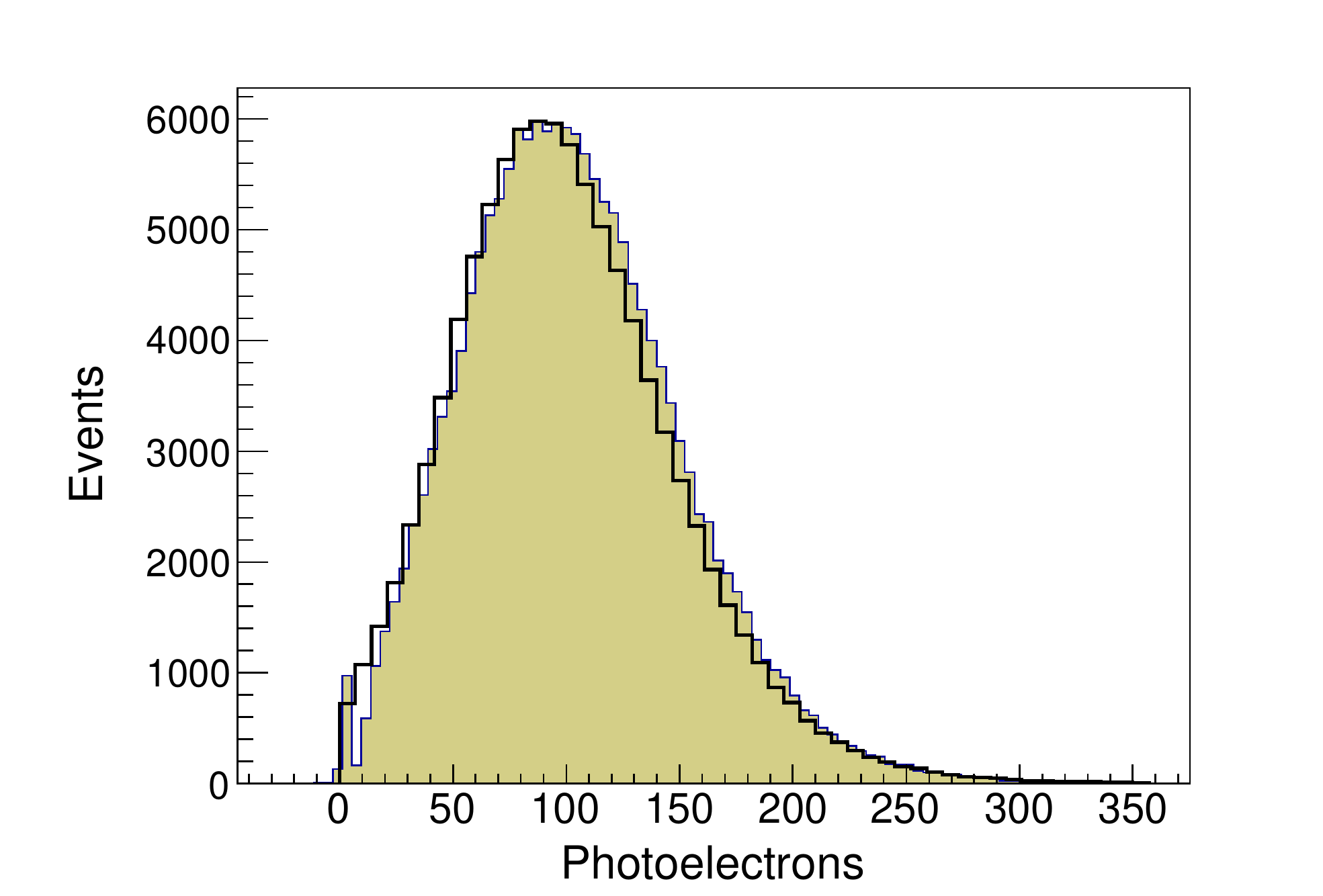}
  \caption{\label{fig:adc_spectrum} 
   Shaded: A typical quartz detector pulse-height spectrum accumulated during event-mode running (see text). The solid curve indicates the prediction from a GEANT4~\cite{GEANT4a} simulation incorporating the geometry of the bar and accounting for all the physical properties of the detector bar and physics processes described in the text. 
  }
\end{figure}

The prediction of a GEANT4~\cite{GEANT4a} simulation is compared to the measured pulse-height observed in a detector bar in   Fig.~\ref{fig:adc_spectrum}.
The measured spectrum corresponds to the sum of the signals from both ends of a pre-radiated detector bar during event mode-running at $\sim$50 pA with the LH$_2$ target, calibrated in photoelectrons. The simulation includes all the details of the bar described above and does an excellent job describing the measured spectrum.  

\subsection{Analysis Software}

Software was developed in C++ to decode 
raw data from the \Qweak \ac{acro-daq} in integrating and event modes for multiple detector systems.
The \ac{acro-daq} integrated over the fundamental 960/s helicity window, and computed  asymmetries for each helicity quartet.

After decoding of the detector signals, basic data quality cuts were applied based on the beam current and position, whether the signal was saturated, and hardware error checks. Asymmetries were then formed for the main detector (blinded), luminosity, and beamline monitors. The results were provided to ROOT~\cite{root} trees and histograms  as well as to a MySQL~\cite{mysql} database. 

\begin{figure}[ht]
  \centering
  \includegraphics[width=0.5\textwidth]{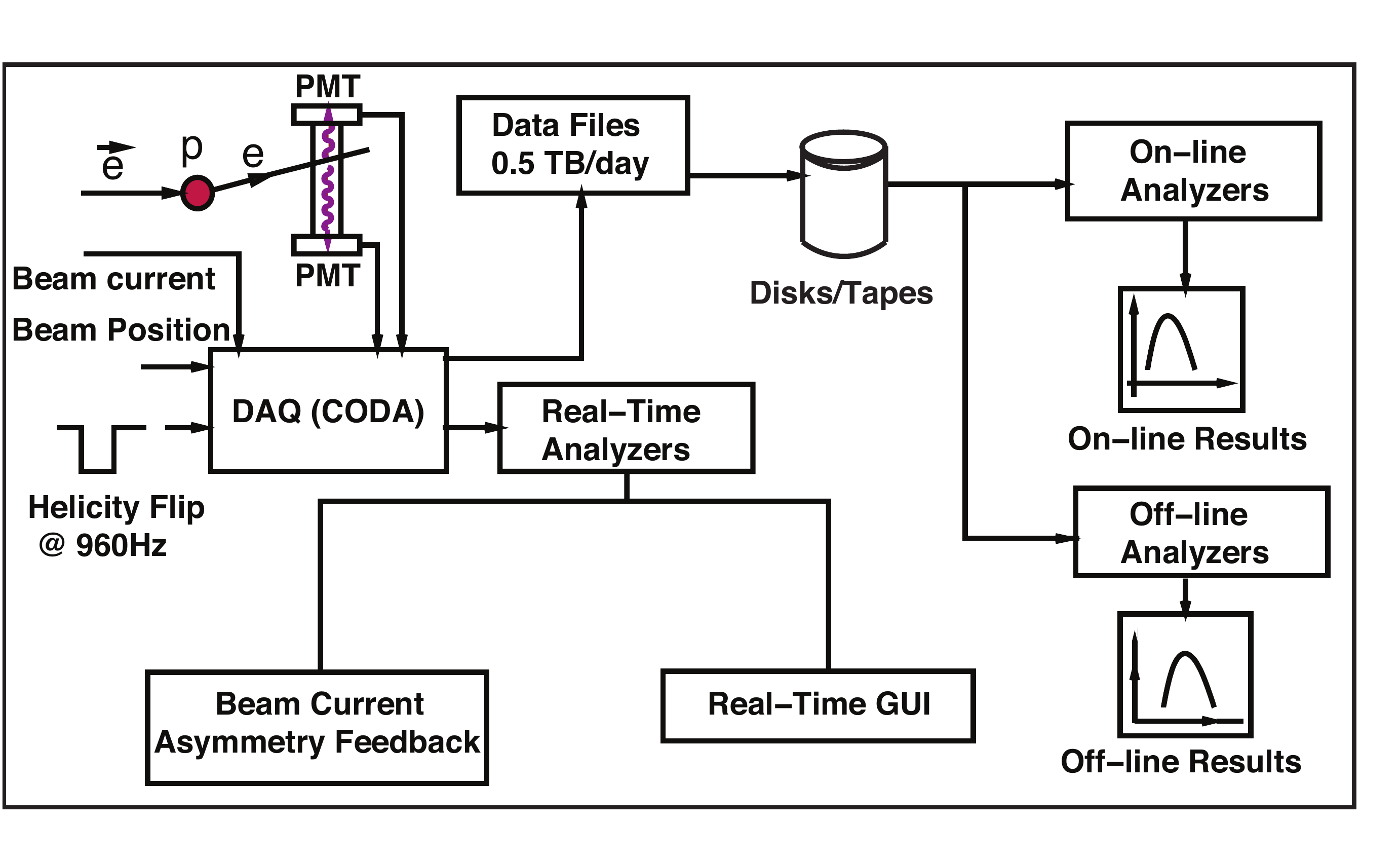}
  \caption{\label{fig:qweak_exp_overview} The basic setup used for the \ac{acro-daq} and analysis framework. 
The \ac{acro-daq} generated the raw data while the analysis framework provided realtime and offline data analysis capability.}
\end{figure}

In event mode, the analysis software decoded scaler, \ac{acro-adc}, and \ac{acro-tdc} information presented by different types of triggers. The output was accessible from a set of ROOT~\cite{root} trees. An overview of the \Qweak data analysis system integrated with the \Qweak \ac{acro-daq} is shown in  Fig.~\ref{fig:qweak_exp_overview}.

\subsubsection{Online Analysis}

The main goal of the online analysis system was to provide a means to assess 
the health of the \Qweak \ac{acro-daq} and the various detector systems, as well as to assess beam quality and provide feedback to the accelerator operators. 
The online analysis was derived from the main analysis framework to access the \ac{acro-daq} in real-time and  monitor the experiment's status.
The monitoring system provided real-time access to a small number of important quantities. A second 
 monitoring system summarized data quality for a much fuller set of observables from the first 5 minutes of each hour long run. Those results were available about half an hour after they were acquired,  passed to a ROOT~\cite{root} file and used to update snap-shots to a web-based log system.
An additional role played by the online analysis was  a 
 beam current asymmetry feedback loop to converge the beam charge asymmetry to zero (see Sec.~\ref{sec:Beam:BCM}).

\subsubsection{Offline Analysis}

The offline analysis chain for the integrating-mode data consisted of two stages.
First, the raw data files were processed to produce helicity-averaged yields and helicity-correlated asymmetries for each quartet.  Then 
the correlations between the detector asymmetries and beam parameter variations were determined and used to correct the 
detector asymmetries, as described in Sec.~\ref{sec:Beam:BMOD}.

 The first step in processing the raw data was to apply
calibrations to the raw detector information 
for each 960/s helicity gate.  Several event cuts were also applied for each helicity gate: 
the beam current had to be above a certain threshold 
(usually 100~$\mu$A) and none of the main detector channels or the most important beam monitor channels could have a fault in the event.  
Next, a 
4000-event ring buffer was used to apply a stability cut on the beam current and on
two main detector channels; if the \ac{acro-rms} of the distribution was outside of limits,
all of the events in the buffer were discarded.  Also, if the beam current dropped
abruptly, a fixed number of events before the drop were discarded.  

Quartet asymmetries and helicity-averaged
yields were calculated for the 
events that
passed the event cuts and stability cuts.  The main detector asymmetries were blinded by an
additive shift in the asymmetry.
The helicity-averaged yields and helicity-correlated asymmetries were then output
to ROOT~\cite{root} files and accumulated totals were stored in a MySQL~\cite{mysql} database for each
5-minute runlet.

Two different techniques were used to determine the correlation of the detector signals to the beam parameters.  
The first used linear regression analysis of the natural beam motion in each 5-minute runlet to determine the correlation matrix between the detector asymmetries and the \ac{acro-hc} variation in a set of beam parameters.  
The matrix was then inverted to extract the sensitivity to those beam parameters for each detector.  
Then the detector asymmetries were corrected for the \ac{acro-hc} variation in the beam parameters, using the sensitivities extracted from that runlet.
A total of 13 different sets of independent parameters were used to evaluate the effect of choosing different beam monitors on the corrected detector asymmetries.

The second method to determine the correlation of the detector signals to the beam parameters used driven beam modulation data, as described in Sec.~\ref{sec:Beam:BMOD}.  
One advantage of these driven modulations was that they were largely uncoupled.  In particular,
the beam modulation allowed separation of the effects of energy and steering changes which were difficult to extract from the natural beam motion data.

\section{Summary}
\label{sec:Summary}

As discussed in this article, high-precision parity-violation measurements offer unique challenges from both a methodology and a technical perspective.  The apparatus is essentially the entire accelerator complex, consisting of the polarized injector, accelerator, beam property measurement apparatus, scattering target, spectrometer, and detector assembly. The small asymmetries and high precision characterizing these experiments inevitably lead to high-luminosity and high-rate environments, where data are typically recorded for each helicity state as opposed to a trigger based on each scattered electron.  Detector multiplicity is important to increase the detected rate, as well as to form an azimuthally symmetric array to reduce systematic errors from \ac{acro-hc} changes in the beam trajectory and potential contamination from transverse asymmetries.

The \Qweak experiment achieved a number of notable technical milestones. These include the highest luminosity, highest beam current on a cryogenic target, and the smallest absolute precision ever achieved in a \ac{acro-pves} measurement.   
It was the first \ac{acro-pves} measurement that required a multi-kilowatt LH$_2$ target that, if not for its unique design via computational fluid dynamics simulations and the use of rapid helicity reversal, would have been a limiting noise source for the measurement.

\begin{figure}[!hhhbtb]
\centering
\includegraphics[width=0.5\textwidth]{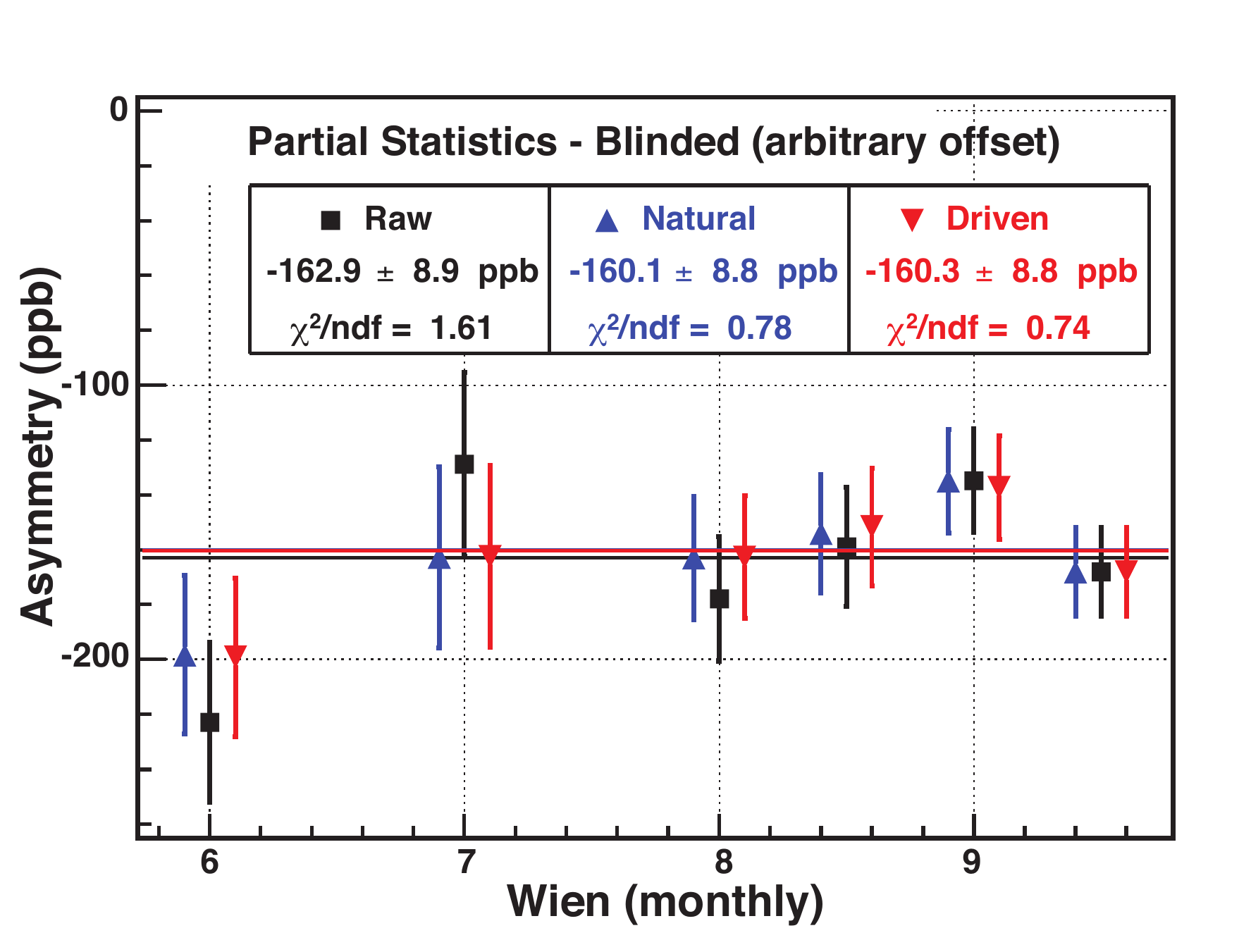}
\caption{\label{Null}
 Subset of the Run 2 production data showing the blinded asymmetry (in ppb) grouped by ($\sim$monthly) Wien state and corrected using two different approaches to determine the  sensitivities of the apparatus to \ac{acro-hc} beam properties that can give rise to  false asymmetries. Other needed corrections are not applied to the data in this figure, as discussed in the text. The results without any correction (solid squares) are compared to the results after correction using the intrinsic random variations in beam properties (Natural motion:  upward pointing triangles) and to the results using the driven beam motion (Driven motion:  
 downward pointing triangles) discussed in Sec.~\ref{sec:Beam:BMOD} where the sensitivities are derived by actively modulating each property of the beam with a magnitude significantly larger than intrinsically carried by the beam. The asymmetries derived using each technique are consistent with each other, and the overall correction for \ac{acro-hcba}s is small.  The data shown here represent the 80\% of Run 2 data for which driven motion was available. Run 1 provides an additional $\sim$1/3 of the total data acquired in the experiment. }
\end{figure}

The final metric of both the instrumentation and methodology is the degree to which all sub-systems were able to perform together at the required level. 
Fig.~\ref{Null} illustrates the consistency of the experiment's measured asymmetry, where
the sign of the various fast and slow helicity flips has been properly accounted for.
These results include a global, additive blinding factor so they cannot as yet be compared to predictions. They also are not yet corrected for beam polarization, target window background, or various other small backgrounds and kinematic corrections. 
However, one clearly sees that the corrections to the asymmetry due to \ac{acro-hc} variations in the beam parameters are quite small, and that they are consistent for the two independent methods used to determine the sensitivities to the beam parameters (natural beam motion and driven beam modulation, as described in Sec.~\ref{sec:Beam:BMOD}). It is also seen that the extracted asymmetry is stable with time over the scale of 
months.
The results are evidence that the \Qweak physics asymmetry measurement is fundamentally sound at the few parts per billion level.
  
\section{Acknowledgments}
\label{sec:Acknowledgments}

This material is based upon work supported by the U.S. Department of Energy, Office of Science, Office of Nuclear Physics under contract DE-AC05-06OR23177,
under which Jefferson Science Associates, LLC operates Thomas Jefferson National Accelerator Facility. Construction and operating funding for the experiment was provided through  the US Department of Energy (DOE), the Natural Sciences and Engineering Research Council of Canada (NSERC), and the National Science Foundation (NSF) with university matching contributions from the College of William and Mary, Virginia Tech, George Washington University, and Louisiana Tech University. We wish to thank the staff of \ac{acro-jlab}, \ac{acro-triumf}, and \ac{acro-bates} for their vital support during this challenging experiment. 
In particular we wish to thank the many technical and operations staff at Jefferson Lab, without whose expertise and support the experiment would not have been possible. We are grateful to A.~Kenyon for his skillful guidance of the experiment's installation. 
We acknowledge helpful contributions from William and Mary students  J.~Bufkin, C.~Capuano, E.~Epperson, G.~Giovanetti, A.~Gvakharia, L.J.~Snow,  B.P.~Walsh,  and A.~Watson, Virginia Tech students J.~Hoffman and J.~Walters, Hendrix College students D.~Cargill, V.~Gammill, K.~Garimella, U.~Garimella, N.~Heiner, E.~Holcomb, R.~Leonard, T.~Pote, G.~Trees,  S.~Webb, and T.~Webb, and University of Manitoba student C.~Koop. 
We are also indebted to  
P.G.~Blunden, J.~Erler, N.L.~Hall,  W.~Melnitchouk, M.J.~Ramsey-Musolf, and A.W.~Thomas  for many useful discussions.

\section{References}
\label{sec:Referencess}


\newcommand{\etal}{\textit{et al.}}
\newcommand{\Journal}[4]{#1 \textbf{#2} (#4) #3} 

\newcommand{\NIMA}{Nuclear Instruments and Methods in Physics Research A}
\newcommand{\NPB}{Nuclear Physics B}
\newcommand{\PLB}{Physics Letters B}
\newcommand{\PRL}{Physical Review Letters}
\newcommand{\PRD}{Physical Review D}
\newcommand{\ZPC}{Zeitschrift f{\"u}r Physik C Particles and Fields}
\newcommand{\AIPC}{AIP Conference Proceedings}
\newcommand{\NUCC}{Il Nuovo Cimento C}
\newcommand{\PRSTAB}{Physical Review Special Topics: Accelerator Beams}
\newcommand{\CONFP}{Conference Proceedings}
\newcommand{\ARNPS}{Annual Reviews of Nuclear Particle Science}
\newcommand{\APB}{Applied Physics B}
\newcommand{\HFI}{Hyperfine Interactions}
\newcommand{\IEEENUC}{IEEE Transactions on Nuclear Science}
\newcommand{\JOSAMA}{Journal of the Optical Society of America A}
\newcommand{\PACCEL}{Particle Accelerators}

\bibliographystyle{nim/model1-num-names}

\end{document}

%% file: authors.tex
\author[JLab]{T.~Allison}
\author[Manitoba]{M.~Anderson\fnref{mitchell}} 
\author[Zagreb]{D.~Androi\'c} 
\author[WM]{D.S.~Armstrong} 
\author[Yerevan]{A.~Asaturyan} 
\author[WM]{T.D.~Averett}
\author[MIT]{R.~Averill} 
\author[MIT]{J.~Balewski} 
\author[JLab]{J.~Beaufait}
\author[Ohio]{R.S.~Beminiwattha} 
% ...

\address[JLab]{Thomas Jefferson National Accelerator Facility, Newport News, VA 23606 USA}
\address[Manitoba]{University of Manitoba, Winnipeg, MB R3T2N2 Canada}
\address[Zagreb]{University of Zagreb, Zagreb, HR 10002 Croatia}
\address[WM]{College of William and Mary, Williamsburg, VA 23185 USA}
\address[Yerevan]{A.~I.~Alikhanyan National Science Laboratory (Yerevan Physics Institute), Yerevan 0036, Armenia}
\address[MIT]{Massachusetts Institute of Technology,  Cambridge, MA 02139 USA} 
\address[Ohio]{Ohio University, Athens, OH 45701 USA}
% ...


\cortext[cor1]{Corresponding author: G.R.~Smith, smithg@jlab.org, +1 757 269-5405}
\fntext[fatiha]{Present address: Duquesne University, Pittsburgh, PA 15282 USA}
\fntext[dalton]{Present address: Thomas Jefferson National Accelerator Facility, Newport News, VA 23606 USA}
\fntext[jason]{Present address: Massachusetts Institute of Technology,  Cambridge, MA 02139 USA}
\fntext[finn]{deceased}
\fntext[grimm]{Present address: Berthold Technologies, Bad Wildbad, Germany}
\fntext[leckey]{Present address: Indiana University, Bloomington, Indiana 47405 USA}
\fntext[lee]{Present address: Institute for Basic Science, Daejeon, South Korea}
\fntext[mahurin]{Present address: Middle Tennessee State University, Murfreesboro, TN 37132}
\fntext[mei]{Present address: CGG Veritas, Houston, TX 77081 USA}
\fntext[mammei]{Present address: University of Manitoba, Winnipeg, MB R3T2N2 Canada}
\fntext[myers]{Present address: Rutgers, the State University of New Jersey, Piscataway, NJ 08854 USA}
\fntext[mitchell]{Present address: ICMP, 
\'{E}cole Polytechnique F\'{e}d\'{e}rale de Lausanne (EPFL), 1015 Lausanne, Switzerland}
\fntext[urban]{Present address: University of California, Berkeley, CA 94720 USA}

%% file: abstract.tex
\begin{abstract}
The Jefferson Lab  $Q{\rm _{weak}}$ experiment determined the weak charge of the proton by measuring the parity-violating elastic scattering asymmetry of longitudinally polarized  electrons from an unpolarized liquid hydrogen target at small momentum transfer.
A custom  apparatus was designed for this experiment to meet the technical challenges presented by the smallest and most precise ${\vec{e}}$p asymmetry ever measured. 
Technical milestones were achieved at Jefferson Lab in target power, beam current, beam helicity reversal rate, polarimetry, detected rates, and control of helicity-correlated beam properties.
The experiment employed 180 \muA of 89\% longitudinally polarized electrons whose helicity was reversed 960 times per second. 
The electrons were accelerated to 1.16\,GeV and directed to a beamline with extensive instrumentation to measure helicity-correlated beam properties that can induce false asymmetries.  M\o{}ller and Compton polarimetry were used to measure the electron beam polarization to 
better than 1\%.  The electron beam was incident on a 34.4 cm liquid hydrogen target. 
After passing through a triple collimator system,  scattered electrons between 5.8$^\circ$ and 11.6$^\circ$ were bent in the toroidal magnetic field of a resistive copper-coil magnet.
The electrons inside this acceptance were
focused onto eight fused silica
\v{C}erenkov detectors arrayed symmetrically around the beam axis. 
A total scattered electron rate of about 7 GHz was incident on the detector array. The detectors were read out in integrating mode by custom-built  low-noise pre-amplifiers and 18-bit sampling ADC modules.  The momentum transfer $Q^2$ = 0.025 GeV$^2$ was determined using dedicated low-current ($\sim$100 pA) measurements with a set of  drift chambers before (and a set of  drift chambers and trigger scintillation counters after) the toroidal magnet.
\end{abstract}

%% file: acronyms.tex
%
\acrodef{acro-ac}[AC]{Alternating Current}
\acrodef{acro-adc}[ADC]{Analog to Digital Converter}
\acrodef{acro-bcm}[BCM]{Beam Charge Monitor}
\acrodef{acro-blast}[BLAST]{MIT/Bates Large Acceptance Spectrometer Toroid}
\acrodef{acro-bpm}[BPM]{Beam Position Monitor}
\acrodef{acro-cad}[CAD]{Computer Aided Design}
\acrodef{acro-ccd}[CCD]{Charge Coupled Device}
\acrodef{acro-cebaf}[CEBAF]{Continuous Electron Beam Accelerator Facility}
\acrodef{acro-cfd}[CFD]{Computational Fluid Dynamics}
\acrodef{acro-chl}[CHL]{Central Helium Liquefier}
\acrodef{acro-coda}[CODA]{Online Data Acquisition}
\acrodef{acro-cpu}[CPU]{Central Processing Unit}
\acrodef{acro-dac}[DAC]{Digital to Analog Converter}
\acrodef{acro-daq}[DAQ]{Data Acquisition}
\acrodef{acro-dc}[DC]{Direct Current}
\acrodef{acro-dcct}[DCCT]{DC Current Transformer}
\acrodef{acro-dd}[DD]{Double Difference}
\acrodef{acro-coda-eb}[EB]{Event Builder}
\acrodef{acro-ecl}[ECL]{Emitter Coupled Logic}
\acrodef{acro-em}[EM]{Electro-Magnetic}
\acrodef{acro-epics}[EPICS]{Experimental Physics and Industrial Control}
\acrodef{acro-coda-er}[ER]{Event Recorder}
\acrodef{acro-esr}[ESR]{End Station Refrigerator}
\acrodef{acro-coda-et}[ET]{Event Transfer}
\acrodef{acro-inj-fc}[FC]{Faraday Cup}
\acrodef{acro-fpga}[FPGA]{Field Programmable Gate Array}
\acrodef{acro-fwhm}[FWHM]{Full Width at Half Maximum}
\acrodef{acro-gui}[GUI]{Graphical User Interface}
\acrodef{acro-hc}[HC]{Helicity-Correlated}
\acrodef{acro-hcba}[HCBA]{Helicity-Correlated Beam Asymmetry}
\acrodef{acro-hdc}[HDC]{Horizontal Drift Chamber}
\acrodef{acro-hv}[HV]{High Voltage}
\acrodef{acro-hx}[HX]{Heat eXchanger}
\acrodef{acro-ia}[IA]{Charge Asymmetry}
\acrodef{acro-iac}[IAC]{Intensity Asymmetry Cell}
\acrodef{acro-ihwp}[IHWP]{Insertable Half Wave Plate}
\acrodef{acro-i2v}[I-to-V]{Current to Voltage}
\acrodef{acro-jlab}[JLab]{Thomas Jefferson National Accelerator Facility}
\acrodef{acro-led}[LED]{Light Emitting Diode}
\acrodef{acro-lh2}[LH2]{Liquid Hydrogen}
\acrodef{acro-linac}[linac]{linear accelerator}
\acrodef{acro-lumis}[lumis]{Luminosity Monitors}
\acrodef{acro-lv1}[LV1]{LeVel 1}
\acrodef{acro-lvds}[LVDS]{Low-Voltage Differential Signal}
\acrodef{acro-mcc}[MCC]{Machine Control Center}
\acrodef{acro-md}[MD]{Main Detector}
\acrodef{acro-bates}[MIT/BATES]{Bates Linear Accelerator Center}
\acrodef{acro-mosfet}[MOSFET]{Metal Oxide Semiconductor Field-Effect Transistor}
\acrodef{acro-mps}[MPS]{Macro Pulses}
\acrodef{acro-mysql}[MySQL]{My Structured Query Language}
\acrodef{acro-pac}[PAC]{Program Advisory Committee}
\acrodef{acro-pe}[p.e.]{Photo-Electron}
\acrodef{acro-pid}[PID]{Proportional-Integral-Differential}
\acrodef{acro-pita}[PITA]{Polarization Induced Transport Asymmetry}
\acrodef{acro-pmt}[PMT]{Photo Multiplier Tube}
\acrodef{acro-ppb}[ppb]{parts per billion}
\acrodef{acro-ppln}[PPLN]{Periodically Poled Lithium Niobate}
\acrodef{acro-ppm}[ppm]{parts per million}
\acrodef{acro-pves}[PVES]{Parity-Violating Electron Scattering}
\acrodef{acro-pzt}[PZT]{Piezoelectric Transducer}
\acrodef{acro-qe}[QE]{Quantum Efficiency}
\acrodef{acro-qed}[QED]{Quantum ElectroDynamics}
\acrodef{acro-qtor}[QTOR]{Q$_{\rm {weak}}$ Toroid}
\acrodef{acro-qwad}[QWAD]{Q$_{\rm {weak}}$ Amplifier-Discriminator}
\acrodef{acro-r1}[R1]{Region 1}
\acrodef{acro-r2}[R2]{Region 2}
\acrodef{acro-r3}[R3]{Region 3}
\acrodef{acro-rf}[RF]{Radio Frequency}
\acrodef{acro-rhwp}[RHWP]{Rotatable Half Wave Plate}
\acrodef{acro-rms}[RMS]{Root-Mean-Square}
\acrodef{acro-roc}[ROC]{Read-Out-Controller}
\acrodef{acro-sc}[SC]{Super-Conducting}
\acrodef{acro-shg}[SHG]{Second Harmonic Generation}
\acrodef{acro-sm}[SM]{Standard Model}
\acrodef{acro-s-n}[S/N]{Signal to Noise ratio}
\acrodef{acro-srf}[SRF]{superconducting radio frequency}
\acrodef{acro-ss}[SSt]{Stainless Steel}
\acrodef{acro-stp}[STP]{Standard Temperature and Pressure}
\acrodef{acro-tdc}[TDC]{Time to Digital Converter}
\acrodef{acro-triumf}[TRIUMF]{TRI-University Meson Facility}
\acrodef{acro-trs}[TRS]{Trigger Supervisor}
\acrodef{acro-ts}[TS]{Trigger Scintillator}
\acrodef{acro-vdc}[VDC]{Vertical Drift Chamber}
\acrodef{acro-vme}[VME]{VERSAModule Eurocard}
\acrodef{acro-vqwk}[VQWK]{Q$_{\rm {weak}}$ 18-bit sampling ADC}